\documentclass[english,american]{article}
\usepackage[T1]{fontenc}
\usepackage[latin9]{inputenc}
\usepackage[unicode=true,pdfusetitle,
 bookmarks=true,bookmarksnumbered=false,bookmarksopen=false,
 breaklinks=false,pdfborder={0 0 1},backref=false,colorlinks=false]
 {hyperref}
\usepackage{geometry}
\usepackage{babel}
\usepackage{array}
\usepackage{amsmath}
\usepackage{amssymb}
\usepackage{graphicx}
\usepackage{esint}
\usepackage[numbers]{natbib}
\usepackage{breakurl}

\newcommand\myref{\refstepcounter{equation}\theequation}
\newcommand{\refmyref}[1]{\newcounter{#1}\setcounter{#1}{\theequation}}

\usepackage{color}
\usepackage[makeroom]{cancel}
\usepackage[normalem]{ulem}
\definecolor{pkcolor}{rgb}{0,0.1,0.7}
\definecolor{ascolor}{rgb}{1,0,1}

\newcommand\pkout{\marginpar{\color{pkcolor}$\clubsuit$}\bgroup\markoverwith{\color{pkcolor}{\rule[0.4ex]{2pt}{0.8pt}}}\ULon}
\newcommand\asout{\marginpar{\color{ascolor}$\heartsuit$}\bgroup\markoverwith{\color{ascolor}{\rule[0.4ex]{2pt}{0.8pt}}}\ULon}

{\catcode`\@=11 \uccode`9=`\l \uccode`8=`\o %
 \uppercase{\gdef\striplong@#1#2#3#4\relax{%
  \ifx9#2\ifx8#3\@xp\@xp\@xp\@xp\@xp\@xp\@xp\zap@to@space\fi\fi}}}

\begin{document}

\title{Wilson lines in the MHV action\date{}}

\author{P. Kotko and A.~M.~ Stasto\\\\ \it\small The Pennsylvania State University,
Physics Department\\ \it\small 104 Davey Lab, University Park, PA
16802, USA}
\maketitle
\begin{abstract}
The MHV action is the Yang-Mills action quantized on the light-front,
where the  two explicit physical gluonic degrees of freedom have been
canonically transformed to a new set of fields. 
This transformation leads to the action with vertices being off-shell continuations of the MHV amplitudes.
We show that the solution to the field transformation expressing one
of the  new fields in terms of the Yang-Mills field is a certain
type of the Wilson line. More precisely, it is a straight infinite gauge link
with a slope extending to the light-cone minus and the transverse
direction. One of the consequences of that fact is that certain MHV
vertices reduced partially on-shell are gauge invariant -- a fact
discovered before using conventional light-front perturbation theory.
We  also analyze the diagrammatic content of the field transformations
leading to the MHV action. We found that the diagrams for the solution
to the transformation (given by the Wilson line) and its inverse differ
only by light-front energy denominators. Further, we investigate the coordinate
space version of the inverse solution to the one given by the Wilson
line. We find an explicit expression given by a power series in fields. We also
 give a geometric interpretation to it by means of a specially defined vector field. Finally, we discuss
the fact that the Wilson line solution to the transformation is directly
related to the all-like helicity gluon wave function, while the inverse
functional is a generating functional for solutions of self-dual
Yang-Mills equations.
\end{abstract}
\newpage 

\tableofcontents{}

\newpage

\section{Introduction}

\label{sec:Intro}

Through the recent decades there has been  great progress in the calculation of the scattering amplitudes in QCD. This is of course indispensable for practical applications, see e.g. \citep{Freitas2016}, and especially in context of the plethora of QCD and electroweak measurements performed at Large Hadron Collider, such as multi-jet production or processes including a production of $W$ and $Z$ bosons. 
However, amplitudes are also interesting  from the theoretical point of view. It turns out that they encode variety of hidden aspects of the theory, even at the tree level.
 For example, all gluonic
amplitudes with two helicity projections different than all other,
so-called maximally helicity violating (MHV) amplitudes, appear to
have an extremely simple one-term form when expressed in terms of spinor
products \citep{Parke:1986gb,Berends:1987me} (see \citep{Mangano:1990by}
for a review of the spinor helicity methods). Since in the ordinary approach
based on the Feynman diagrams this simplicity is  not evident 
(the number of diagrams grows significantly with the increase of the
number of legs), the intuition suggests that there has to be a
hidden structure or symmetry. Indeed, in the last two decades the
understanding of this result in terms of abstract geometry has enormously
developed: from the formulation of amplitudes in the twistor space
\citep{Witten2004} to new mathematical structures like the amplituhedron
\citep{Arkani-Hamed2014}. Somewhat in parallel, more `standard' but
powerful methods based on a generic analytic structure of the Feynman
diagrams have also been developed (see e.g. \citep{Henn} for a review
of tree-level and loop-level techniques). One of those is the Britto-Cachazo-Feng-Witten
(BCFW) recursion relation  \citep{Britto:2004ap,Britto:2005fq}. In this recursion, which utilizes analytical properties of the scattering amplitudes,  one uses lower on-shell amplitudes with deformed complex momenta to compute the on-shell amplitudes with higher number of legs.  Another, somewhat
related method but truly motivated by the geometry is the one proposed
by Cachazo, Svrcek and Witten (CSW) \citep{Cachazo2004}. Since this
is the method that provides the motivation to the present work, let
us briefly remind its main elements.

The building blocks of the CSW method are so-called MHV vertices,
i.e. MHV amplitudes which are continued off-shell. It turns out that any amplitude
can be constructed by combining such vertices using scalar propagators.
This twistor-space motivated conjecture has been directly proved by
means of the BCFW recursion \citep{Britto:2005fq,Risager2005}. In
parallel, another beautiful derivation of the method was found; in
\citep{Mansfield2006} a canonical transformation of fields was proposed
that transforms the light-front quantized Yang-Mills action into a
new action which explicitly contains the off-shell MHV vertices. In what
follows, we shall refer to that new action as the `MHV action'. One
of the early issues of that formalism was that all-like helicity amplitudes,
which are zero at the tree level but non-zero in general, were not accessible
from the action. This issue was later solved in \citep{Ettle2006b,Ettle2007,Brandhuber2007,Ettle2008}.
Further developments  in this direction included an explicit construction of the twistor action which leads to the generation of the MHV diagrams \citep{Boels:2007qn}. 

In the present work we reconsider the field transformation used in
\citep{Mansfield2006} and show that its solution can be expressed
by a particular form of the Wilson line (for a review of Wilson lines
in quantum field theory see \citep{Cherednikov2014}). This seems
to be overlooked in the literature, although the momentum space solution
given by a power expansion in the fields is known. This might be due to the fact
that the most common usage of the Wilson lines utilizes the light-like (or transverse
at the light-cone infinity) straight paths. Here, however, the straight
path is defined in the complexified Minkowski space and extends to the
light-cone minus and transverse directions. Also, the slope of this Wilson line is integrated over. Since the field transformation
is reversible it makes sense to ask about the inverse solution. It
is also known in momentum space \citep{Ettle2006b}, but since the
Wilson line is defined in the position space it is interesting to construct
also its position space inverse. The inverse solution is interesting
also because it is a generating functional for solutions of the self-dual
Yang-Mills equations \citep{Bardeen1996,Rosly1997,Gorsky2006}.

An immediate consequence of the fact that the MHV action contains
infinite Wilson lines is that one can construct gauge invariant off-shell
currents which have the MHV form. Interestingly, such objects were
found independently in \citep{Cruz-Santiago2013} using the ordinary
light-front perturbation theory \citep{Kogut1970,Brodsky1998}. Later
it was shown, without referring to the CSW method or MHV action, that
such gauge invariant MHV currents originate in a matrix element of
the Wilson line \citep{Cruz-Santiago2015,Kotko2016}. It turns out that
it is the same Wilson line as the one derived here as a solution to
canonical field transformation.  We shall discuss these and related topics in detail in this work. Also, using the light-front perturbation theory we shall elucidate the diagrammatic content of the Wilson line solution as well as the inverse transformation.

The structure of the paper is as follows.  In Section~\ref{sec:Notation} we set up the notations and conventions used throughout this work. In section~\ref{sec:MHV_action}  we recall the field transformation used to derive the MHV Lagrangian. This section contains mostly known results and is thus introductory. The following sections explore
the concepts described above. The next  Section~\ref{sec:WilsonLineSolution}
contains one of the main results of the work, i.e. the proof of the
Wilson line solution. In Section~\ref{sec:GaugeInvAmps} we discuss the construction of the gauge invariant off-shell currents which satisfy the Ward identities, using the Wilson line  defined previously. In Section~\ref{sec:Diagramma} we discuss the   diagrammatic content of both the transformation and its inverse. In Section~\ref{sec:Inverse_transf} we discuss the inverse transformation in the coordinate space. The geometric interpretation of both transformations in terms of a specially defined vector field in two dimensional space is discussed in Section~\ref{sec:Geometry}. Finally, in Section ~\ref{sec:Summary} we summarize the results and present our conclusions.

\section{Notation}

\label{sec:Notation}

We start by introducing the notation and conventions used throughout
the paper.

The light cone basis is defined by the two lightlike four-vectors
\begin{gather}
\eta=\frac{1}{\sqrt{2}}\left(1,0,0,-1\right)\,,\,\,\,\,\tilde{\eta}=\frac{1}{\sqrt{2}}\left(1,0,0,1\right)\, ,\label{eq:etavec}
\end{gather}
and two space like complex four-vectors spanning the transverse plane
\begin{equation}
\varepsilon_{\perp}^{\pm}=\frac{1}{\sqrt{2}}\left(0,1,\pm i,0\right)\,.\label{eq:epsPlMin}
\end{equation}
The contravariant coordinates of a four-vector $v$ in that basis
are denoted as follows
\begin{gather}
v^{+}=v\cdot\eta\,,\,\,\,\, v^{-}=v\cdot\tilde{\eta}\,,\label{eq:plusmindef}\\
v^{\bullet}=v\cdot\varepsilon_{\bot}^{+}\,,\,\,\,\, v^{\star}=v\cdot\varepsilon_{\bot}^{-}\,.\label{eq:zzbardef}
\end{gather}
The covariant components are
\begin{gather}
v_{+}=v^{-},\,\,\, v_{-}=v^{+}\,,\\
v_{\bullet}=-v^{\star},\,\,\, v_{\star}=-v^{\bullet}\,.
\end{gather}
The explicit expansion of a four-vector $v$ reads
\begin{equation}
v=v^{+}\tilde{\eta}+v^{-}\eta-v^{\star}\varepsilon_{\perp}^{+}-v^{\bullet}\varepsilon_{\perp}^{-}\,.
\end{equation}
The scalar product can be thus written as%
\footnote{Our notation corresponds to the one used in \citep{Mansfield2006}
as follows: $v^{\bullet}\leftrightarrow v_{z}$, $v^{\star}\leftrightarrow v_{\overline{z}}$, while the plus and minus components stay the same. We found the notation
with $z$ and $\overline{z}$ as subscripts somewhat cumbersome, especially when
raising and lowering the indices. %
}
\begin{equation}
u\cdot v=u^{+}w^{-}+u^{-}w^{+}-u^{\bullet}w^{\star}-u^{\star}w^{\bullet}\,.\label{eq:scalarprod}
\end{equation}

The gauge field in the fundamental representation is denoted as 
\begin{equation}
\hat{A}^{\mu}=A_{a}^{\mu}t^{a}\,,
\end{equation}
where $t^{a}$ are color generators satisfying 
\begin{equation}
\left[t^{a},t^{b}\right]=i\sqrt{2}f^{abc}t^{c}\,,
\end{equation}
and 
\begin{equation}
\mathrm{Tr}\left(t^{a}t^{b}\right)=\delta^{ab}\,.
\end{equation}
We shall often use the rescaled coupling constant
\begin{equation}
g'=\frac{g}{\sqrt{2}}\,.
\end{equation}

Since we shall work within a theory quantized on equal light-cone
time surface, defined by $x^{+}$, it is convenient to introduce a three
vector
\begin{equation}
\mathbf{x}\equiv\left(x^{-},x^{\bullet},x^{\star}\right)\, ,
\end{equation}
and the corresponding measure
\begin{equation}
d^{3}\mathbf{x}=dx^{-}dx^{\bullet}dx^{\star}\,.
\end{equation}
We reserve $\mathbf{x},\,\mathbf{y},\,\mathbf{z}$ (with possible
subscripts) to denote three-vectors in position space. Similarly,
we introduce momentum space three-vectors $\mathbf{p},\,\mathbf{q},\,\mathbf{r}$
(with possible subscripts):
\begin{equation}
\mathbf{p}\equiv\left(p^{+},p^{\bullet},p^{\star}\right)\, ,
\end{equation}
and the corresponding measure
\begin{equation}
d^{3}\mathbf{p}=\frac{dp^{+}dp^{\bullet}dp^{\star}}{\left(2\pi\right)^{3}}\,.
\end{equation}
Above, we have included the $\left(2\pi\right)^{3}$ factor in the
measure in order to avoid a proliferation of these factors when making
the Fourier transforms.

We shall use the following notation and definition of the Fourier
transform of fields:
\begin{equation}
A\left(x^{+},\mathbf{x}\right)=\int d^{3}\mathbf{p}\, e^{-i\mathbf{x}\cdot\mathbf{p}}\tilde{A}\left(x^{+},\mathbf{p}\right)\,,\label{eq:FT_def}
\end{equation}
for fixed $x^{+}$. We shall almost always suppress the $x^{+}$ argument
and write simply
\begin{equation}
A\left(x^{+},\mathbf{x}\right)\equiv A\left(\mathbf{x}\right),\,\,\,\textrm{or}\,\,\,\tilde{A}\left(x^{+},\mathbf{p}\right)\equiv\tilde{A}\left(\mathbf{p}\right).\,
\end{equation}

Let us now introduce the essential part of the notation which is equivalent
to the spinor algebra. For given four momenta $p,q$ we define the
following variables \citep{Motyka2009}:
\begin{equation}
v_{\left(q\right)\left(p\right)}=\frac{q^{\star}}{q^{+}}-\frac{p^{\star}}{p^{+}},\;\;\;\; v_{\left(q\right)\left(p\right)}^{*}=\frac{q^{\bullet}}{q^{+}}-\frac{p^{\bullet}}{p^{+}}\,.\label{eq:v_def}
\end{equation}
Note, that these symbols are equally well defined for three-vectors
$\mathbf{p},\mathbf{q}$. We shall thus not distinguish those situations. The real version of these variables have a very natural interpretation within the light-front perturbation theory. Since $p^+$ component on the light-front can be interpreted as a `mass', the variable  $\frac{p_{\perp}}{p^+}$ can be therefore thought of as the `velocity' in the 2-dimensional space. 
In addition, it will appear useful to consider the following non anti-symmetric
modification of those symbols \citep{Cruz-Santiago2015}:
\begin{equation}
\tilde{v}_{\left(q\right)\left(p\right)}=q^{+}v_{\left(p\right)\left(q\right)}=-q^{\star}+q^{+}\frac{p^{\star}}{p^{+}} \, ,\,\label{eq:vtild_def}
\end{equation}
and similar for $\tilde{v}_{\left(q\right)\left(p\right)}^{*}$. In
particular, these variables can be expressed as follows
\begin{equation}
\tilde{v}_{\left(q\right)\left(p\right)}=q\cdot\varepsilon_{p}^{+},\,\,\,\,\tilde{v}_{\left(q\right)\left(p\right)}=q\cdot\varepsilon_{p}^{-},\,
\end{equation}
where $\varepsilon_{p}^{\pm}$ are polarization vectors defined as
\begin{equation}
\varepsilon_{p}^{\pm}=\varepsilon_{\perp}^{\pm}-\frac{p\cdot\varepsilon_{\perp}^{\pm}}{p^{+}}\eta\,.\label{eq:PolarizationVect}
\end{equation}
In our applications we shall typically deal with a set of momenta
$p_{1},\dots,p_{i}$ (or $\mathbf{p}_{1},\dots,\mathbf{p}_{i}$).
For that case we will write, for example,
\begin{equation}
\tilde{v}_{\left(p_{i}\right)\left(p_{j}\right)}\equiv\tilde{v}_{ij},\,\,\,\tilde{v}_{\left(p_{1\dots i}\right)\left(p_{1\dots j}\right)}\equiv\tilde{v}_{\left(1\dots i\right)\left(1\dots j\right)},\,\,\,\textrm{etc.}\,,
\end{equation}
where
\begin{equation}
p_{1\dots i}=\sum_{k=1}^{i}p_{k}\,.\label{eq:momentum_sum_def}
\end{equation}
 The symbols $\tilde{v}_{ij}$ satisfy several useful identities which
we collect in Appendix~\ref{sec:App_ident}.

As mentioned, the above symbols are directly related to the spinor
products:
\begin{equation}
\left\langle ij\right\rangle =-\sqrt{\frac{2p_{j}^{+}}{p_{i}^{+}}}\,\tilde{v}_{ij},\;\;\;\left[ij\right]=-\sqrt{\frac{2p_{j}^{+}}{p_{i}^{+}}}\,\tilde{v}_{ij}^{*}\,,\label{eq:spinor_prod}
\end{equation}
where
\begin{equation}
\left\langle ij\right\rangle =\epsilon^{\alpha\beta}\lambda_{\alpha}\left(k_{i}\right)\lambda_{\beta}\left(k_{j}\right),\,\,\,\left[ij\right]=\epsilon^{\dot{\alpha}\dot{\beta}}\tilde{\lambda}_{\dot{\alpha}}\left(k_{i}\right)\tilde{\lambda}_{\dot{\beta}}\left(k_{j}\right)\,.
\end{equation}
Above $\lambda_{\alpha}\left(k\right)$ and $\tilde{\lambda}_{\dot{\alpha}}\left(k\right)$
are Weyl spinors of positive and negative chirality. The reason for
using the symbols $\tilde{v}_{ij}$ instead of spinor products will
become clear later.

\section{The MHV action}

\label{sec:MHV_action}

As mentioned in the Introduction, the CSW method \citep{Cachazo2004}
was motivated by the twistor theory and later proved by means of the
more standard methods based on analytic structure of amplitudes \citep{Britto:2005fq,Risager2005}.
It was however desirable to construct an explicit action for the MHV
vertices. This task was accomplished in \citep{Mansfield2006} by
means of certain field transformation performed on the light-front
Yang-Mills action. Since the properties of the transformation and
the resulting MHV action are central to the present work, we shall
now review the main concepts. In order to keep everything consistent
(and for educational purposes) we have rederived the MHV action from
the scratch, using the notation described in the previous section.
For interested readers less familiar with the subject we present more
detailed elements of our derivation in Appendix~\ref{sec:App_MHVaction}.

The Yang-Mills action is defined as
\begin{equation}
S_{\mathrm{Y-M}}=-\frac{1}{4}\int d^{4}x\,\mathrm{Tr}\, F^{\mu\nu}F_{\mu\nu}\,.
\end{equation}
Here, $F^{\mu\nu}=\frac{i}{g'}\left[\mathcal{D}^{\mu},\mathcal{D}^{\nu}\right]$
with $\mathcal{D}^{\mu}=\partial^{\mu}-ig'\hat{A}^{\mu}$. The action can be rewritten
in terms of the physical degrees of freedom involving two transverse gluon
polarizations alone \citep{Scherk1975}. It is done as follows (see
also \citep{Mansfield2006} for this operation in the present context
of the MHV vertices). First we rewrite the Yang-Mills action using
the light cone coordinates (\ref{eq:plusmindef}) and transverse coordinates
(\ref{eq:zzbardef}) (the fact that we use the complexified transverse
plane is not essential for this step; one could use just Cartesian
components as well). Next, we choose the light-cone
gauge $A\cdot\eta=A^{+}=0$. The Yang-Mills action becomes quadratic in
$A^{-}$, which allows to integrate this field out of the path integral.
The resulting action can be written as 
\begin{equation}
S_{\mathrm{Y-M}}^{\left(\mathrm{LC}\right)}\left[A^{\bullet},A^{\star}\right]=\int dx^{+}\left(\mathcal{L}_{+-}^{\left(\mathrm{LC}\right)}+\mathcal{L}_{++-}^{\left(\mathrm{LC}\right)}+\mathcal{L}_{+--}^{\left(\mathrm{LC}\right)}+\mathcal{L}_{++--}^{\left(\mathrm{LC}\right)}\right)\,,\label{eq:actionLC}
\end{equation}
where
\begin{gather}
\mathcal{L}_{+-}^{\left(\mathrm{LC}\right)}\left[A^{\bullet},A^{\star}\right]=-\int d^{3}\mathbf{x}\,\mathrm{Tr}\,\hat{A}^{\bullet}\square\hat{A}^{\star}\,,\\
\mathcal{L}_{++-}^{\left(\mathrm{LC}\right)}\left[A^{\bullet},A^{\star}\right]=-2ig'\,\int d^{3}\mathbf{x}\,\mathrm{Tr}\,\gamma_{\mathbf{x}}\hat{A}^{\bullet}\left[\partial_{-}\hat{A}^{\star},\hat{A}^{\bullet}\right]\,,\\
\mathcal{L}_{--+}^{\left(\mathrm{LC}\right)}\left[A^{\bullet},A^{\star}\right]=-2ig'\,\int d^{3}\mathbf{x}\,\mathrm{Tr}\,\overline{\gamma}_{\mathbf{x}}\hat{A}^{\star}\left[\partial_{-}\hat{A}^{\bullet},\hat{A}^{\star}\right]\,,\\
\mathcal{L}_{++--}^{\left(\mathrm{LC}\right)}\left[A^{\bullet},A^{\star}\right]=-g^{2}\int d^{3}\mathbf{x}\,\mathrm{Tr}\,\left[\partial_{-}\hat{A}^{\bullet},\hat{A}^{\star}\right]\partial_{-}^{-2}\left[\partial_{-}\hat{A}^{\star},\hat{A}^{\bullet}\right]\,.
\end{gather}
Above we have used the following differential operators acting on
functions of $\mathbf{x}$:
\begin{equation}
\gamma_{\mathbf{x}}=\partial_{-}^{-1}\partial_{\bullet},\,\,\,\,\,\overline{\gamma}_{\mathbf{x}}=\partial_{-}^{-1}\partial_{\star}\,.\label{eq:gamma_op_def}
\end{equation}
The inverse operator $\partial_{-}^{-1}$ is realized through the
indefinite integral (see \citep{Scherk1975} for a particular implementation
dealing with the integration constant ambiguity). 

The CSW action is now constructed from (\ref{eq:actionLC}) using
a canonical transformation of fields $\left(A^{\bullet},A^{\star}\right)\rightarrow\left(B^{\bullet},B^{\star}\right)$,
so that the Jacobian is field independent and such that \citep{Mansfield2006}
\begin{equation}
\mathcal{L}_{+-}^{\left(\mathrm{LC}\right)}\left[A^{\bullet},A^{\star}\right]+\mathcal{L}_{++-}^{\left(\mathrm{LC}\right)}\left[A^{\bullet},A^{\star}\right]=\mathcal{L}_{+-}^{\left(\mathrm{LC}\right)}\left[B^{\bullet},B^{\star}\right]\,.\label{eq:transf_req1}
\end{equation}
This eliminates the vertex $\mathcal{L}_{++-}$ which has only one  negative helicity and thus does not belong to the CSW action.
 The relevant canonical transformation of the $A^{\star}$ field has
the form 
\begin{equation}
\partial_{-}A_{a}^{\star}\left(\mathbf{x}\right)=\int d^{3}\mathbf{y}\,\frac{\delta B_{c}^{\bullet}\left(\mathbf{y}\right)}{\delta A_{a}^{\bullet}\left(\mathbf{x}\right)}\partial_{-}B_{c}^{\star}\left(\mathbf{y}\right).\label{eq:Transformation_A-}
\end{equation}
 The requirement (\ref{eq:transf_req1}) gives the following transformation
for the $A^{\bullet}$ field 
\begin{equation}
\int d^{3}\mathbf{y}\,\mathrm{Tr}\Bigg\{\left[D_{\star},\gamma_{\mathbf{y}}\hat{A}^{\bullet}\left(\mathbf{y}\right)\right]t^{c}\Bigg\}\frac{\delta B_{a}^{\bullet}\left(\mathbf{x}\right)}{\delta A_{c}^{\bullet}\left(\mathbf{y}\right)}=\omega_{\mathbf{x}}B_{a}^{\bullet}\left(\mathbf{x}\right)\,.\label{eq:Transformation_A+}
\end{equation}
where the differential operator $\omega_{\mathbf{x}}$ is defined
as
\begin{equation}
\omega_{\mathbf{x}}=\partial_{\bullet}\partial_{\star}\partial_{-}^{-1}\,.\label{eq:omegaop}
\end{equation}
The relation (\ref{eq:Transformation_A+}) holds on the constant $x^{+}$
hyper-surface.

In order to obtain the new action $S_{\mathrm{Y-M}}^{\left(\mathrm{LC}\right)}\left[B^{\bullet},B^{\star}\right]$,
in principle, one has to solve the equations (\ref{eq:Transformation_A-}),(\ref{eq:Transformation_A+})
for $A^{\bullet}$ and $A^{\star}$ and insert the solutions to the
action (\ref{eq:actionLC}). However, in the original work \citep{Mansfield2006}
the new action was constructed using only analytic properties of the
transformations and equivalence theorem for the S-matrix. The explicit
solution for $A^{\bullet}$ and $A^{\star}$ fields was found in \citep{Ettle2006b}
in momentum space. Within our conventions the solution for $A^{\bullet}$  reads (see also Section \ref{sub:A[B]}) 
\begin{equation}
\tilde{A}_{a}^{\bullet}\left[\tilde{B}^{\bullet}\right]\left(\mathbf{P}\right)=\tilde{B}_{a}^{\bullet}\left(\mathbf{P}\right)+\sum_{i=2}^{\infty}\int d^{3}\mathbf{p}_{1}\dots d^{3}\mathbf{p}_{i}\,\tilde{\Psi}_{i}^{a\left\{ b_{1}\dots b_{i}\right\} }\left(\mathbf{P};\left\{ \mathbf{p}_{1},\dots,\mathbf{p}_{i}\right\} \right)\tilde{B}_{b_{1}}^{\bullet}\left(\mathbf{p}_{1}\right)\dots\tilde{B}_{b_{i}}^{\bullet}\left(\mathbf{p}_{i}\right)\,,\label{eq:Ap[B]}
\end{equation}
with
\begin{multline}
\tilde{\Psi}_{n}^{a\left\{ b_{1}\dots b_{n}\right\} }\left(\mathbf{P};\left\{ \mathbf{p}_{1},\dots,\mathbf{p}_{n}\right\} \right)=-\left(-g'\right)^{n-1}\,\frac{\tilde{v}_{\left(1\dots n\right)1}^{*}}{\tilde{v}_{1\left(1\dots n\right)}^{*}}\,\frac{1}{\tilde{v}_{n\left(n-1\right)}^{*}\dots\tilde{v}_{32}^{*}\,\tilde{v}_{21}^{*}}\,\\
\times\delta^{3}\left(\mathbf{p}_{1}+\dots+\mathbf{p}_{n}-\mathbf{P}\right)\,\mathrm{Tr}\left(t^{a}t^{b_{1}}\dots t^{b_{n}}\right)\,.\label{eq:Psi_n}
\end{multline}
Above, the curly brackets indicate that the function is symmetric
with respect to the pairs $\left(b_{i},\mathbf{p}_{i}\right)$ enclosed
by the brackets \textit{and} that the result for $\tilde{\Psi}_{n}$
was obtained utilizing the fact that the enclosed arguments are integrated/summed
with the fields. Sometimes we shall refer to such equality as the
equality in the `weak sense'. For example, the following expression
\begin{equation}
\tilde{\Psi}_{n}^{ab_{1}\dots b_{n}}\left(\mathbf{P};\left\{ \mathbf{p}_{1},\dots,\mathbf{p}_{n}\right\} \right)=\sum_{\sigma\in S_{n}}\mathrm{Tr}\left(t^{a}t^{b_{\sigma\left(1\right)}}\dots t^{b_{\sigma\left(n\right)}}\right)\tilde{\Psi}_{n}\left(\mathbf{P};\mathbf{p}_{\sigma\left(1\right)},\dots,\mathbf{p}_{\sigma\left(n\right)}\right)\,,\label{eq:Psi_color_decomp}
\end{equation}
where $S_{n}$ is the permutation group, and
\begin{equation}
\tilde{\Psi}_{n}\left(\mathbf{P};\mathbf{p}_{1},\dots,\mathbf{p}_{n}\right)=-\frac{1}{n!}\left(-g'\right)^{n-1}\,\frac{\tilde{v}_{\left(1\dots n\right)1}^{*}}{\tilde{v}_{1\left(1\dots n\right)}^{*}}\,\frac{1}{\tilde{v}_{n\left(n-1\right)}^{*}\dots\tilde{v}_{32}^{*}\,\tilde{v}_{21}^{*}}\,\delta^{3}\left(\mathbf{p}_{1}+\dots+\mathbf{p}_{n}-\mathbf{P}\right)\,,\label{eq:Psi_colorord}
\end{equation}
is different than (\ref{eq:Psi_n}). However 
\begin{multline}
\int d^{3}\mathbf{p}_{1}\dots d^{3}\mathbf{p}_{n}\tilde{B}_{b_{1}}^{\bullet}\left(\mathbf{p}_{1}\right)\dots\tilde{B}_{b_{n}}^{\bullet}\left(\mathbf{p}_{n}\right)\Bigg\{\tilde{\Psi}_{n}^{a\left\{ b_{1}\dots b_{n}\right\} }\left(\mathbf{P};\left\{ \mathbf{p}_{1},\dots,\mathbf{p}_{n}\right\} \right)\\
-\tilde{\Psi}_{n}^{ab_{1}\dots b_{n}}\left(\mathbf{P};\left\{ \mathbf{p}_{1},\dots,\mathbf{p}_{n}\right\} \right)\Bigg\}=0\,,
\end{multline}
thus they are equal in the `weak sense'. 

The solution for $A^{\star}$ (see Appendix~\ref{sec:App_MHVaction}) is
\begin{multline}
\tilde{A}_{a}^{\star}\left[\tilde{B}^{\bullet},\tilde{B}^{\star}\right]\left(\mathbf{P}\right)=\tilde{B}_{a}^{\star}\left(\mathbf{P}\right)\\
+\sum_{i=2}^{\infty}\int d^{3}\mathbf{p}_{1}\dots d^{3}\mathbf{p}_{i}\,\tilde{\Omega}_{i}^{ab_{1}\left\{ b_{2}\dots b_{i}\right\} }\left(\mathbf{P};\mathbf{p}_{1},\left\{ \mathbf{p}_{2},\dots,\mathbf{p}_{i}\right\} \right)\tilde{B}_{b_{1}}^{\star}\left(\mathbf{p}_{1}\right)\tilde{B}_{b_{2}}^{\bullet}\left(\mathbf{p}_{2}\right)\dots\tilde{B}_{b_{i}}^{\bullet}\left(\mathbf{p}_{i}\right)\,,\label{eq:Am[B]}
\end{multline}
with
\begin{equation}
\tilde{\Omega}_{n}^{ab_{1}\left\{ b_{2}\dots b_{n}\right\} }\left(\mathbf{P};\mathbf{p}_{1},\left\{ \mathbf{p}_{2},\dots,\mathbf{p}_{n}\right\} \right)=n\left(\frac{p_{1}^{+}}{p_{1\dots n}^{+}}\right)^{2}\tilde{\Psi}_{n}^{ab_{1}\dots b_{n}}\left(\mathbf{P};\mathbf{p}_{1},\dots,\mathbf{p}_{n}\right)\,.\label{eq:Omega_n}
\end{equation}

The above transformations, when applied to $S_{\mathrm{Y-M}}^{\left(\mathrm{LC}\right)}\left[\tilde{A}^{\bullet},\tilde{A}^{\star}\right]$
give the MHV action:
\begin{equation}
S_{\mathrm{Y-M}}^{\left(\mathrm{LC}\right)}\left[\tilde{B}^{\bullet},\tilde{B}^{\star}\right]=\int dx^{+}\left(\mathcal{L}_{+-}^{\left(\mathrm{LC}\right)}+\mathcal{L}_{--+}^{\left(\mathrm{LC}\right)}+\dots+\mathcal{L}_{--+\dots+}^{\left(\mathrm{LC}\right)}+\dots\right)\,,\label{eq:MHV_action}
\end{equation}
where the part of the Lagrangian for interaction of $n$ fields is
\begin{multline}
\mathcal{L}_{--+\dots+}^{\left(\mathrm{LC}\right)}=\int d^{3}\mathbf{p}_{1}\dots d^{3}\mathbf{p}_{n}\delta^{3}\left(\mathbf{p}_{1}+\dots+\mathbf{p}_{n}\right)\\ \tilde{B}_{b_{1}}^{\star}\left(\mathbf{p}_{1}\right)\tilde{B}_{b_{2}}^{\star}\left(\mathbf{p}_{2}\right)\tilde{B}_{b_{3}}^{\bullet}\left(\mathbf{p}_{3}\right)\dots\tilde{B}_{b_{n}}^{\bullet}\left(\mathbf{p}_{n}\right)\tilde{\mathcal{V}}_{--+\dots+}^{b_{1}\dots b_{n}}\left(\mathbf{p}_{1},\dots,\mathbf{p}_{n}\right)\,.\label{eq:MHV_n_point}
\end{multline}
This interaction term corresponds to the MHV vertex with  two gluons with minus helicities, corresponding to the  $\tilde{B}^{\star}$ fields, and the rest of the particles, corresponding to $\tilde{B}^{\bullet}$ fields,  having plus helicites.
It is convenient to decompose the MHV vertex into color-ordered objects,
similar to what we did in (\ref{eq:Psi_color_decomp}) (this technique
is often used for multi leg QCD amplitudes, see \citep{Mangano:1990by}
for a review)
\begin{equation}
\tilde{\mathcal{V}}_{--+\dots+}^{b_{1}\dots b_{n}}\left(\mathbf{p}_{1},\dots,\mathbf{p}_{n}\right)=\sum_{\sigma\in S_{n}/Z_{n}}\mathrm{Tr}\left(t^{b_{\sigma\left(1\right)}}\dots t^{b_{\sigma\left(n\right)}}\right)\tilde{\mathcal{V}}_{--+\dots+}\left(\mathbf{p}_{\sigma\left(1\right)},\mathbf{p}_{\sigma\left(2\right)},\dots,\mathbf{p}_{\sigma\left(n\right)}\right)\,.\label{eq:MHV_color}
\end{equation}
Then one finds 
\begin{equation}
\tilde{\mathcal{V}}_{--+\dots+}\left(\mathbf{p}_{1},\dots,\mathbf{p}_{n}\right)=\frac{1}{n!}\left(g'\right)^{n-1}\left(\frac{p_{1}^{+}}{p_{2}^{+}}\right)^{2}\frac{\tilde{v}_{21}^{*4}}{\tilde{v}_{1n}^{*}\tilde{v}_{n\left(n-1\right)}^{*}\tilde{v}_{\left(n-1\right)\left(n-2\right)}^{*}\dots\tilde{v}_{21}^{*}}\,.\label{eq:MHV_vertex_colororder}
\end{equation}
When written in terms of spinors using (\ref{eq:spinor_prod}) we
see that they look like the Mangano-Parke amplitudes, except that
they are fully off-shell objects. This is one of the reasons we prefer
to use the $\tilde{v}_{ij}$ symbols instead of spinors. The latter
are normally defined for on-shell momenta and need a prescription
to be continued off-shell, in general situation.

In the original formulation \citep{Cachazo2004} the spinors in the
MHV vertices were defined using an analytic continuation to the off-shell
momenta by means of  an additional auxiliary on-shell momentum. In the context
of the action (\ref{eq:MHV_action}) the off-shell continuation means
that we can use any $p^{-}$ for fixed  $\left(p^{+},p^{\bullet},p^{\star}\right)\equiv\mathbf{p}$.
components of the momenta. They are the $\mathbf{p}$ components that
determine the vertex. In fact, the spinor products can be constructed
without a reference to $p^{-}$, see Eq.~(\ref{eq:spinor_prod}). Obviously,
the off-shell continued spinors will coincide with the on-shell spinors
with the same $p^{-}=p^{\bullet}p^{\star}/p^{+}$,
fixed by the on-shell condition $p^{2}=0$. One can also look at this
from another point of view. In light front quantized theory the internal
lines in the Feynman diagrams are on-shell (see e.g. \citep{Cruz-Santiago2015a})
at the cost of the energy denominators involving $p^{-}$ components.
Thus in that case the spinors can be defined for any internal line.
This fact was used for instance in \citep{Motyka2009} to express
the vertexes in terms of spinors.

In the next section we shall take a closer look at the transformation
(\ref{eq:Transformation_A+}). In particular we shall consider a solution
expressing the $B^{\bullet}$ field in terms of $A^{\bullet}$ fields
and see that it gives a new interpretation to the MHV vertices.

\section{Wilson line solution to the field transformation}

\label{sec:WilsonLineSolution}

In the previous section we have recalled the procedure to obtain the
MHV action. The essential step is to solve the transformation (\ref{eq:Transformation_A+})
and find the functional $A^{\bullet}\left[B^{\bullet}\right]$, so
that the new action can be expressed in terms of $B$ fields only, with the infinite set of MHV vertices.
Now, it is extremely interesting to study the true physical meaning
of the new fields, in particular the $B^{\bullet}$ field. As we will
show in the following section it turns out that it is a certain type
of straight infinite Wilson line.

Let us introduce a four-vector
\begin{equation}
\varepsilon_{\alpha}^{+}=\varepsilon_{\perp}^{+}-\alpha\eta\,.\label{eq:SlopeDef0}
\end{equation}
where $\alpha$ is a real parameter and the vectors $\varepsilon_{\perp}$ and $\eta$ were defined in Section~\ref{sec:Notation}. In the coordinates $\left(x^{-},x^{\bullet},x^{\star}\right)$
it reads
\begin{equation}
\left(-\alpha,-1,0\right)\equiv\mathbf{e}_{\alpha}\,.\label{eq:SlopeDef}
\end{equation}
Let us then postulate the solution to the equation (\ref{eq:Transformation_A+})
in the following form
\begin{equation}
B_{a}^{\bullet}\left(\mathbf{x}\right)=\int_{-\infty}^{\infty}d\alpha\,\mathrm{Tr}\left\{ \frac{1}{2\pi ig'}t^{a}\partial_{-}\mathbb{P}\exp\left[ig'\int_{-\infty}^{\infty}ds\,\hat{A}^{\bullet}\left(\mathbf{x}+s\mathbf{e}_{\alpha}\right)\right]\right\} \, ,\label{eq:WilsonLineAnsatz}
\end{equation}
where $\mathbb{P}$ denotes the path ordering of fields with increase of the parameter $s$. This path exponential is in fact the Wilson line (for a modern
review of the Wilson lines in quantum field theory see \citep{Cherednikov2014}).
Indeed, taking the path to be
\begin{equation}
z^{\mu}\left(s\right)=x^{\mu}+s\varepsilon_{\alpha}^{\mu},\,\,\, s\in\left(-\infty,+\infty\right)\, ,\label{eq:path}
\end{equation}
we have
\begin{equation}
\mathbb{P}\exp\left[ig'\int dz_{\mu}\,\hat{A}^{\mu}\left(z\right)\right]=\mathbb{P}\exp\left[ig'\int_{-\infty}^{\infty}ds\,\hat{A}^{\bullet}\left(x^{+},\mathbf{x}+s\mathbf{e}_{\alpha}\right)\right]\,,
\end{equation}
where we have used the gauge condition $A^{+}=0$. The idea of using
non-lightlike Wilson lines with a slope defined to be a vector reminiscent of the polarization vector (compare (\ref{eq:SlopeDef0}) and (\ref{eq:PolarizationVect});
see also explanation below) is motivated by a procedure presented in \citep{Kotko2014a}, where 
such objects were used to construct off-shell amplitudes which satisfy
Ward identities. 

Notice the integral over $d\alpha$ in (\ref{eq:WilsonLineAnsatz})
which means that the summation over all slopes has to be taken into
account. 

We now prove that (\ref{eq:WilsonLineAnsatz}) is the solution to
(\ref{eq:Transformation_A+}). It is convenient to perform the proof  in the momentum
space. Let us thus write the expansion of this solution in momentum
space
\begin{multline}
\tilde{B}_{a}^{\bullet}\left(\mathbf{P}\right)=\int d^{3}\mathbf{x}\, e^{i\mathbf{x}\cdot\mathbf{P}}B_{a}^{\bullet}\left(\mathbf{x}\right)=\tilde{A}^{\bullet}\left(\mathbf{P}\right)+\int d^{3}\mathbf{p}_{1}d^{3}\mathbf{p}_{2}\,\tilde{\Theta}_{2}^{ab_{1}b_{2}}\left(\mathbf{P};\mathbf{p}_{1},\mathbf{p}_{2}\right)\,\tilde{A}_{b_{1}}^{\bullet}\left(\mathbf{p}_{1}\right)\tilde{A}_{b_{2}}^{\bullet}\left(\mathbf{p}_{2}\right)+\dots\\
+\int d^{3}\mathbf{p}_{1}\dots d^{3}\mathbf{p}_{n}\,\tilde{\Theta}_{n}^{ab_{1}\dots b_{n}}\left(\mathbf{P};\mathbf{p}_{1},\dots,\mathbf{p}_{n}\right)\,\tilde{A}_{b_{1}}^{\bullet}\left(\mathbf{p}_{1}\right)\dots\tilde{A}_{b_{2}}^{\bullet}\left(\mathbf{p}_{n}\right)+\dots\,.\label{eq:WLexpansion}
\end{multline}
Let us calculate the $n$-th coefficient of the expansion $\tilde{\Theta}_{n}$.
To this end, consider the $n$-th term in the expansion of the Wilson
line 
\begin{multline}
\tilde{B}_{a}^{\bullet\left(n\right)}\left(\mathbf{P}\right)=\frac{1}{2\pi}\left(ig'\right)^{n-1}\int d^{3}\mathbf{x}\, e^{i\mathbf{x}\cdot\mathbf{P}}\int d\alpha\,\partial_{-}\int_{-\infty}^{+\infty}ds_{1}\int_{-\infty}^{s_{1}}ds_{2}\dots\int_{-\infty}^{s_{n-1}}ds_{n}\,\\
A_{b_{1}}^{\bullet}\left(\mathbf{x}+s_{1}\mathbf{e}_{\alpha}\right)\dots A_{b_{n}}^{\bullet}\left(\mathbf{x}+s_{n}\mathbf{e}_{\alpha}\right)\,\mathrm{Tr}\left(t^{a}t^{b_{1}}\dots t^{b_{n}}\right)\\
=\frac{1}{2\pi}\left(ig'\right)^{n-1}\int d^{3}\mathbf{x}\, e^{i\mathbf{x}\cdot\mathbf{P}}\int d^{3}\mathbf{p}_{1}\dots d^{3}\mathbf{p}_{n}\int d\alpha\,\partial_{-}\int_{-\infty}^{+\infty}ds_{1}\int_{-\infty}^{s_{1}}ds_{2}\dots\int_{-\infty}^{s_{n-1}}ds_{n}\\
e^{-i\mathbf{x}\cdot\left(\mathbf{p}_{1}+\dots+\mathbf{p}_{n}\right)}\, e^{-is_{1}\mathbf{e}_{\alpha}\cdot\mathbf{p}_{1}}\dots e^{-is_{n}\mathbf{e}_{\alpha}\cdot\mathbf{p}_{n}}\tilde{A}_{b_{1}}^{\bullet}\left(\mathbf{p}_{1}\right)\dots A_{b_{n}}^{\bullet}\left(\mathbf{p}_{n}\right)\,\mathrm{Tr}\left(t^{a}t^{b_{1}}\dots t^{b_{n}}\right)\label{eq:NthWLcoef}
\end{multline}
For the ordered integrals we get
\begin{multline}
\int_{-\infty}^{+\infty}ds_{1}\int_{-\infty}^{s_{1}}ds_{2}\dots\int_{-\infty}^{s_{n-1}}ds_{n}\,
 e^{-is_{1}\mathbf{e}\left(\alpha\right)\cdot\mathbf{p}_{1}}\dots e^{-is_{n}\mathbf{e}\left(\alpha\right)\cdot\mathbf{p}_{n}}\\
=2\pi\,\delta\left(\mathbf{e}_{\alpha}\cdot\mathbf{p}_{1\dots n}\right)\frac{i^{n-1}}{\left(\mathbf{e}_{\alpha}\cdot\mathbf{p}_{2\dots n}+i\epsilon\right)\left(\mathbf{e}_{\alpha}\cdot\mathbf{p}_{3\dots n}+i\epsilon\right)\dots\left(\mathbf{e}_{\alpha}\cdot\mathbf{p}_{n}+i\epsilon\right)}\,.
\label{eq:eikonals}
\end{multline}
Above we have used the notation $\mathbf{p}_{i}+\mathbf{p}_{i+1}+\dots+\mathbf{p}_{m-1}+\mathbf{p}_{m}\equiv\mathbf{p}_{i\dots m}$.
The result is a generalized function as is evident from the $i\epsilon$
prescription. The existence of such prescription can be proved by
considering a limit of a non-straight almost infinite path \citep{Kotko2014a}. 

Coming back to (\ref{eq:NthWLcoef}) using the form of $\mathbf{e}_{\alpha}$
and definitions (\ref{eq:vtild_def}) we arrive at 
\begin{multline}
\tilde{B}_{a}^{\bullet\left(n\right)}\left(\mathbf{P}\right)=\left(g'\right)^{n-1}\int d^{3}\mathbf{p}_{1}\dots d^{3}\mathbf{p}_{n}\delta^{3}\left(\mathbf{p}_{1\dots n}-\mathbf{P}\right)\tilde{A}_{b_{1}}^{\bullet}\left(\mathbf{p}_{1}\right)\dots A_{b_{n}}^{\bullet}\left(\mathbf{p}_{n}\right)\,\mathrm{Tr}\left(t^{a}t^{b_{1}}\dots t^{b_{n}}\right)\\
\frac{1}{\left(\tilde{v}_{\left(2\dots n\right)\left(1\dots n\right)}^{*}+i\epsilon\right)\dots\left(\tilde{v}_{\left(n-1n\right)\left(1\dots n\right)}^{*}+i\epsilon\right)\left(\tilde{v}_{n\left(1\dots n\right)}^{*}+i\epsilon\right)}\,\,.\label{eq:NthWLcoef1}
\end{multline}
In what follows we shall omit the $i\epsilon$ prescription.

Comparing (\ref{eq:WLexpansion}) with (\ref{eq:NthWLcoef1}) we get
\begin{multline}
\tilde{\Theta}_{n}^{ab_{1}\dots b_{n}}\left(\mathbf{P};\mathbf{p}_{1},\dots,\mathbf{p}_{n}\right)=\left(g'\right)^{n-1}\delta^{3}\left(\mathbf{p}_{1\dots n}-\mathbf{P}\right)\frac{1}{\tilde{v}_{\left(2\dots n\right)\left(1\dots n\right)}^{*}\dots\tilde{v}_{\left(n-1n\right)\left(1\dots n\right)}^{*}\tilde{v}_{n\left(1\dots n\right)}^{*}}\,\mathrm{Tr}\left(t^{a}t^{b_{1}}\dots t^{b_{n}}\right)\\
=\left(-g'\right)^{n-1}\delta^{3}\left(\mathbf{p}_{1\dots n}-\mathbf{P}\right)\frac{1}{\tilde{v}_{1\left(1\dots n\right)}^{*}\tilde{v}_{\left(12\right)\left(1\dots n\right)}^{*}\dots\tilde{v}_{\left(1\dots n-1\right)\left(1\dots n\right)}^{*}}\,\mathrm{Tr}\left(t^{a}t^{b_{1}}\dots t^{b_{n}}\right)\,,\label{eq:Thetan}
\end{multline}
where we have used the property \eqref{enu:vtild_prop_5} from Appendix~\ref{sec:App_ident}
to get the last equality.

Now we will prove that it satisfies the equation (\ref{eq:Transformation_A+}).
In Section~\ref{sub:Diagramm_B+} we shall show that assuming (\ref{eq:Transformation_A+}),
the expansion coefficients $\tilde{\Theta}_{n}$ satisfy the following
recursion (see also \citep{Mansfield2006} for a similar recursion
in position space) 
\begin{multline}
\tilde{\Theta}_{n}^{ab_{1}\dots b_{n}}\left(\mathbf{P};\mathbf{p}_{1},\dots,\mathbf{p}_{n}\right)=\frac{-1}{D_{1\dots n}}\,\\
\Bigg\{\tilde{V}_{++-}^{cb_{n-1}b_{n}}\left(\mathbf{p}_{n-1},\mathbf{p}_{n}\right)\frac{1}{p_{\left(n-1\right)n}^{+}}\tilde{\Theta}_{n-1}^{ab_{1}\dots b_{n-2}c}\left(\mathbf{P};\mathbf{p}_{1},\dots,\mathbf{p}_{n-1}+\mathbf{p}_{n}\right)\\
+\tilde{V}_{++-}^{cb_{n-2}b_{n-1}}\left(\mathbf{p}_{n-2},\mathbf{p}_{n-1}\right)\frac{1}{p_{\left(n-2\right)\left(n-1\right)}^{+}}\tilde{\Theta}_{n-1}^{ab_{1}\dots b_{n-3}cb_{n}}\left(\mathbf{P};\mathbf{p}_{1},\dots,\mathbf{p}_{n-2}+\mathbf{p}_{n-1},\mathbf{p}_{n}\right)+\dots\Bigg\}\,.\label{eq:Theta_n_recrel}
\end{multline}
In the equation above we introduced  the quantity
\begin{equation}
D_{1\dots n}=\frac{1}{p_{1\dots n}^{+}}\sum_{i,j=1}^{n}\tilde{v}_{ij}\tilde{v}_{ji}^{*} \; .\label{eq:D_vivj}
\end{equation}
It is a so-called energy denominator known from the perturbation
theory on the light front \citep{Lepage1980} (see Section~\ref{sub:Diagramm_B+} for
a justification of that statement). Equation (\ref{eq:Theta_n_recrel})
should be understood in the `weak' sense as explained in the previous section, i.e. upon integrating
with the $A^{\bullet}$ fields over $\mathbf{p}_{i}$ momenta (and
summing over color indices).  To complete the proof we first note,
that for $n=1$ we have (from (\ref{eq:NthWLcoef}))
\begin{equation}
\tilde{\Theta}_{1}^{ab_{1}}\left(\mathbf{P};\mathbf{p}_{1}\right)=\delta^{ab_{1}}\delta^{3}\left(\mathbf{p}_{1}-\mathbf{P}\right)\,.
\end{equation}
For $n=2$ we have (from (\ref{eq:Thetan}))
\begin{equation}
\tilde{\Theta}_{2}^{ab_{1}b_{2}}\left(\mathbf{P};\mathbf{p}_{1},\mathbf{p}_{2}\right)=-g'\delta^{3}\left(\mathbf{p}_{12}-\mathbf{P}\right)\frac{1}{\tilde{v}_{1\left(12\right)}^{*}}\,\mathrm{Tr}\left(t^{a}t^{b_{1}}t^{b_{2}}\right)\,.
\end{equation}
Let us compare this with what one gets from recursion (\ref{eq:Theta_n_recrel}).
We have
\begin{multline}
\frac{-1}{D_{12}}\,\tilde{V}_{++-}^{ab_{1}b_{2}}\left(\mathbf{p}_{1},\mathbf{p}_{2}\right)\frac{1}{p_{12}^{+}}\,\delta^{3}\left(\mathbf{p}_{12}-\mathbf{P}\right)=\frac{-1}{2\tilde{v}_{21}\tilde{v}_{12}^{*}}\,\left(-ig\right)f^{ab_{1}b_{2}}\frac{p_{12}^{+}}{p_{1}^{+}}\tilde{v}_{12}\delta^{3}\left(\mathbf{p}_{12}-\mathbf{P}\right)\\
=-\frac{1}{2}igf^{ab_{1}b_{2}}\,\frac{1}{\tilde{v}_{1\left(12\right)}^{*}}\delta^{3}\left(\mathbf{p}_{12}-\mathbf{P}\right)=\frac{1}{2}\, g'\,\frac{1}{\tilde{v}_{1\left(12\right)}^{*}}\,\left[\mathrm{Tr}\left(t^{b_{2}}t^{b_{1}}t^{a}\right)-\mathrm{Tr}\left(t^{a}t^{b_{1}}t^{b_{2}}\right)\right]\delta^{3}\left(\mathbf{p}_{12}-\mathbf{P}\right)\,.
\end{multline}
Remembering that it is to be integrated with $A_{b_{1}}^{\bullet}\left(\mathbf{p}_{1}\right)A_{b_{2}}^{\bullet}\left(\mathbf{p}_{2}\right)$
we can change variables $1\leftrightarrow2$ for the first term and
using $\tilde{v}_{2\left(12\right)}=-\tilde{v}_{1\left(12\right)}$
arrive at
\begin{equation}
-g'\,\frac{1}{\tilde{v}_{1\left(12\right)}^{*}}\,\mathrm{Tr}\left(t^{a}t^{b_{1}}t^{b_{2}}\right)\delta^{3}\left(\mathbf{p}_{12}-\mathbf{P}\right)\,,
\end{equation}
which equals $\tilde{\Theta}_{2}^{ab_{1}b_{2}}$. Now we proceed to
an arbitrary $n$. We consider the r.h.s of (\ref{eq:Theta_n_recrel})
and insert (\ref{eq:Thetan}) and obtain
\begin{multline}
\left(-ig\right)\left(-g'\right)^{n-2}\delta^{3}\left(\mathbf{p}_{1\dots n}-\mathbf{P}\right)\frac{-1}{D_{1\dots n}}\,\frac{1}{\tilde{v}_{1\left(1\dots n\right)}^{*}\tilde{v}_{\left(12\right)\left(1\dots n\right)}^{*}\dots\tilde{v}_{\left(1\dots n-1\right)\left(1\dots n\right)}^{*}}\\
\Bigg\{\frac{\tilde{v}_{\left(n-1\right)n}}{p_{n-1}^{+}}\,\tilde{v}_{\left(1\dots n-1\right)\left(1\dots n\right)}^{*}\, f^{cb_{n-1}b_{n}}\mathrm{Tr}\left(t^{a}t^{b_{1}}\dots t^{b_{n-2}}t^{c}\right)\\
+\frac{\tilde{v}_{\left(n-2\right)\left(n-1\right)}}{p_{n-2}^{+}}\,\tilde{v}_{\left(1\dots n-2\right)\left(1\dots n\right)}^{*}\, f^{cb_{n-2}b_{n-1}}\mathrm{Tr}\left(t^{a}t^{b_{1}}\dots t^{b_{n-3}}t^{c}t^{b_{n}}\right)+\dots\Bigg\}\,.
\end{multline}
We turn each term in the bracket into pure traces. For example, for
the first term we have
\begin{equation}
f^{cb_{n-1}b_{n}}\mathrm{Tr}\left(t^{a}t^{b_{1}}\dots t^{b_{n-2}}t^{c}\right)=\frac{1}{i\sqrt{2}}\left[\mathrm{Tr}\left(t^{a}t^{b_{1}}\dots t^{b_{n-2}}t^{b_{n-1}}t^{b_{n}}\right)-\mathrm{Tr}\left(t^{a}t^{b_{1}}\dots t^{b_{n-2}}t^{b_{n}}t^{b_{n-1}}\right)\right]\,,
\end{equation}
and so on. Next, we use the fact that it has to be integrated with
the gauge fields. Thus, we can change variables for each of the second
term arriving from the $f^{abc}$ tensor expansion: $n-1\leftrightarrow n$,
$n-2\leftrightarrow n-1$ and so on. Using
\begin{equation}
\frac{\tilde{v}_{\left(n-1\right)n}}{p_{n-1}^{+}}=v_{n\left(n-1\right)}=-v_{\left(n-1\right)n}\; ,
\end{equation}
we get
\begin{multline}
\left(-g'\right)\left(-g'\right)^{n-2}\delta^{3}\left(\mathbf{p}_{1\dots n}-\mathbf{P}\right)\frac{-2}{D_{1\dots n}}\,\frac{1}{\tilde{v}_{1\left(1\dots n\right)}^{*}\tilde{v}_{\left(12\right)\left(1\dots n\right)}^{*}\dots\tilde{v}_{\left(1\dots n-1\right)\left(1\dots n\right)}^{*}}\\
\Bigg\{ v_{n\left(n-1\right)}\,\tilde{v}_{\left(1\dots n-1\right)\left(1\dots n\right)}^{*}+v_{\left(n-1\right)\left(n-2\right)}\,\tilde{v}_{\left(1\dots n-2\right)\left(1\dots n\right)}^{*}+\dots\Bigg\}\,.
\end{multline}
Using now the relation (\ref{eq:MotykaStasto_rel1}) we arrive at
\begin{equation}
\left(-g'\right)^{n-1}\delta^{3}\left(\mathbf{p}_{1\dots n}-\mathbf{P}\right)\,\frac{1}{\tilde{v}_{1\left(1\dots n\right)}^{*}\tilde{v}_{\left(12\right)\left(1\dots n\right)}^{*}\dots\tilde{v}_{\left(1\dots n-1\right)\left(1\dots n\right)}^{*}}\,,
\end{equation}
which is exactly $\tilde{\Theta}_{n}$ obtained in Eq.~(\ref{eq:Thetan}).
Thus, indeed Wilson line (\ref{eq:WilsonLineAnsatz}) is a solution to the transformation
(\ref{eq:Transformation_A+}). 

It is convenient to introduce the Feynman rules for the Wilson line. Following
the standard treatment \citep{Collins:1981uw} the propagators are
given by the denominators in (\ref{eq:eikonals}) while the vertex
is just $ig'\hat{A}^{\bullet}$:

\begin{flushleft}
\begin{tabular}{>{\centering}m{0.92\columnwidth}>{\centering}m{0.05\columnwidth}}
\bigskip{}

\raggedright{}\centerline{\includegraphics[height=0.029\paperheight]{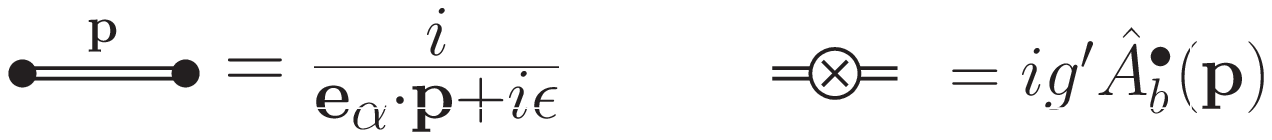}} & \centering{}\centering{(\myref )}
\refmyref{WL_diag_1}\tabularnewline
\end{tabular}
\par\end{flushleft}

\noindent Using these elements, the $n$-th term in the expansion of the $B^{\bullet}$
field can be represented by the following diagram:

\begin{flushleft}
\begin{tabular}{>{\centering}m{0.92\columnwidth}>{\centering}m{0.05\columnwidth}}
\bigskip{}

\raggedright{}\centerline{\includegraphics[height=0.045\paperheight]{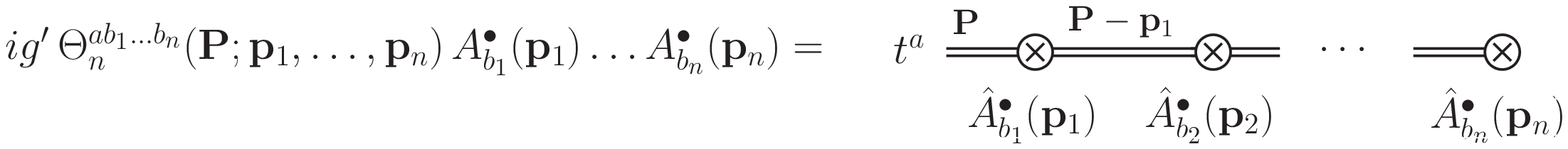}} & \centering{}\centering{(\myref )}
\refmyref{WL_diag_2}\tabularnewline
\end{tabular}
\par\end{flushleft}

\noindent Above, the trace over color matrices is understood. The
$B^{\bullet}$ field thus has the following simple diagrammatic expansion:

\begin{flushleft}
\begin{tabular}{>{\centering}m{0.92\columnwidth}>{\centering}m{0.05\columnwidth}}
\bigskip{}

\raggedright{}\centerline{\includegraphics[height=0.02\paperheight]{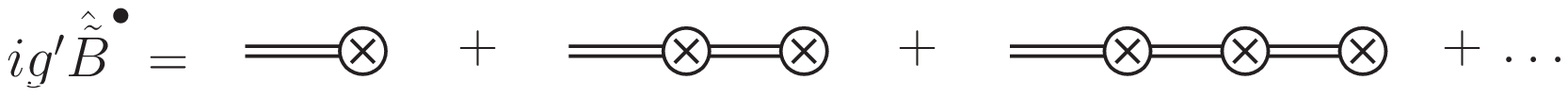}} & \centering{}\centering{(\myref )}
\refmyref{WL_diag_3}\tabularnewline
\end{tabular}
\par\end{flushleft}

In the next section we shall study an immediate consequence of this
result.

\section{Gauge invariant off-shell currents}

\label{sec:GaugeInvAmps}

As demonstrated in the previous section, the interaction terms \eqref{eq:MHV_n_point} in the MHV Lagrangian contain the infinite Wilson lines encoded in the $B^{\bullet}$ fields. This has interesting consequences which we shall explore below in the current section.

Let us consider the MHV vertex, where all fields except one with negative
helicity (corresponding here to the field $B^{\star}$) become on-shell.
Thus, more formally, we consider a current for a transition of an
off-shell particle given by the field $\tilde{B}^{\bullet}\left(p_{1\dots n}\right)$ (a
`gluon' in the action $S_{\mathrm{Y-M}}^{\left(\mathrm{LC}\right)}\left[\tilde{B}^{\bullet},\tilde{B}^{\star}\right]$)
to a set of on-shell `gluons' given by the  fields $\tilde{B}^{\bullet}\left(p_{1}\right),\,\tilde{B}^{\star}\left(p_{2}\right),\dots,\,\tilde{B}^{\star}\left(p_{n}\right)$
\begin{equation}
\mathcal{J}^{ab_{1}\dots b_{n}}\left(p_{1\dots n};p_{1},\dots,p_{n}\right)=\lim_{p_{1}^{2}\rightarrow0}\dots\lim_{p_{n}^{2}\rightarrow0}\, ip_{1}^{2}\dots ip_{n}^{2}\,\tilde{G}^{ab_{1}\dots b_{n}}\left(-p_{1\dots n},p_{1},\dots,p_{n}\right)\,,
\end{equation}
where the momentum space Green's function reads
\begin{multline}
\tilde{G}_{a_{1}\dots a_{n}}\left(k_{1},\dots,k_{N}\right)=\int d^{4}x_{1}\dots d^{4}x_{N}\, e^{i\left(x_{1}\cdot k_{1}+\dots+x_{N}\cdot k_{N}\right)}\\
\times\left\langle 0\right|\mathcal{T}\, B_{a_{1}}^{\bullet}\left(x_{1}\right)B_{a_{2}}^{\bullet}\left(x_{2}\right)B_{a_{3}}^{\star}\left(x_{3}\right)\dots B_{a_{N}}^{\star}\left(x_{N}\right)\left|0\right\rangle \,.
\end{multline}
It is understood that, in the above expression  only connected contributions are taken into account. Elementary
tree level calculation leads to the anticipated result 
\begin{equation}
\mathcal{J}^{ab_{1}\dots b_{n}}\left(p_{1\dots n};p_{1},\dots,p_{n}\right)=\frac{-i}{p_{1\dots n}^{2}}\, n!\,\tilde{\mathcal{V}}_{--+\dots+}^{ab_{1}\dots b_{n}}\left(-\mathbf{p}_{1\dots n},\mathbf{p}_{1},\dots,\mathbf{p}_{n}\right)\delta\left(p_{1}^{-}+\dots+p_{n}^{-}-p_{1\dots n}^{-}\right)\,,\label{eq:CSW_current1}
\end{equation}
with $\tilde{\mathcal{V}}_{--+\dots+}$ being the MHV vertex (\ref{eq:MHV_color}). Above,  the momenta $p_1,\dots ,p_n$ are on-shell.

Now, let us consider the same current in terms of $A^{\bullet}$,
$A^{\star}$ fields. Since we work at tree level, we may replace all
on-shell $B^{\bullet}$, $B^{\star}$ fields by $A^{\bullet}$, $A^{\star}$
due to the $S$-matrix equivalence theorem (stating that a reversible
field transformation does not change the $S$-matrix up to a wave function
renormalization, which, however, does not matter at the tree level). Thus
we have
\begin{multline}
\mathcal{J}^{ab_{1}\dots b_{n}}\left(p_{1\dots n};p_{1},\dots,p_{n}\right)=\lim_{p_{1}^{2}\rightarrow0}\dots\lim_{p_{n}^{2}\rightarrow0}\, ip_{1}^{2}\dots ip_{n}^{2}\,\int d^{4}x\, d^{4}x_{1}\dots d^{4}x_{n}\, e^{i\left(x\cdot p_{1\dots n}+x_{1}\cdot p_{1}+\dots+x_{n}\cdot p_{n}\right)}\\
\times\left\langle 0\right|\mathcal{T}\, B_{a}^{\bullet}\left[A^{\bullet}\right]\left(x\right)A_{b_{1}}^{\bullet}\left(x_{1}\right)A_{b_{2}}^{\star}\left(x_{2}\right)\dots A_{b_{n}}^{\star}\left(x_{n}\right)\left|0\right\rangle \, .
\end{multline}
This can be interpreted as a matrix element of the Wilson line operator
\begin{equation}
\mathcal{J}^{ab_{1}\dots b_{n}}\left(p_{1\dots n};p_{1},\dots,p_{n}\right)=\int d^{4}x\, e^{ix\cdot p_{1\dots n}}\left\langle 0\right|B_{a}^{\bullet}\left[A^{\bullet}\right]\left(x\right)\left|p_{1},+;p_{2},-;\dots;p_{n},-\right\rangle \,,\label{eq:WilsonLine_ME}
\end{equation}
where the kets represent on-shell gluons with definite helicity. Since
the Wilson line extends from $-\infty$ to $+\infty$ it is gauge
invariant with respect to the small gauge transformations. This fact together
with on-shellness of the remaining gluons means that the whole object
is gauge invariant. It was explicitly demonstrated using the Slavnov-Taylor
identities for a similar link operator in \citep{Kotko2014a}. In
terms of color ordered Feynman diagrams, (\ref{eq:WilsonLine_ME})
can be expressed as

\begin{flushleft}
\begin{tabular}{>{\raggedleft}m{0.92\columnwidth}>{\centering}m{0.05\columnwidth}}
\bigskip{}

\raggedright{}\centerline{\includegraphics[height=0.06\paperheight]{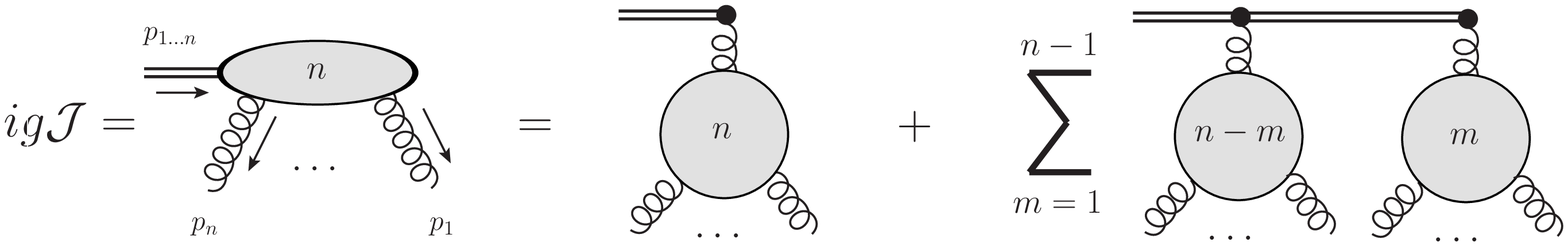}} & \centering{}\tabularnewline
\raggedright{}\flushright{\includegraphics[height=0.06\paperheight]{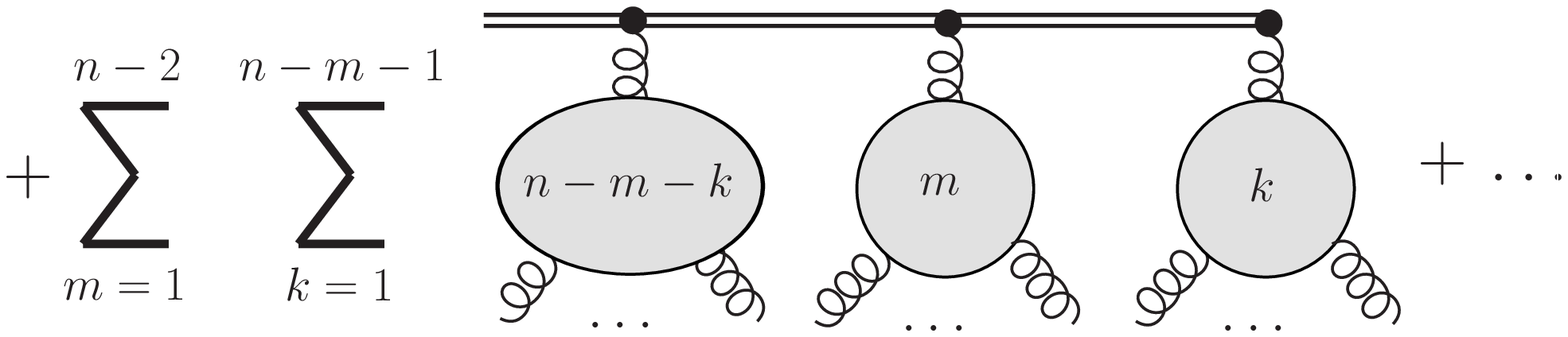}} & \centering{}\centering{(\myref )}
\refmyref{WLME_diag_1}\tabularnewline
\end{tabular}
\par\end{flushleft}

\noindent where the blobs represent the ordinary gluonic off-shell
currents given by the sum of standard Feynman diagrams. Let us denote
these currents as

\begin{flushleft}
\begin{tabular}{>{\centering}m{0.92\columnwidth}>{\centering}m{0.05\columnwidth}}
\bigskip{}

\raggedright{}\centerline{\includegraphics[height=0.05\paperheight]{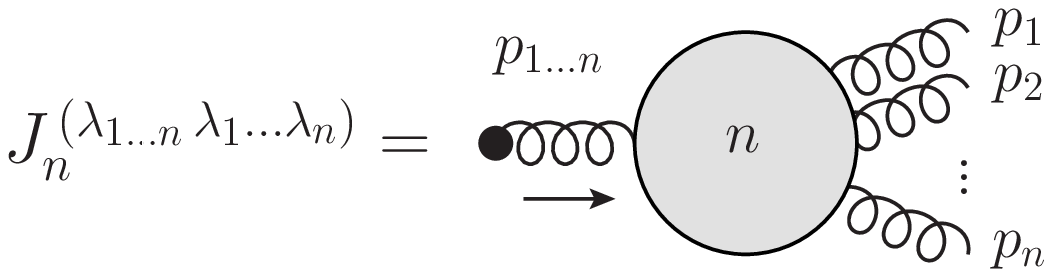}} & \centering{}\centering{(\myref )}
\refmyref{offshell_current_diag}\tabularnewline
\end{tabular}
\par\end{flushleft}

\noindent where $\lambda_{i}=\pm$ denote the helicity corresponding
to the leg $i$. Note, that the currents considered here have no free
vector index, but are projected on a polarization vector corresponding
to $\lambda_{1\dots n}$.

The off-shell currents with the MHV helicity configuration defined
as (\ref{eq:WilsonLine_ME}) were considered before in the context
of the recurrence relations for amplitudes within the light-front perturbation
theory \citep{Cruz-Santiago2015,Kotko2016} (for a review see \citep{Cruz-Santiago2015a}).
Let us briefly recall the main results, as they are tightly connected
to the present work. Suppose one wants to solve the light-front version
of the Berends-Giele relation for the MHV configuration, that is,
one wants to find an off-shell current $J_{n}^{\left(++-\dots-\right)}$
for any $n$. The task was accomplished in \citep{Cruz-Santiago2013}
and the result reads (we work with color amplitudes for brevity)%
\footnote{The results in \citep{Cruz-Santiago2013,Cruz-Santiago2015,Kotko2016}
consider `mostly-plus' MHV configuration while the present work would
correspond to `mostly-minus' MHV amplitudes in these works. Here we have
adapted the cited results to the present helicity convention.%
}
\begin{multline}
J_{n}^{\left(++-\dots-\right)}\left(p_{1\dots n}\right)=\tilde{J}_{n}^{\left(++-\dots-\right)}\left(p_{1\dots n}\right)\\
-ig'\sum_{j=2}^{n-1}\tilde{J}_{j}^{\left(++-\dots-\right)}\left(p_{1\dots j}\right)\frac{p_{1\dots n}^{+}}{p_{j+1\dots n}^{+}\tilde{v}_{\left(1\dots j\right)\left(j+1\right)}^{*}}\, J_{n-j}^{\left(+-\dots-\right)}\left(p_{j+1\dots n}\right)\,,\label{eq:LF_recrel}
\end{multline}
where the structure on the r.h.s. appeared somewhat naturally, with
both quantities $\tilde{J}_{m}^{\left(++-\dots-\right)}$ and $J_{m}^{\left(+-\dots-\right)}$
derived by summing certain light-front diagrams. First, $J_{m}^{\left(+-\dots-\right)}$
is a current for transition of a minus helicity off-shell gluon to
all plus on-shell gluons and reads
\begin{equation}
J_{m}^{\left(+-\dots-\right)}\left(p_{1\dots m}\right)=-\left(-g'\right)^{m-1}\frac{\tilde{v}_{\left(1\dots m\right)1}^{*}}{\tilde{v}_{1\left(1\dots m\right)}^{*}}\,\frac{1}{\tilde{v}_{m\left(m-1\right)}^{*}\dots\tilde{v}_{32}^{*}\tilde{v}_{21}^{*}}\,.
\end{equation}
Note that this is identical to $\tilde{\Psi}_{m}$ -- a fact to which
we come back later in Section~\ref{sub:A[B]}. Second, the auxiliary quantity
$\tilde{J}_{m}^{\left(++-\dots-\right)}$ turned out to have the MHV
form, despite being off-shell. In fact
\begin{equation}
\tilde{J}_{m}^{\left(++-\dots-\right)}\left(p_{1\dots m}\right)=\left(g'\right)^{m-1}\frac{-i}{p_{1\dots n}^{2}}\left(\frac{p_{1}^{+}}{p_{2}^{+}}\right)^{2}\frac{\tilde{v}_{1\left(1\dots m\right)}^{*3}}{\tilde{v}_{\left(1\dots m\right)m}^{*}\tilde{v}_{m\left(m-1\right)}^{*}\tilde{v}_{\left(m-1\right)\left(m-2\right)}^{*}\dots\tilde{v}_{21}^{*}}\,,
\end{equation}
which is exactly \eqref{eq:CSW_current1} (up to the momentum conservation
delta function)
\begin{equation}
\tilde{J}_{n}^{\left(++-\dots-\right)}\sim\mathcal{J}\,.
\end{equation}
 Later, in \citep{Cruz-Santiago2015,Kotko2016} it was shown that
this result has indeed its origin in a matrix element of Wilson line,
identical to (\ref{eq:WilsonLine_ME}), which here we have derived
from the MHV Lagrangian. Actually, in \citep{Cruz-Santiago2015,Kotko2016}
the Wilson line slope was fixed and taken to be (with the present
notation) $\varepsilon_{1\dots n}^{+}=\varepsilon_{\perp}^{+}-(p_{1\dots n}^{\bullet}/p_{1\dots n}^{+})\eta$.
Here the slope is $\varepsilon_{\alpha}^{+}=\varepsilon_{\perp}^{+}-\alpha\eta$,
but as seen in the detailed computation from Section~\ref{sec:WilsonLineSolution}
the integration over $\alpha$ in (\ref{eq:WilsonLineAnsatz}) sets
$\alpha=p_{1\dots n}^{\bullet}/p_{1\dots n}^{+}$ when the Fourier
transform is taken.

To conclude, when considering recurrence relations within the light-front
perturbation theory, the partially reduced CSW vertex appears naturally.
Moreover, it is a matrix element of the straight infinite Wilson line,
thus it is gauge invariant which makes the resulting expressions simple,
cf. (\ref{eq:LF_recrel}). In the present paper, we have proved that
property directly from the MHV Lagrangian showing that in fact, the new
field $B^{\bullet}$ is expressed by the Wilson line.

Finally, let us notice that another class of gauge invariant currents
(or amplitudes) can be constructed by retaining both $B^{\bullet}$
fields off-shell. The resulting off-shell amplitude will have simple
MHV form and will be again gauge invariant (see the demonstration
with two Wilson lines in Ref.~\citep{Kotko2014a}). This property
for similar amplitudes with two off-shell legs was also observed in
\citep{vanHameren:2014iua} in the context of the high energy scattering, where the off-shell gluons have only longitudinal polarizations.

\section{Inverse transformation and energy denominators in the light-front theory}

\label{sec:Diagramma}

In this section we take a closer look at the solutions to the field
transformation (\ref{eq:Transformation_A+}), $B^{\bullet}=B^{\bullet}\left[A^{\bullet}\right]$
and $A^{\bullet}=A^{\bullet}\left[B^{\bullet}\right]$, using the Feynman
diagrams in momentum space. The main motivation to do this comes from
 calculations of the wave-functions and fragmentation functions \citep{Motyka2009}  within the light-front perturbation
theory \citep{Kogut1970,Brodsky1998}. Let us recall that in the light-front approach all lines
are treated on-shell (in the sense of having $p^{2}=0$) with the
transverse polarizations and one has to sum over various light-cone
time orderings. Instead of propagators, the intermediate states are
assigned energy denominators, similar to the non-relativistic quantum
mechanics. The energy denominator is the difference between the sum of light-cone energies $E_p=p^{\bullet} p^{\star}/p^+$ of the intermediate states, and the light-cone energy of the reference state (being the final or initial state).

We want to draw attention to two interesting results derived in \citep{Motyka2009}
using the light-front perturbation theory: (i) the calculation of the
$n$-gluon component of the gluon wave function with the helicity configuration
$\left(+\rightarrow+\dots+\right)$; (ii) the calculation of the fragmentation
function of a gluon with the helicity configuration $\left(+\rightarrow+\dots+\right)$.
Those two cases differ by the interpretation of the initial and final
states. For (i) the initial gluon is on-shell and it splits to the $n$-gluon
\textit{intermediate} state (being thus off-shell). For (ii) an intermediate
gluon state splits to bunch of final state gluons which are on-shell.
Thus the cases (i) and (ii) differ only by the usage of energy denominators,
more precisely, they differ only by a reference state which defines
the energy denominators: final or initial state. Yet, it turns out
that the result for both cases are very different. Now, the interesting
point is that the wave function (i) corresponds exactly to the expansion
coefficients $\tilde{\Theta}_{n}$ of the momentum space solution
$\tilde{B}^{\bullet}=\tilde{B}^{\bullet}\left[\tilde{A}^{\bullet}\right]$,
while the fragmentation function (ii) corresponds exactly to the expansion
coefficients $\tilde{\Psi}_{n}$ of the solution $\tilde{A}^{\bullet}=\tilde{A}^{\bullet}\left[\tilde{B}^{\bullet}\right]$.
Thus the coefficients of the inverse transformation can be obtained
by taking the same diagrams, but slightly different energy denominators. 

Below, we derive the coefficients $\tilde{\Theta}_{n}$ and $\tilde{\Psi}_{n}$
using the light-front diagrams and confirm the above statements.

\subsection{The diagrammatic content of $B^{\bullet}\left[A^{\bullet}\right]$}

\label{sub:Diagramm_B+}

In Section~\ref{sec:WilsonLineSolution} we have obtained the solution
$B^{\bullet}\left[A^{\bullet}\right]$ using the Wilson line. Here
we derive the solution by working directly in momentum space and introducing
light-front  diagrams. This is somewhat similar to what was
done in \citep{Mansfield2006}, but there the iterative solution was
given in position space and no reference to light-front  diagrams
was given. 

Similar to Section~\ref{sec:WilsonLineSolution} we express the field
$B^{\bullet}$ as a power expansion in $A^{\bullet}$ fields. In momentum
space we have
\begin{multline}
\tilde{B}_{a}^{\bullet}\left(\mathbf{p}\right)=\tilde{A}_{a}^{\bullet}\left(\mathbf{p}\right)+\int d^{3}\mathbf{q}_{1}d^{3}\mathbf{q}_{2}\,\tilde{\Gamma}_{2}^{a\left\{ b_{1}b_{2}\right\} }\left(\mathbf{p};\left\{ \mathbf{q}_{1},\mathbf{q}_{2}\right\} \right)\tilde{A}_{b_{1}}^{\bullet}\left(\mathbf{q}_{1}\right)\tilde{A}_{b_{2}}^{\bullet}\left(\mathbf{q}_{2}\right)+\dots\\
+\int d^{3}\mathbf{q}_{1}\dots d^{3}\mathbf{q}_{n}\,\tilde{\Gamma}_{n}^{a\left\{ b_{1}\dots b_{n}\right\} }\left(\mathbf{P};\left\{ \mathbf{q}_{1},\dots,\mathbf{q}_{n}\right\} \right)\tilde{A}_{b_{1}}^{\bullet}\left(\mathbf{q}_{1}\right)\dots\tilde{A}_{b_{n}}^{\bullet}\left(\mathbf{q}_{n}\right)+\dots\label{eq:BplusFT-1}
\end{multline}
where the symbols $\tilde{\Gamma}^{a\left\{ b_{1}\dots n_{n}\right\} }\left(\mathbf{P};\left\{ \mathbf{q}_{1},\dots,\mathbf{q}_{n}\right\} \right)$
are symmetric in pairs of indices $\left(b_{i},\mathbf{q}_{i}\right)$.
We use different symbols for the expansion coefficients than in Section~\ref{sec:WilsonLineSolution}
not to pre-assume they are identical.

It can be demonstrated  (see Appendix~\ref{sec:App_B[A]})  that the coefficients $\tilde{\Gamma}_{n}$  satisfy the following recursion relation
\begin{multline}
\tilde{\Gamma}_{n}^{a\left\{ b_{1}\dots b_{n}\right\} }\left(\mathbf{P};\left\{ \mathbf{p}_{1},\dots,\mathbf{p}_{n}\right\} \right)=-\frac{1}{2\left(E_{P}-E_{p_{1}}-\dots-E_{p_{n}}\right)}\,\frac{1}{s_{n}}\\
\Bigg\{\tilde{V}_{++-}^{cb_{n-1}b_{n}}\left(\mathbf{p}_{n-1},\mathbf{p}_{n}\right)\frac{1}{p_{n-1}^{+}+p_{n}^{+}}\tilde{\Gamma}_{n-1}^{a\left\{ b_{1}\dots b_{n-2}c\right\} }\left(\mathbf{P};\left\{ \mathbf{p}_{1},\dots,\mathbf{p}_{n-1}+\mathbf{p}_{n}\right\} \right)\\
+\tilde{V}_{++-}^{cb_{n-2}b_{n}}\left(\mathbf{p}_{n-2},\mathbf{p}_{n}\right)\frac{1}{p_{n-2}^{+}+p_{n}^{+}}\tilde{\Gamma}_{n-1}^{a\left\{ b_{1}\dots b_{n-3}b_{n-1}c\right\} }\left(\mathbf{P};\left\{ \mathbf{p}_{1},\dots,\mathbf{p}_{n-1},\mathbf{p}_{n-2}+\mathbf{p}_{n}\right\} \right)+\dots\\
+\tilde{V}_{++-}^{cb_{n-2}b_{n-1}}\left(\mathbf{p}_{n-2},\mathbf{p}_{n-1}\right)\frac{1}{p_{n-2}^{+}+p_{n-1}^{+}}\tilde{\Gamma}_{n-1}^{a\left\{ b_{1}\dots b_{n-3}cb_{n}\right\} }\left(\mathbf{P};\left\{ \mathbf{p}_{1},\dots,\mathbf{p}_{n-2}+\mathbf{p}_{n-1},\mathbf{p}_{n}\right\} \right)+\dots\Bigg\}\,,\label{eq:Gamma_n_recrel-1}
\end{multline}
where
\begin{equation}
\tilde{V}_{++-}^{a_{1}a_{2}a_{3}}\left(\mathbf{p}_{1},\mathbf{p}_{2}\right)=-igf^{a_{1}a_{2}a_{3}}\left(\frac{p_{1}^{\star}}{p_{1}^{+}}-\frac{p_{2}^{\star}}{p_{2}^{+}}\right)\left(-p_{12}^{+}\right)\,,\label{eq:vertex_++-}
\end{equation}
is the vertex appearing in the Lagrangian (see Eq.~(\ref{eq:App_vertex3g})).
The light-cone energy is
\begin{equation}
E_{p}=\frac{p^{\star}p^{\bullet}}{p^{+}}\,.\label{eq:Ep_def}
\end{equation}
The symmetry factor $s_{n}$ reads
\begin{equation}
s_{n}=\frac{n}{2}\,.
\end{equation}
The equation (\ref{eq:Gamma_n_recrel-1}) can be graphically represented
as follows

\begin{tabular}{>{\centering}m{0.87\columnwidth}>{\centering}m{0.05\columnwidth}}
\bigskip{}

\raggedright{}\includegraphics[height=0.06\paperheight]{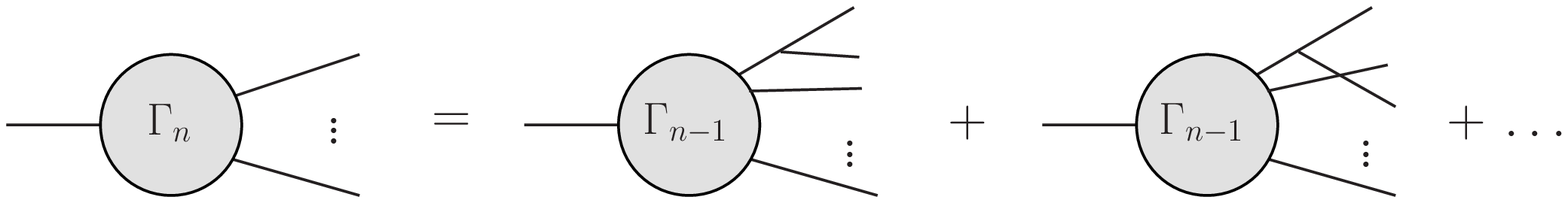} & \tabularnewline
\raggedleft{}\includegraphics[height=0.06\paperheight]{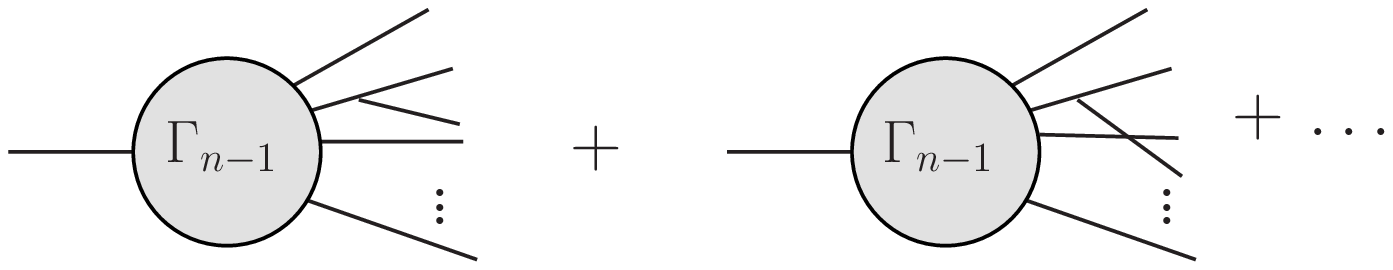} & \centering{}\tabularnewline
\[
+\dots
\]
 & \centering{}\centering{(\myref )}
\refmyref{Gamma_diag1}\tabularnewline
\end{tabular}

It is convenient to introduce the color decomposition for $\tilde{\Gamma}_{n}$
\begin{equation}
\tilde{\Gamma}_{n}^{a\left\{ b_{1}\dots b_{n}\right\} }\left(\mathbf{P};\left\{ \mathbf{p}_{1},\dots,\mathbf{p}_{n}\right\} \right)=\sum_{\sigma\in S_{n}}\mathrm{Tr}\left(t^{a}t^{b_{\sigma\left(1\right)}}\dots t^{b_{\sigma\left(n\right)}}\right)\tilde{\Gamma}_{n}\left(\mathbf{P};\mathbf{p}_{\sigma\left(1\right)},\dots,\mathbf{p}_{\sigma\left(n\right)}\right)\,,\label{eq:Gamma_colordecomp-1}
\end{equation}
where $S_{n}$ is the  permutation group. Solving the color-ordered analog
of the recursion (\ref{eq:Gamma_n_recrel-1}) (see Appendix~\ref{sec:App_B[A]})
we obtain
\begin{equation}
\tilde{\Gamma}_{n}\left(\mathbf{P};\mathbf{p}_{1},\dots,\mathbf{p}_{n}\right)=\frac{1}{n!}\,\left(-g'\right)^{n-1}\,\frac{1}{\tilde{v}_{1\left(1\dots n\right)}^{*}\tilde{v}_{\left(12\right)\left(1\dots n\right)}^{*}\dots\tilde{v}_{\left(1\dots n-1\right)\left(1\dots n\right)}^{*}}\,\delta^{3}\left(\mathbf{p}_{1}+\dots+\mathbf{p}_{n}-\mathbf{P}\right)\,.\label{eq:Gamma_n_sol-1-1}
\end{equation}
It is easy to see that (\ref{eq:Gamma_colordecomp-1}) with (\ref{eq:Gamma_n_sol-1-1})
is exactly (\ref{eq:Thetan}) in the `weak sense'. 
On the other hand, calculating a few first terms in the recursion
(\ref{eq:Gamma_n_recrel-1}) (see Appendix~\ref{sec:App_B[A]}) we
see that the expansion of $B^{\bullet}$ field has the following diagrammatic
content: 

\begin{flushleft}
\begin{tabular}{>{\centering}m{0.92\columnwidth}>{\centering}m{0.05\columnwidth}}
\bigskip{}

\centerline{\includegraphics[width=0.9\textwidth]{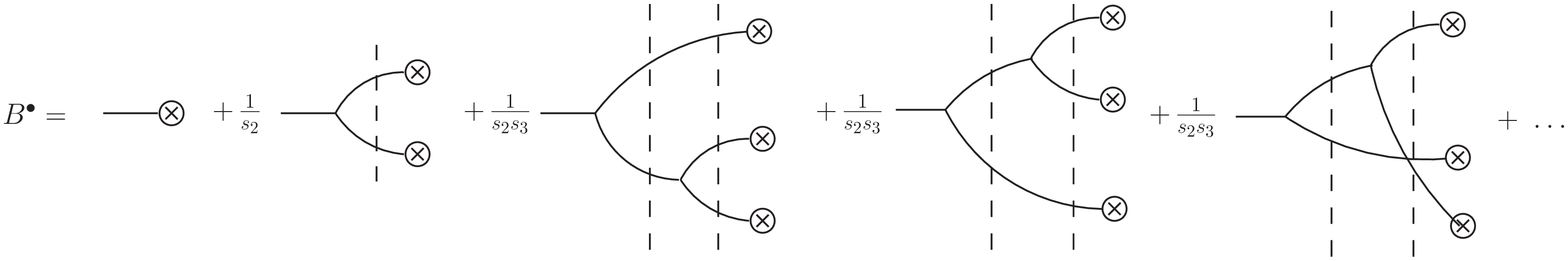}} & \centering{}\centering{(\myref )}
\refmyref{B+_diags}\tabularnewline
\end{tabular}
\par\end{flushleft}

\noindent Above the vertical dashed lines represent the energy denominators
$1/D_{1\dots i}$ where
\begin{equation}
D_{1\dots i}=2\left(E_{1\dots n}-\sum_{l=1}^{i}E_{l}\right)\,.\label{eq:Energy_denom}
\end{equation}
Above $E_{1\dots n}\equiv E_{p_{1\dots n}}$ is the energy of the
initial state and $\sum_{l=1}^{i}E_{l}$ is the sum of energies of
the lines that the cut crosses through (the intermediate states). In addition,
as usual in the light-front perturbation theory, a factor $1/p^{+}$
is assigned to each intermediate state.

It is interesting that the sum of the light-front diagrams containing
the triple gluon splitting given by (\ref{eq:vertex_++-}) and energy denominators
collapses into factors like (\ref{eq:Gamma_n_sol-1-1}). As mentioned,
this was for the first time shown in \citep{Motyka2009} without any
reference to the MHV action. In Section~\ref{sec:WilsonLineSolution}
we have shown, that they in fact are the expansion coefficients of
the Wilson line (\ref{eq:WilsonLineAnsatz}) in momentum space. Thus,
the gluon wave function  with gluon components with all-like helicities  is given  by the
Wilson line (\ref{eq:WilsonLineAnsatz}).

\subsection{The diagrammatic content of $A^{\bullet}\left[B^{\bullet}\right]$}
\label{sub:A[B]}
Consider now the inverse transformation, that is an expansion of $A^{\bullet}$
field in terms of $B^{\bullet}$ in momentum space:
\begin{multline}
\tilde{A}_{a}^{\bullet}\left(\mathbf{p}\right)=\tilde{B}_{a}^{\bullet}\left(\mathbf{p}\right)+\int d^{3}\mathbf{p}_{1}d^{3}\mathbf{p}_{2}\,\tilde{\Psi}_{2}^{a\left\{ b_{1}b_{2}\right\} }\left(\mathbf{p};\left\{ \mathbf{p}_{1},\mathbf{p}_{2}\right\} \right)\tilde{B}_{b_{1}}^{\bullet}\left(\mathbf{p}_{1}\right)\tilde{B}_{b_{2}}^{\bullet}\left(\mathbf{p}_{2}\right)+\dots\\
+\int d^{3}\mathbf{p}_{1}\dots d^{3}\mathbf{p}_{n}\,\tilde{\Psi}_{n}^{a\left\{ b_{1}\dots b_{n}\right\} }\left(\mathbf{p};\left\{ \mathbf{p}_{1},\dots,\mathbf{p}_{n}\right\} \right)\tilde{B}_{b_{1}}^{\bullet}\left(\mathbf{p}_{1}\right)\dots\tilde{B}_{b_{n}}^{\bullet}\left(\mathbf{p}_{n}\right)+\dots\,\label{eq:Az_expansion_mom}
\end{multline}
The expansion coefficient functions $\tilde{\Psi}^{a\left\{ b_{1}\dots b_{n}\right\} }\left(\mathbf{P};\left\{ \mathbf{p}_{1},\dots,\mathbf{p}_{n}\right\} \right)$
are symmetric in pairs of indices $\left(b_{i},\mathbf{p}_{i}\right)$.
These functions have been derived in \citep{Ettle2006b}. Here we
will concentrate on their diagrammatic content and the relation to $\tilde{\Gamma}_{n}$
 coefficients.

Inserting the expansion (\ref{eq:Az_expansion_mom}) to (\ref{eq:BplusFT-1})
and comparing coefficients with the same number of $B^{\bullet}$
fields we can derive a set of equations. For the two-point function
we have
\begin{multline}
\int d^{3}\mathbf{p}_{1}d^{3}\mathbf{p}_{2}\,\tilde{\Psi}_{2}^{a\left\{ b_{1}b_{2}\right\} }\left(\mathbf{p};\left\{ \mathbf{p}_{1},\mathbf{p}_{2}\right\} \right)\tilde{B}_{b_{1}}^{\bullet}\left(\mathbf{p}_{1}\right)\tilde{B}_{b_{2}}^{\bullet}\left(\mathbf{p}_{2}\right)\\
=-\int d^{3}\mathbf{q}_{1}d^{3}\mathbf{q}_{2}\,\tilde{\Gamma}_{2}^{a\left\{ b_{1}b_{2}\right\} }\left(\mathbf{p};\left\{ \mathbf{q}_{1},\mathbf{q}_{2}\right\} \right)\tilde{B}_{b_{1}}^{\bullet}\left(\mathbf{q}_{1}\right)\tilde{B}_{b_{2}}^{\bullet}\left(\mathbf{q}_{2}\right)\,.
\end{multline}
For the three-point function we get
\begin{multline}
\int d^{3}\mathbf{p}_{1}d^{3}\mathbf{p}_{2}d^{3}\mathbf{p}_{3}\,\tilde{\Psi}_{3}^{a\left\{ b_{1}b_{2}b_{3}\right\} }\left(\mathbf{p};\left\{ \mathbf{p}_{1},\mathbf{p}_{2},\mathbf{p}_{3}\right\} \right)\tilde{B}_{b_{1}}^{\bullet}\left(\mathbf{p}_{1}\right)\tilde{B}_{b_{2}}^{\bullet}\left(\mathbf{p}_{2}\right)\tilde{B}_{b_{3}}^{\bullet}\left(\mathbf{p}_{3}\right)\\
=-\int d^{3}\mathbf{q}_{1}d^{3}\mathbf{q}_{2}\int d^{3}\mathbf{p}_{1}d^{3}\mathbf{p}_{2}\,\tilde{\Gamma}_{2}^{a\left\{ b_{1}b_{2}\right\} }\left(\mathbf{p};\left\{ \mathbf{q}_{1},\mathbf{q}_{2}\right\} \right)\,\tilde{\Psi}_{2}^{b_{2}\left\{ c_{1}c_{2}\right\} }\left(\mathbf{q}_{2};\left\{ \mathbf{p}_{1},\mathbf{p}_{2}\right\} \right)\tilde{B}_{b_{1}}^{\bullet}\left(\mathbf{q}_{1}\right)\tilde{B}_{c_{1}}^{\bullet}\left(\mathbf{p}_{1}\right)\tilde{B}_{c_{2}}^{\bullet}\left(\mathbf{p}_{2}\right)\\
-\int d^{3}\mathbf{q}_{1}d^{3}\mathbf{q}_{2}\,\int d^{3}\mathbf{p}_{1}d^{3}\mathbf{p}_{2}\tilde{\Gamma}_{2}^{a\left\{ b_{1}b_{2}\right\} }\left(\mathbf{p};\left\{ \mathbf{q}_{1},\mathbf{q}_{2}\right\} \right)\,\tilde{\Psi}_{2}^{b_{1}\left\{ c_{1}c_{2}\right\} }\left(\mathbf{q}_{1};\left\{ \mathbf{p}_{1},\mathbf{p}_{2}\right\} \right)\tilde{B}_{c_{1}}^{\bullet}\left(\mathbf{p}_{1}\right)\tilde{B}_{c_{2}}^{\bullet}\left(\mathbf{p}_{2}\right)\tilde{B}_{b_{2}}^{\bullet}\left(\mathbf{q}_{2}\right)\\
-\int d^{3}\mathbf{q}_{1}d^{3}\mathbf{q}_{2}d^{3}\mathbf{q}_{3}\,\tilde{\Gamma}_{3}^{a\left\{ b_{1}b_{2}b_{3}\right\} }\left(\mathbf{P};\left\{ \mathbf{q}_{1},\mathbf{q}_{2},\mathbf{q}_{3}\right\} \right)\tilde{B}_{b_{1}}^{\bullet}\left(\mathbf{q}_{1}\right)\tilde{B}_{b_{2}}^{\bullet}\left(\mathbf{q}_{2}\right)\tilde{B}_{b_{3}}^{\bullet}\left(\mathbf{q}_{3}\right)\,,
\end{multline}
and so on. It is convenient to express the above recursion in a diagrammatic
form. Showing the planar diagrams only we have for $\tilde{\Psi}_{3}$:

\begin{flushleft}
\begin{tabular}{>{\centering}m{0.87\columnwidth}>{\centering}m{0.05\columnwidth}}
\bigskip{}

\raggedright{}\centerline{\includegraphics[height=0.06\paperheight]{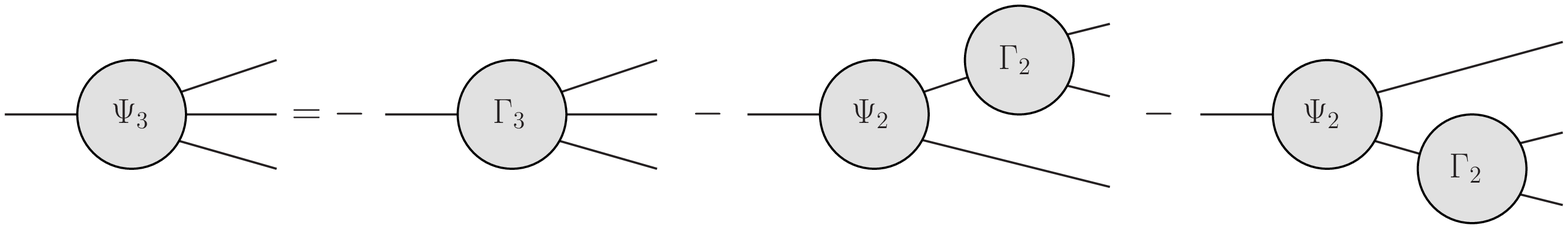}} & \centering{}\centering{(\myref )}
\refmyref{Psi3_diag}\tabularnewline
\end{tabular}
\par\end{flushleft}

\noindent For $\tilde{\Psi}_{4}$ we obtain 

\begin{flushleft}
\begin{tabular}{>{\centering}m{0.87\columnwidth}>{\centering}m{0.05\columnwidth}}
\bigskip{}

\raggedright{}\centerline{\includegraphics[height=0.06\paperheight]{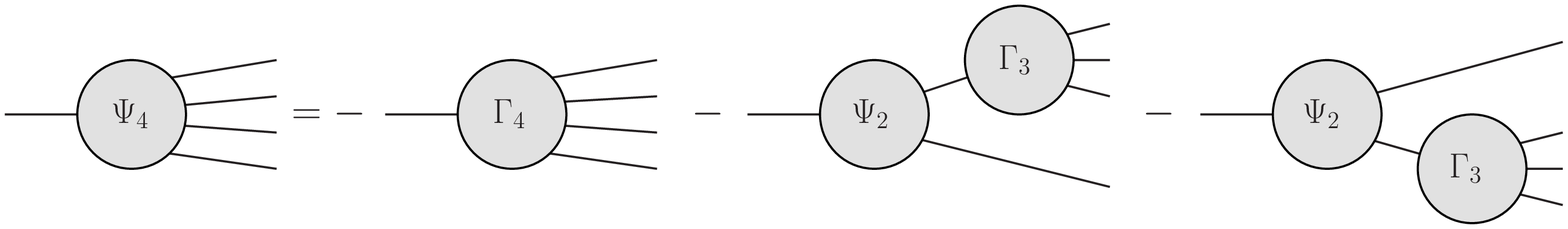}} & \tabularnewline
\end{tabular}
\par\end{flushleft}

\begin{flushleft}
\begin{tabular}{>{\centering}m{0.87\columnwidth}>{\centering}m{0.05\columnwidth}}
\raggedleft{}\includegraphics[height=0.06\paperheight]{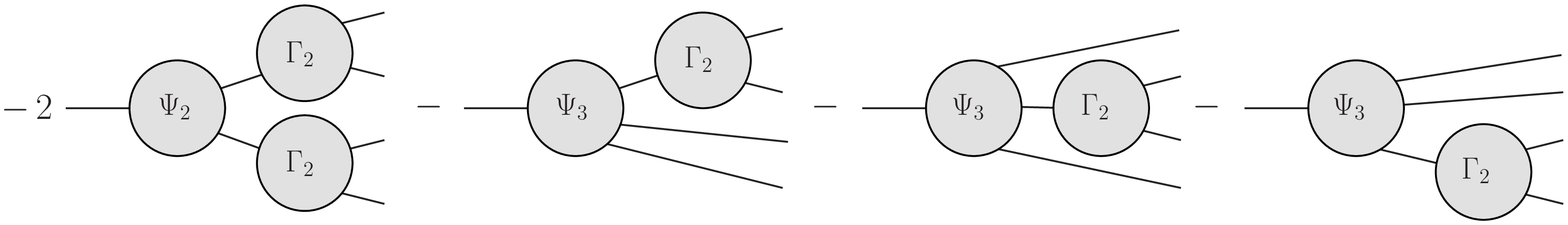} & \centering{}\centering{(\myref )}
\refmyref{Psi4_diag}\tabularnewline
\end{tabular}
\par\end{flushleft}

\noindent and so on. This can be expressed entirely in terms of the known
$\tilde{\Gamma}$ functions. Consider, for example, the $\tilde{\Psi}_{3}$
function. Using the fact that $\tilde{\Psi}_{2}=-\tilde{\Gamma}_{2}$
and utilizing the result for $\tilde{\Gamma}_{2}$ we have

\begin{flushleft}
\begin{tabular}{>{\centering}m{0.87\columnwidth}>{\centering}m{0.05\columnwidth}}
\bigskip{}

\raggedright{}\centerline{\includegraphics[height=0.08\paperheight]{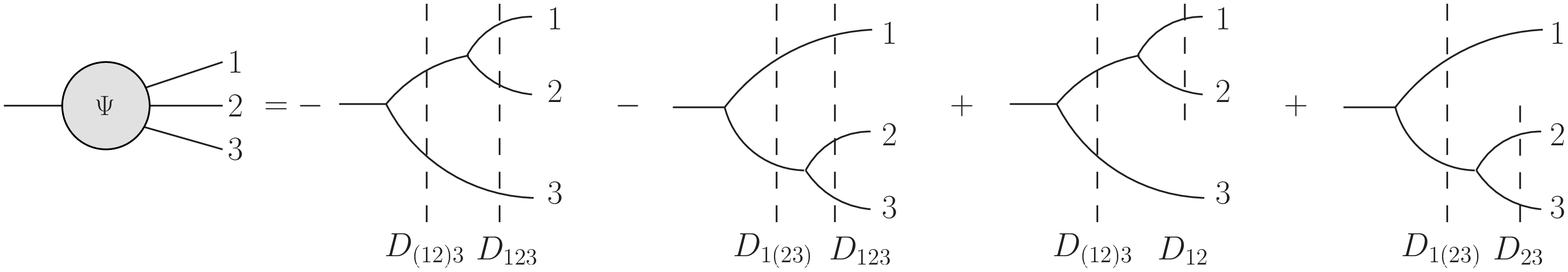}} & \centering{}\centering{(\myref )}
\refmyref{Psi3_diag_1}\tabularnewline
\end{tabular}
\par\end{flushleft}

\noindent The first (second) diagram differs from the third (fourth)
only by the one of the energy denominators. Noticing that 
\begin{gather}
D_{123}-D_{12}=D_{\left(12\right)3}\,,\\
D_{123}-D_{23}=D_{1\left(23\right)}\,,
\end{gather}
where $D_{\left(12\right)3}=2\left(E_{p_{123}}-E_{p_{12}}-E_{p_{3}}\right)$, 
etc., we can write (\arabic{Psi3_diag_1}) as

\begin{flushleft}
\begin{tabular}{>{\centering}m{0.87\columnwidth}>{\centering}m{0.05\columnwidth}}
\bigskip{}

\raggedright{}$\,\,\,\,\,\,\,\,\,\,\,\,\,\,\,\,\,\,\,\,\,\,\,\,\,\,\,\,\,\,\,\,\,\,\,\:\,\,\,\,\,\,\,\,\,\,\,\,\,\,\,\,\,\,\,\,\,\,$\includegraphics[height=0.08\paperheight]{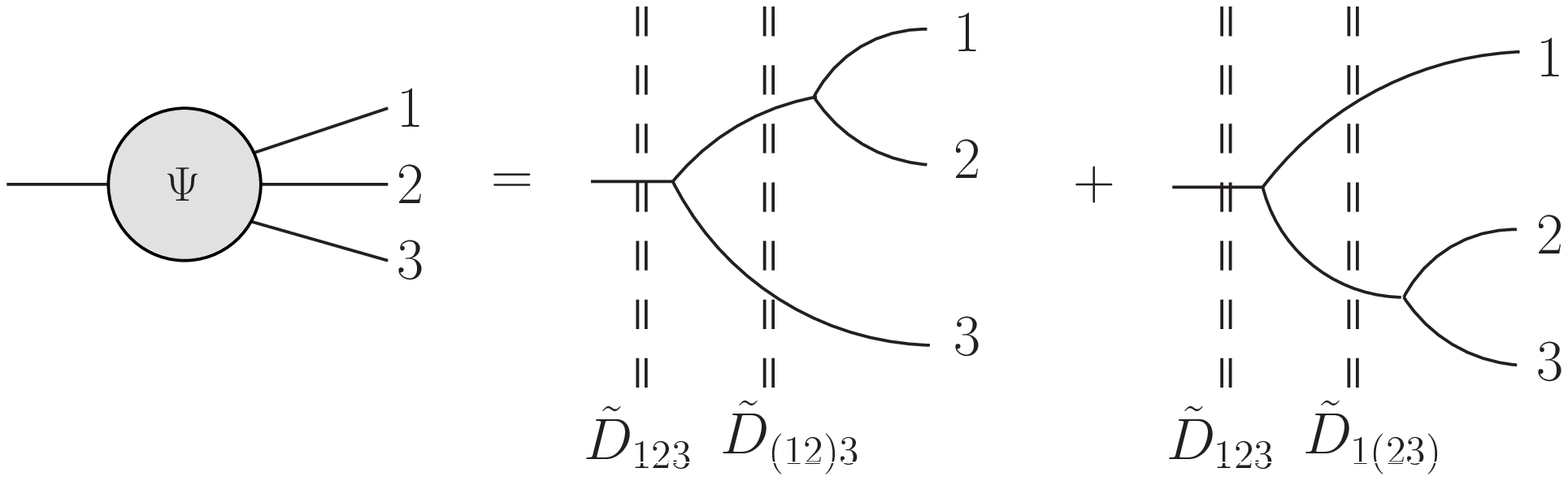} & \centering{}\centering{(\myref )}
\refmyref{Psi3_diag_2}\tabularnewline
\end{tabular}
\par\end{flushleft}

\noindent The double dashed line now denotes the energy denominators defined
with respect to the final states:
\begin{equation}
\tilde{D}_{1\dots i}=2\left(\sum_{k=1}^{n}E_{i}-\sum_{j=1}^{i}E_{j}\right)\,,\label{eq:Dtilde_def}
\end{equation}
where the first sum runs over the energies of the final states, while the second sum runs over the energies of the intermediate states. In our example, some of the relations to the old denominators are 
\begin{gather}
\tilde{D}_{123}=-D_{123}\,,\\
\tilde{D}_{\left(12\right)3}=-D_{12}\,,
\end{gather}
and so on. 

It is easy to see, that $\tilde{\Psi}_{4},\tilde{\Psi}_{5},\dots$
can be all calculated from similar diagrams. In fact, one can calculate
$\tilde{\Psi}_{n}$ from the following recursion relation (being in
fact the light-front equivalent of the Berends-Giele recursion \citep{Berends:1987me},  the so-called cluster decomposition \citep{Brodsky1986})

\begin{tabular}{>{\raggedleft}m{0.87\columnwidth}>{\centering}m{0.05\columnwidth}}
\medskip{}

\raggedright{}\centerline{\includegraphics[height=0.08\paperheight]{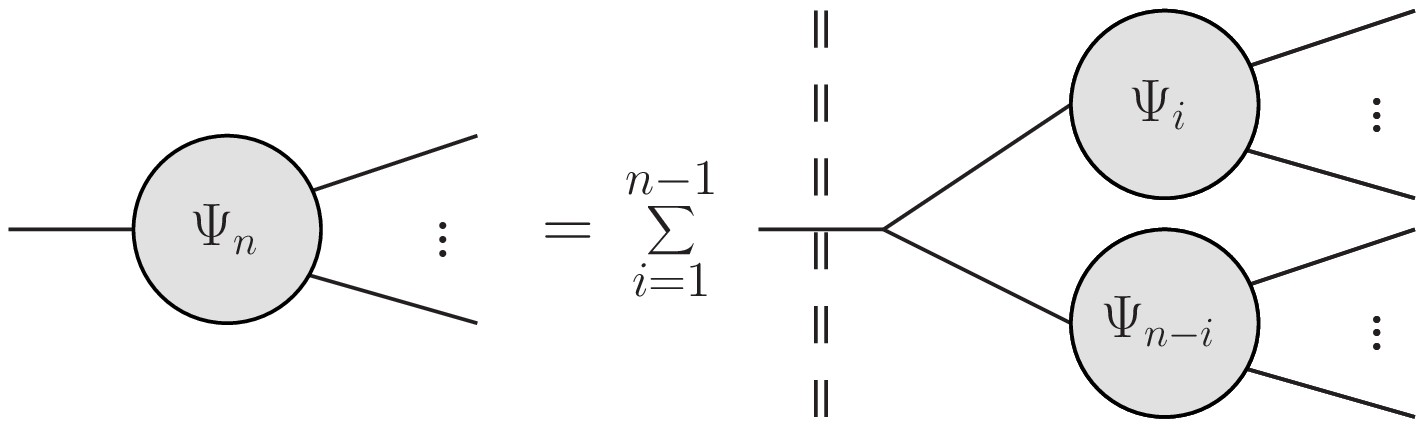}} & \centering{}\centering{(\myref )}
\refmyref{Psin_BerendsGiele}\tabularnewline
\end{tabular}

\noindent The standard Berends-Giele recursion for all-like helicity currents 
was solved long time ago in their original paper \citep{Berends:1987me}.
The above light-front version has been solved in \citep{Motyka2009}
(see also \citep{Cruz-Santiago2015,Kotko2016} for different methods
to solve it, using gauge invariance). The nature of the above diagrams, in particular the
nature of the energy denominators, allows to identify $\tilde{\Psi}_{n}$
with the fragmentation functions considered in \citep{Motyka2009}.
In fact, the resulting expression for color-ordered $\tilde{\Psi}_{n}$
reads 
\begin{equation}
\tilde{\Psi}_{n}\left(\mathbf{P};\mathbf{p}_{1},\dots,\mathbf{p}_{n}\right)=-\frac{1}{n!}\left(-g'\right)^{n-1}\,\frac{\tilde{v}_{\left(1\dots n\right)1}^{*}}{\tilde{v}_{1\left(1\dots n\right)}^{*}}\,\frac{1}{\tilde{v}_{n\left(n-1\right)}^{*}\dots\tilde{v}_{32}^{*}\,\tilde{v}_{21}^{*}}\,\delta^{3}\left(\mathbf{p}_{1}+\dots+\mathbf{p}_{n}-\mathbf{P}\right)\,\label{eq:Psi_n_sol-color_ordered}
\end{equation}
and it agrees with the results of \citep{Motyka2009} (modulo pre-factors
due to the normalization conventions). It also agrees with \citep{Ettle2006b}. 

Thus we have shown that mere change of energy denominators in the
diagrams contributing to $\tilde{\Gamma}_{n}$ transforms them into
$\tilde{\Psi}_{n}$, which are the coefficients of the inverse functional.
The coefficients $\tilde{\Gamma}_{n}$ are the expansion coefficients of the  Wilson line (\ref{eq:WilsonLineAnsatz}) in momentum space.  Thus a natural question arises as to what is the   inverse functional in coordinate space, i.e. an inverse of the Wilson line. 

\section{Inverse transformation in position space}

\label{sec:Inverse_transf}

We saw that the solution 
\begin{equation}
B^{\bullet}\left[A^{\bullet}\right]=\mathcal{W}\left[A^{\bullet}\right]\,,\label{eq:Functional_W}
\end{equation}
 is given by the Wilson line defined in (\ref{eq:WilsonLineAnsatz}).
On the other hand, the power expansion coefficients of that solution
and of the inverse solution 
\begin{equation}
A^{\bullet}\left[B^{\bullet}\right]=\mathcal{W}^{-1}\left[B^{\bullet}\right]\,\label{eq:functional_Winv}
\end{equation}
have interesting symmetry. As seen in the previous section, they contain
the same light-front Feynman diagrams, but the energy denominators for the intermediate states
are calculated with respect to different reference light-front energy, i.e. corresponding to either initial or final state.

Because of that observation, it is tempting to investigate how those
two  solutions differ in the position space. Thus,
in this section we construct an inverse functional to the Wilson line
(\ref{eq:WilsonLineAnsatz}). Looking at $\tilde{\Gamma}_{n}$ (recall
that $\tilde{\Gamma}_{n}=\tilde{\Theta}_{n}$) and comparing with
$\tilde{\Psi}_{n}$ gives an idea of how to construct the inverse
position space solution. Namely in $\tilde{\Gamma}_{n}$ we see a
chain of denominators containing $\tilde{v}_{\left(1\dots i\right)\left(1\dots n\right)}$
terms. They are easily understandable from the Feynman rules, see (\arabic{WL_diag_2}).
Since, the slope of the Wilson line is $\varepsilon_{1\dots n}^{+}=\varepsilon_{\perp}^{+}-\left(p_{1\dots n}^{\bullet}/p_{1\dots n}^{+}\right)\eta$
the consecutive double line propagators contribute 
\begin{multline}
\frac{1}{\left(p_{1\dots n}-p_{1}\right)\cdot\varepsilon_{1\dots n}}\,\frac{1}{\left(p_{1\dots n}-p_{12}\right)\cdot\varepsilon_{1\dots n}}\dots\frac{1}{\left(p_{1\dots n}-p_{1\dots n-1}\right)\cdot\varepsilon_{1\dots n}}\\
=\frac{\left(-1\right)^{n-1}}{\tilde{v}_{\left(1\dots n-1\right)\left(1\dots n\right)}\tilde{v}_{\left(1\dots n-2\right)\left(1\dots n\right)}\dots\tilde{v}_{1\left(1\dots n\right)}}\,.
\end{multline}
Now, looking at the expression for $\tilde{\Psi}_{n}$ \eqref{eq:Psi_n_sol-color_ordered}, we see that one could think of
a `Wilson line' which changes the slope after \textit{every} emission
to the polarization vector of the emitted gluon. 

Using the above observations as a guidance, we find a suitable functional
and prove that it has expansion in terms of $\tilde{\Psi}_{n}$ in
momentum space.

Consider a generic matrix-valued field $\hat{\phi}$. Let us define
the following functional
\begin{equation}
\mathcal{U}\left[\hat{\phi}\right]\left(\mathbf{x}\right)=\sum_{n=1}^{\infty}\int ds_{1}d\alpha_{1}\,\hat{\phi}\left(\mathbf{x}+s_{1}\mathbf{e}_{\alpha_{1}}\right)\prod_{i=2}^{n}\int ds_{i}d\alpha_{i}\int_{-\infty}^{0}d\tau_{i-1}\partial_{-}\hat{\phi}\left(\mathbf{x}+\tau_{i-1}\mathbf{e}_{\alpha_{i-1}}+s_{i}\mathbf{e}_{\alpha_{i}}\right)\,,\label{eq:U_functional}
\end{equation}
where the vector $\mathbf{e_{\alpha}}$ is defined in Eq.~(\ref{eq:SlopeDef}).
The variables $s_{i}$ and $\alpha_{i}$ run over the whole $\mathbb{R}$
space. We claim that 
\begin{equation}
\mathcal{W}^{-1}\left[\hat{B}^{\bullet}\right]\left(\mathbf{x}\right)=\frac{i}{g'}\,\partial_{-}\mathcal{U}\left[\frac{g'}{2\pi}\hat{B}^{\bullet}\right]\left(\mathbf{x}\right)\,.\label{eq:perturbiner_pos_space}
\end{equation}
In order to prove it, let us consider the $n$-th term in (\ref{eq:perturbiner_pos_space})
and let us pass to the momentum space:
\begin{multline}
\mathcal{W}^{-1}\left[\hat{B}^{\bullet}\right]^{\left(n\right)}=\frac{i\left(g'\right)^{n-1}}{\left(2\pi\right)^{n}}\,\partial_{-}\int ds_{1}d\alpha_{1}\dots ds_{n}d\alpha_{n}\int_{-\infty}^{0}d\tau_{1}\dots d\tau_{n-1}\,\hat{B}^{\bullet}\left(\mathbf{x}+s_{1}\mathbf{e}\left(\alpha_{1}\right)\right)\\
\partial_{-}\hat{B}^{\bullet}\left(\mathbf{x}+\tau_{1}\mathbf{e}\left(\alpha_{1}\right)+s_{2}\mathbf{e}\left(\alpha_{2}\right)\right)\dots\partial_{-}\hat{B}^{\bullet}\left(\mathbf{x}+\tau_{n-1}\mathbf{e}\left(\alpha_{n-1}\right)+s_{n}\mathbf{e}\left(\alpha_{n}\right)\right)\\
=\frac{i\left(g'\right)^{n-1}}{\left(2\pi\right)^{n}}\,\left(-i\right)^{n}\int d^{3}\mathbf{p}_{1}\dots d^{3}\mathbf{p}_{n}e^{-i\mathbf{x}\cdot\mathbf{p}_{1\dots n}}\tilde{\hat{B}}^{\bullet}\left(\mathbf{p}_{1}\right)\dots\tilde{\hat{B}}^{\bullet}\left(\mathbf{p}_{n}\right)p_{1\dots n}^{+}p_{2}^{+}\dots p_{n}^{+}\\
\int d\alpha_{1}\dots d\alpha_{n}\,\int ds_{1}\dots ds_{n}\, e^{-is_{1}\mathbf{p}_{1}\cdot\mathbf{e}\left(\alpha_{1}\right)}\dots e^{-is_{n}\mathbf{p}_{n}\cdot\mathbf{e}\left(\alpha_{n}\right)}\\
\int_{-\infty}^{0}d\tau_{1}\dots d\tau_{n-1}\, e^{-i\tau_{1}\mathbf{p}_{2}\cdot\mathbf{e}\left(\alpha_{1}\right)}\dots e^{-i\tau_{n-1}\mathbf{p}_{n}\cdot\mathbf{e}\left(\alpha_{n-1}\right)}\,.
\end{multline}
Performing elementary calculations similar to those in Section~\ref{sec:WilsonLineSolution}
we get
\begin{equation}
\mathcal{W}^{-1}\left[\hat{B}^{\bullet}\right]^{\left(n\right)}\left(\mathbf{x}\right)=-\left(-g'\right)^{n-1}\int d^{3}\mathbf{p}_{1}\dots d^{3}\mathbf{p}_{n}e^{-i\mathbf{x}\cdot\mathbf{p}_{1\dots n}}\frac{\tilde{v}_{\left(1\dots n\right)1}}{\tilde{v}_{1\left(1\dots n\right)}}\frac{1}{\tilde{v}_{n\left(n-1\right)}^{*}\dots\tilde{v}_{21}^{*}}\,\tilde{\hat{B}}^{\bullet}\left(\mathbf{p}_{1}\right)\dots\tilde{\hat{B}}^{\bullet}\left(\mathbf{p}_{n}\right)\,.\label{eq:Winv_n}
\end{equation}
The final formula reads
\begin{equation}
\tilde{A}_{a}^{\bullet}\left(\mathbf{P}\right)=\int d^{3}\mathbf{x}\, e^{i\mathbf{x}\cdot\mathbf{P}}A_{a}^{\bullet}\left(\mathbf{x}\right)=\int d^{3}\mathbf{x}\, e^{i\mathbf{x}\cdot\mathbf{P}}\,\mathrm{Tr}\left\{ \frac{i}{g'}\, t^{a}\partial_{-}\mathcal{U}\left[\frac{g'}{2\pi}\hat{B}^{\bullet}\right]\left(\mathbf{x}\right)\right\} \,.
\end{equation}
We see, that (\ref{eq:Winv_n}) exactly reproduces $\tilde{\Psi}_{n}$.

In the end, let us mention that the functional $\mathcal{W}^{-1}$
is actually the generating functional for the position space solutions
to the self-dual Yang-Mills equations \citep{Bardeen1996} (see also
\citep{Chalmers1996,Korepin1996,Cangemi1997,Rosly1997}). Namely
\begin{equation}
\Psi_{n}^{a\left\{ b_{1}\dots b_{n}\right\} }\left(\mathbf{x};\mathbf{y}_{1},\dots,\mathbf{y}_{n}\right)=\left.\frac{\delta A_{a}^{\bullet}\left[\phi\right]\left(\mathbf{x}\right)}{\delta\phi_{b_{1}}\left(\mathbf{y}_{1}\right)\dots\delta\phi_{b_{n}}\left(\mathbf{y}_{n}\right)}\right|_{\phi=0}\,.
\end{equation}
This gives $\Psi_{n}$ expressed in terms of delta functions and Heaviside
step functions. 

\section{On the geometry of functionals $\mathcal{W}$ and $\mathcal{W}^{-1}$}

\label{sec:Geometry}

In the previous section we have constructed the functional inverse
to (\ref{eq:WilsonLineAnsatz}), given by (\ref{eq:perturbiner_pos_space})
with (\ref{eq:U_functional}). The Wilson line solution has a very
clear geometrical interpretation. Namely, the gauge fields lie on
a line given by the path (\ref{eq:path}) with a slope $\alpha$,
which is ultimately integrated over. In this section we suggest a geometrical interpretation for the inverse solution. To do so we shall
construct a 2D space on which we will represent expansion (\ref{eq:U_functional})
using  certain vector field. We shall apply the same construction for
both functionals $\mathcal{W}$ and $\mathcal{W}^{-1}$ to represent
them both in a unified way. 

To this end let us consider a generic field $\phi$ (can be $A^{\bullet}$
or $B^{\bullet}$). Let us define the following object:
\begin{equation}
\mathfrak{p}_{\alpha}\left(\tau,\alpha'\right)=\partial_{-}\int_{-\infty}^{+\infty}ds\,\phi\left(\mathbf{x}+\tau\mathbf{e}_{\alpha'}+s\mathbf{e}_{\alpha}\right)\,,\label{eq:l_def_1}
\end{equation}
where the vector $\mathbf{e}_{\alpha}$ is defined in Eq.~(\ref{eq:SlopeDef}).
Notice that for $\phi=B^{\bullet}$ these are the objects that appear
in (\ref{eq:U_functional}). Above the point $\mathbf{x}$ is fixed
once and for all. The parameters $\tau,\alpha,\alpha'\in\mathbb{R}$.
Next, let us note that the three-vectors $\mathbf{e}_{\alpha}$ actually
have only two non-zero components, see (\ref{eq:SlopeDef}). Therefore
\begin{equation}
y^{*}=s+\tau,\,\,\,\, y^{-}=-\alpha s-\alpha'\tau\,
\end{equation}
can be thought of as a real coordinates on a 2D plane.

\begin{figure}
\begin{centering}
\includegraphics[width=6cm]{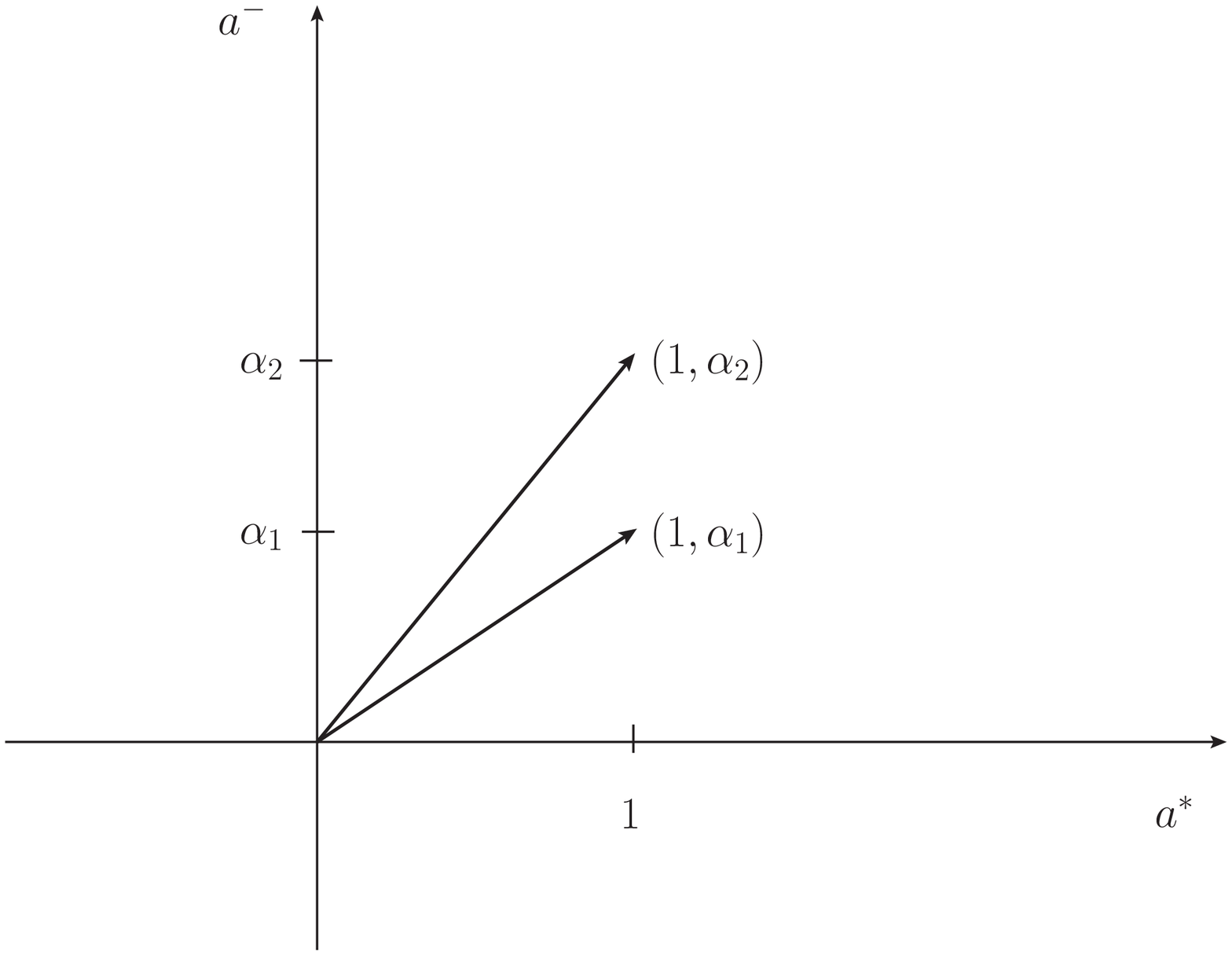}$\,\,\,\,\,\,\,$\includegraphics[width=6cm]{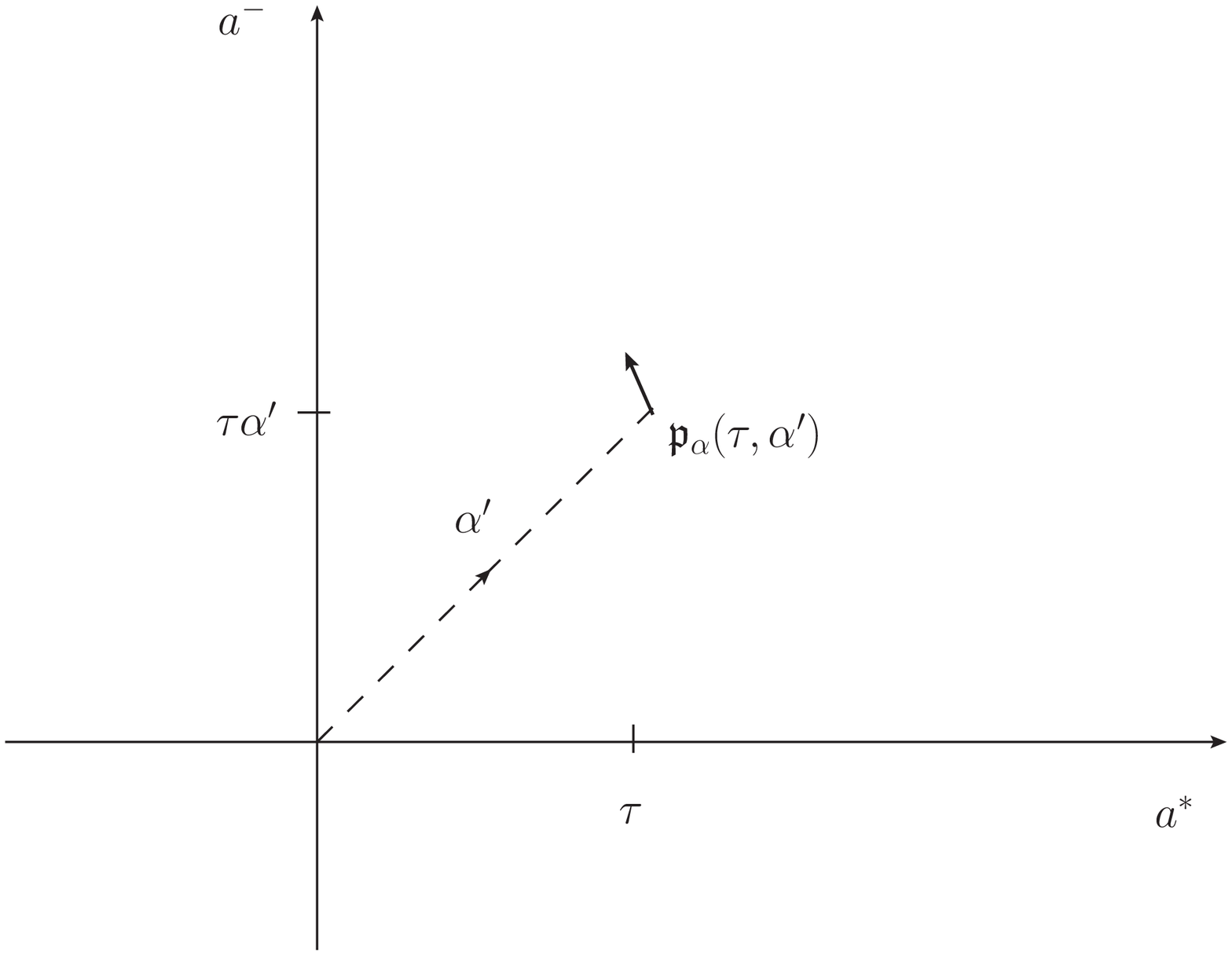}
\par\end{centering}

\caption{Left: representation of vectors $\mathbf{e}_{\alpha_{1}}$, $\mathbf{e}_{\alpha_{2}}$
defined in Eq.~(\ref{eq:SlopeDef}) in the 2D space. Right: Representation
of a vector \foreignlanguage{english}{$\mathfrak{p}_{\alpha}\left(\tau,\alpha'\right)$}
in that space.\label{fig:slope_def}}
\end{figure}

Therefore, consider an affine space consisting of the points 
\begin{equation}
a=\left(a^{*},a^{-}\right),\,\, a^{*},a^{-}\in\mathbb{R}\,.
\end{equation}
For our purposes, it is convenient to parametrize any point $c$ in
that space by the `slope' $\alpha'$ and the parameter $\tau$ 
\begin{equation}
c=\left(\tau,\tau\alpha'\right)\,.
\end{equation}
In particular, a point corresponding to the vector $\mathbf{e}_{\alpha}$
has coordinates
\begin{equation}
\mathbf{e}_{\alpha}\rightarrow\left(1,\alpha\right)\,.
\end{equation}
This is illustrated in left of Fig.~\ref{fig:slope_def}.

The object (\ref{eq:l_def_1}) can be thought of as a vector field
attached to the point $c=\left(\tau,\tau\alpha'\right)$ with a direction
given by $\mathbf{e}_{\alpha}$. The magnitude is given by the integral
in (\ref{eq:l_def_1}) and of course can vary from point to point. For simplicity we shall represent \foreignlanguage{english}{$\mathfrak{p}_{\alpha}\left(\tau,\alpha'\right)$}
as an arrow with some fixed length attached in the point $c$ and
oriented along $\mathbf{e}_{\alpha}$, see right of Fig.~\ref{fig:slope_def}.

\begin{figure}
\begin{centering}
\includegraphics[width=7.5cm]{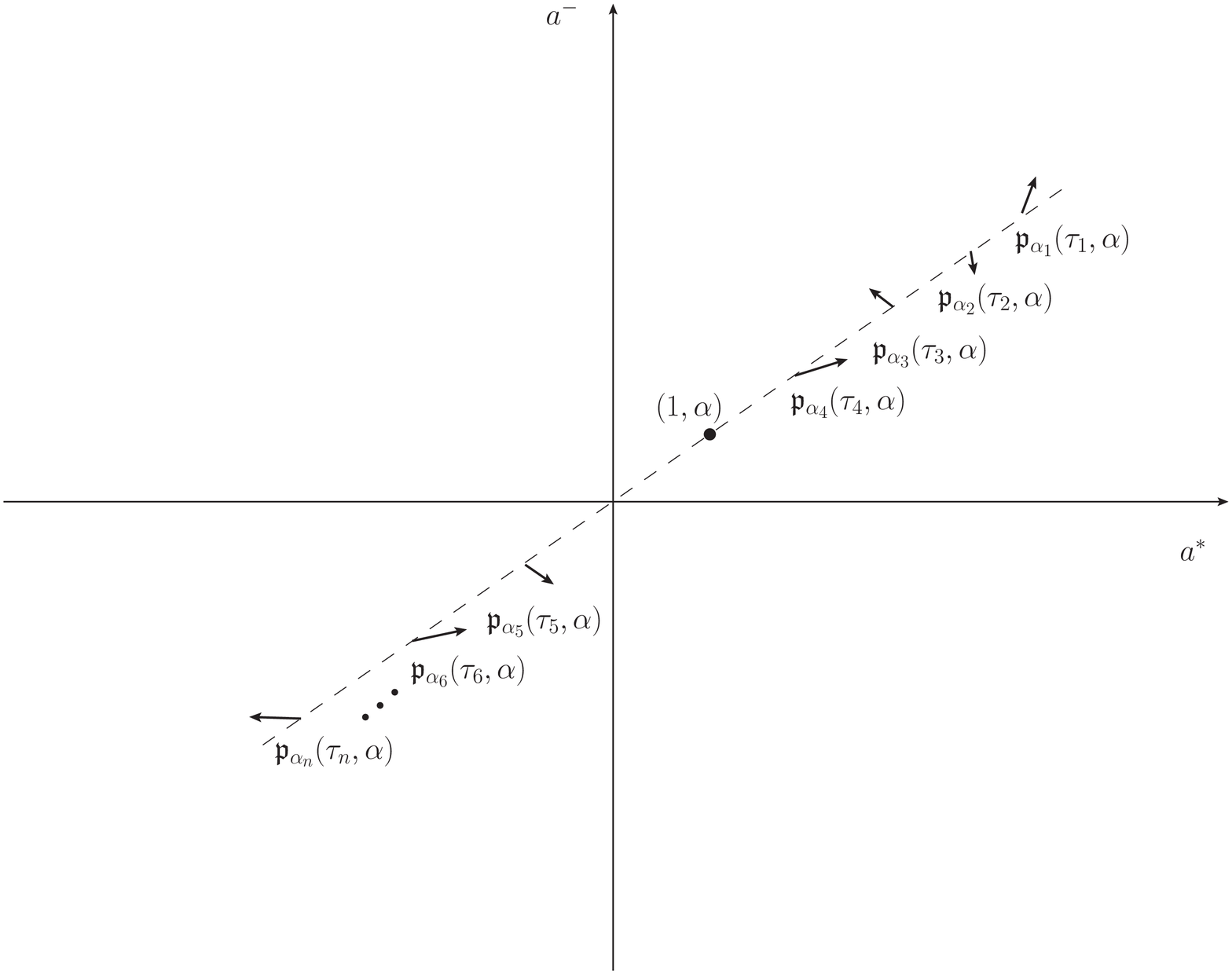}$\,\,\,$\includegraphics[width=7.5cm]{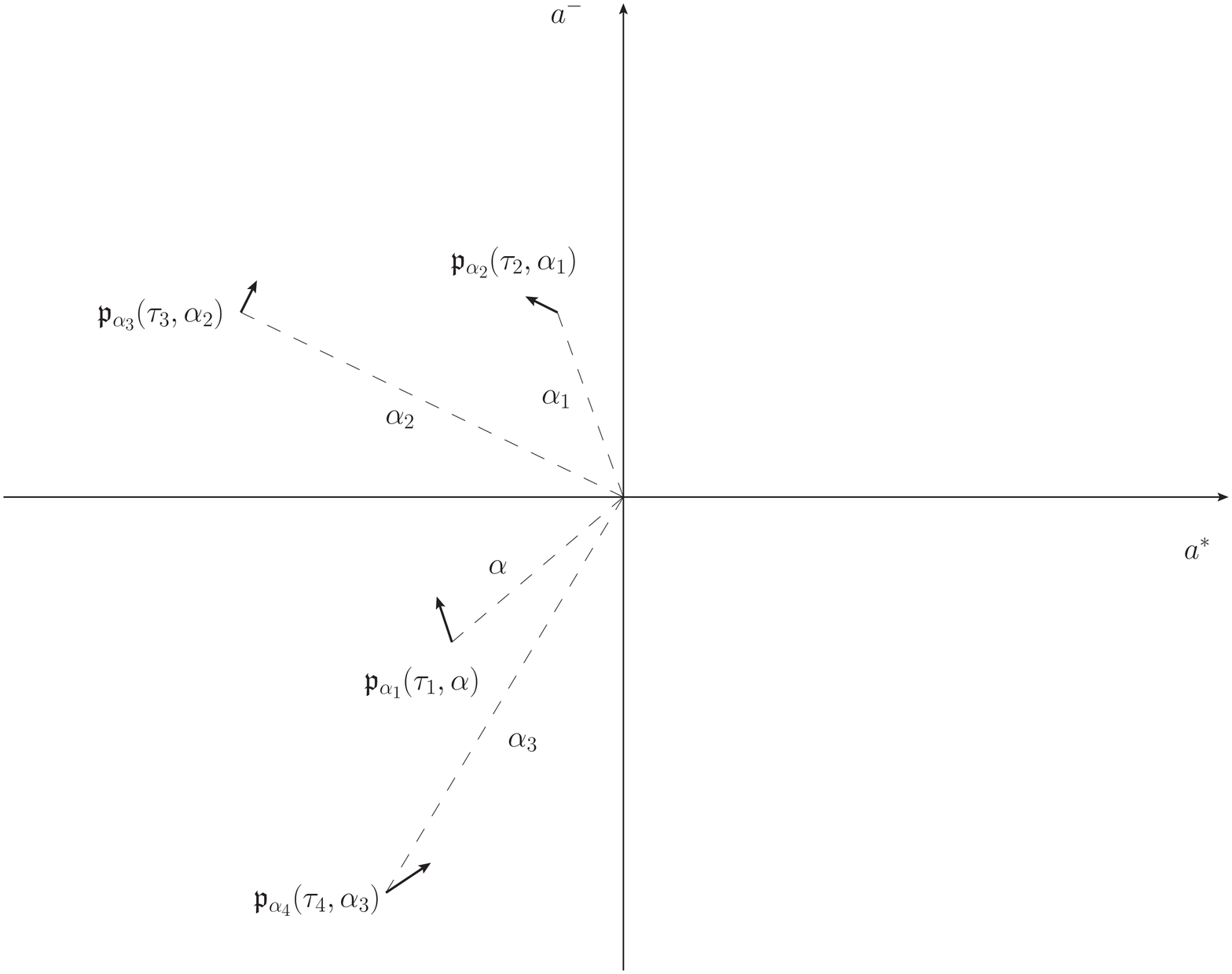}
\par\end{centering}

\caption{Left: representation of the object $\ell_{\alpha}^{\left(n\right)}$.
The vectors $\mathfrak{p}_{\alpha_{i}}\left(\tau_{i},\alpha\right)$
lie on the line given by $\mathbf{e}_{\alpha}$. Right: representation
of the object $\mathfrak{L}_{\alpha}^{\left(n\right)}$. The vectors
$\mathfrak{p}_{\alpha_{i}}\left(\tau_{i},\alpha\right)$ are scattered
on the plane along a piecewise path  which is determined as follows: the position of the 
following vector depends on the {\it direction} of the {\it preceding} vector. Thus, for example,
 the ray given by $\alpha_2$ on which the $\mathfrak{p}_{\alpha_3}(\tau_3,\alpha_2)$ is positioned 
 is parallel to the direction of the vector $\mathfrak{p}_{\alpha_2}(\tau_2,\alpha_1)$.
\label{fig:l_line}}
\end{figure}

Consider now the following object 
\begin{equation}
\ell_{\alpha}^{\left(n\right)}=\mathfrak{p}_{\alpha_{1}}\left(\tau_{1},\alpha\right)\mathfrak{p}_{\alpha_{2}}\left(\tau_{2},\alpha\right)\dots\mathfrak{p}_{\alpha_{n}}\left(\tau_{n},\alpha\right)\,,
\end{equation}
which -- according to the construction above -- may be identified
with a set depicted in left of Fig.~\ref{fig:l_line}. So far, the
objects we considered were defined for an unspecified scalar field
$\phi$. Here we want to work with matrix-valued gauge fields $\hat{A}$
or $\hat{B}$, thus we need to be careful about the exact order of objects
in expressions like the one above. To underline that we now deal with
the matrix-valued objects we put the hats:
\begin{equation}
\ell_{\alpha}^{\left(n\right)}=\hat{\mathfrak{p}}_{\alpha_{1}}\left(\tau_{1},\alpha\right)\hat{\mathfrak{p}}_{\alpha_{2}}\left(\tau_{2},\alpha\right)\dots\hat{\mathfrak{p}}_{\alpha_{n}}\left(\tau_{n},\alpha\right)\,.
\end{equation}

Let us now show, that the following integral 
\begin{equation}
\int d\alpha_{1}\dots d\alpha_{n}\int_{-\infty}^{+\infty}d\tau_{1}\int_{-\infty}^{\tau_{1}}d\tau_{2}\dots
\int_{-\infty}^{\tau_{n-1}}d\tau_{n}\,\,\hat{\mathfrak{p}}_{\alpha_{1}}\left(\tau_{1},\alpha\right)\hat{\mathfrak{p}}_{\alpha_{2}}\left(\tau_{2},\alpha\right)\dots\hat{\mathfrak{p}}_{\alpha_{n}}\left(\tau_{n},\alpha\right)\label{eq:Wilson_pvec}
\end{equation}
is directly related to the $n$-th coefficient in the expansion of
the Wilson line in (\ref{eq:WilsonLineAnsatz}) (when $\hat{\phi}=\hat{A}^{\bullet}$
of course). This is most readily performed in the momentum space. We use (\ref{eq:l_def_1})
and perform the Fourier transform
\begin{multline}
\int d\alpha_{1}\dots d\alpha_{n}\int ds_{1}\dots ds_{n}\int_{-\infty}^{+\infty}d\tau_{1}\int_{-\infty}^{\tau_{1}}d\tau_{2}\dots\int_{-\infty}^{\tau_{n-1}}d\tau_{n}\,\\
\partial_{-}\hat{A}^{\bullet}\left(\mathbf{x}+s_{1}\mathbf{e}_{\alpha_{1}}+\tau_{1}\mathbf{e}_{\alpha}\right)\dots\partial_{-}\hat{A}^{\bullet}\left(\mathbf{x}+s_{n}\mathbf{e}_{\alpha_{n}}+\tau_{n}\mathbf{e}_{\alpha}\right)\\
=\int d^{3}\mathbf{p}_{1}\dots d^{3}\mathbf{p}_{n}e^{-i\mathbf{x}\cdot\mathbf{p}_{1\dots n}}p_{1}^{+}\dots p_{n}^{+}\hat{\tilde{A}}^{\bullet}\left(\mathbf{p}_{1}\right)\dots\hat{\tilde{A}}^{\bullet}\left(\mathbf{p}_{n}\right)\\
\int d\alpha_{1}\dots d\alpha_{n}\int ds_{1}\dots ds_{n}e^{-is_{1}\mathbf{p}_{1}\cdot\mathbf{e}_{\alpha_{1}}}\dots e^{-is_{n}\mathbf{p}_{n}\cdot\mathbf{e}_{\alpha_{n}}}\\
\int_{-\infty}^{+\infty}d\tau_{1}\int_{-\infty}^{\tau_{1}}d\tau_{2}\dots\int_{-\infty}^{\tau_{n-1}}d\tau_{n}\, e^{-i\tau_{1}\mathbf{p}_{1}\cdot\mathbf{e}_{\alpha}}\dots e^{-i\tau_{n}\mathbf{p}_{n}\cdot\mathbf{e}_{\alpha}}\,.
\end{multline}
Note, that 
\begin{equation}
p_{1}^{+}\dots p_{n}^{+}\int d\alpha_{1}\dots d\alpha_{n}\int ds_{1}\dots ds_{n}e^{-is_{1}\mathbf{p}_{1}\cdot\mathbf{e}_{\alpha_{1}}}\dots e^{-is_{n}\mathbf{p}_{n}\cdot\mathbf{e}_{\alpha_{n}}}=1\,
\end{equation}
and we are left with
\begin{multline}
\int d^{3}\mathbf{p}_{1}\dots d^{3}\mathbf{p}_{n}e^{-i\mathbf{x}\cdot\mathbf{p}_{1\dots n}}p_{1}^{+}\dots p_{n}^{+}\hat{\tilde{A}}^{\bullet}\left(\mathbf{p}_{1}\right)\dots\hat{\tilde{A}}^{\bullet}\left(\mathbf{p}_{n}\right)\\
\int_{-\infty}^{+\infty}d\tau_{1}\int_{-\infty}^{\tau_{1}}d\tau_{2}\dots\int_{-\infty}^{\tau_{n-1}}d\tau_{n}\, e^{-i\tau_{1}\mathbf{p}_{1}\cdot\mathbf{e}_{\alpha}}\dots e^{-i\tau_{n}\mathbf{p}_{n}\cdot\mathbf{e}_{\alpha}}\,,
\end{multline}
which is exactly the same chain of integrals as in (\ref{eq:NthWLcoef}) up to normalization prefactors which are not essential for the present discussion.

Thus, to conclude, the solution given by the Wilson line can be represented
by the infinite sum of the objects like the one in Fig.~\ref{fig:l_line}.

\begin{figure}
\begin{centering}
\includegraphics[width=7cm,height=6cm]{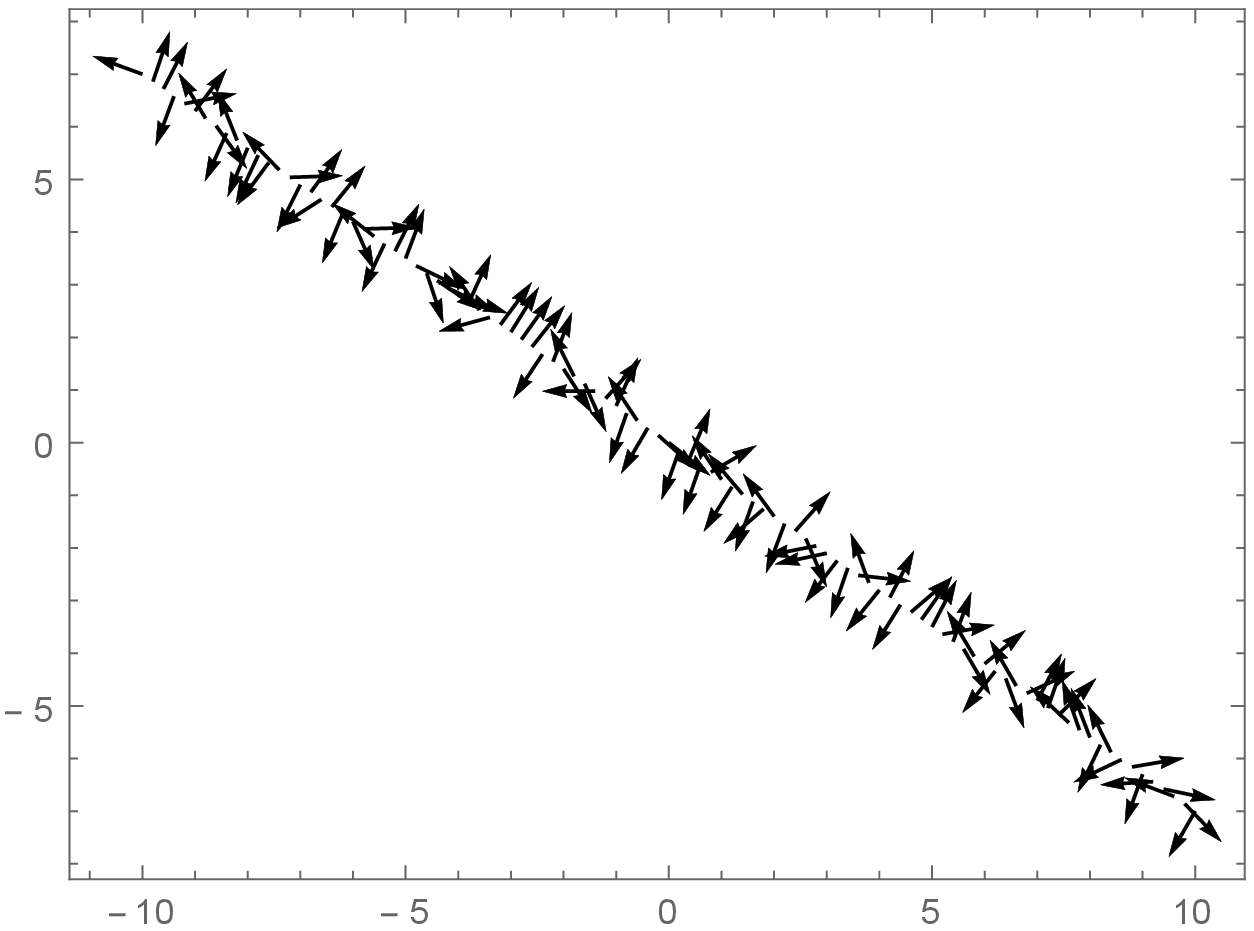}$\,\,\,\,\,\,$\includegraphics[width=7cm,height=6cm]{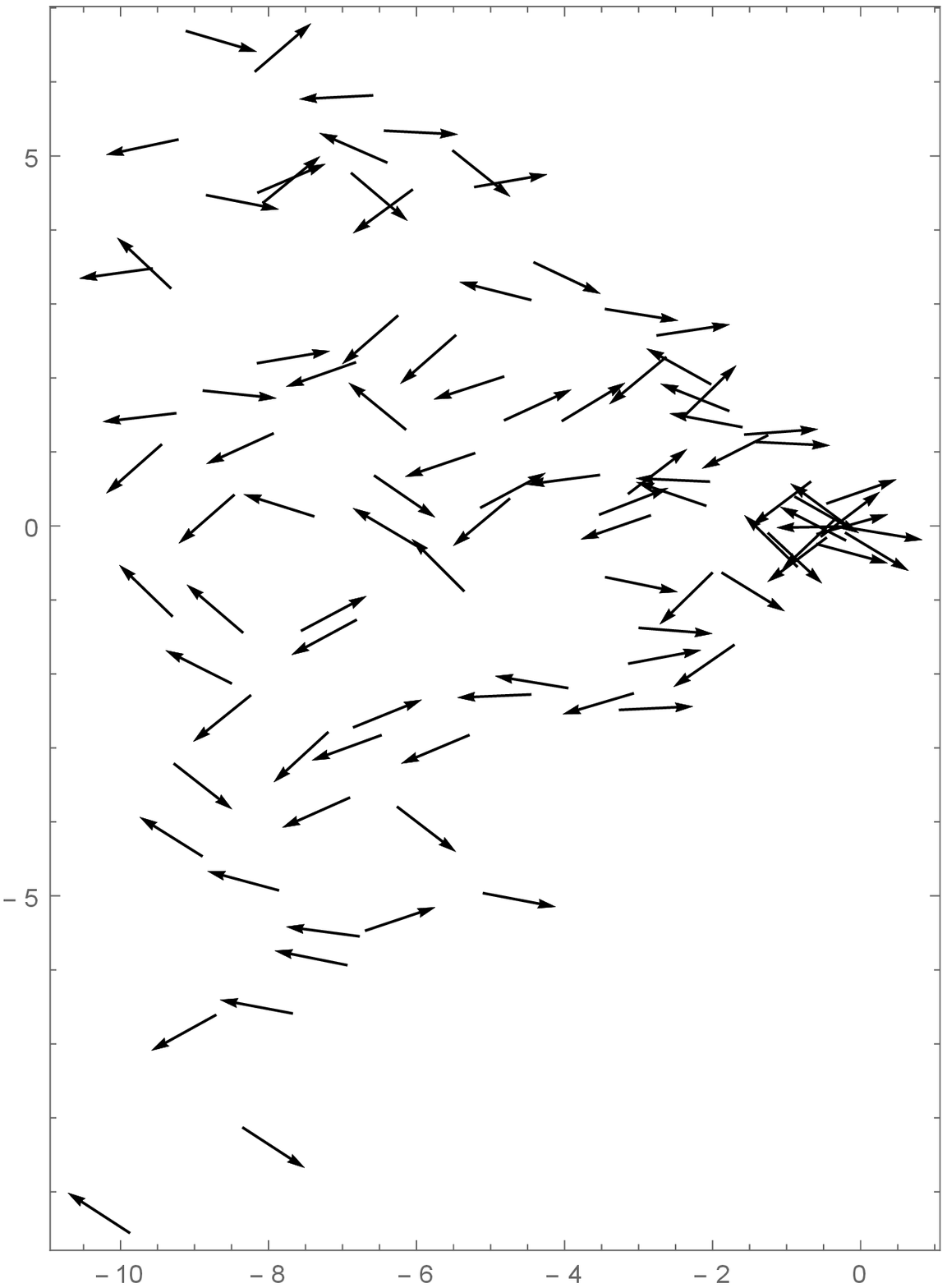}
\par\end{centering}

\caption{Similar to Fig.~\ref{fig:l_line}. Left: the functional $\mathcal{W}$,
(\ref{eq:Functional_W}), is represented as a vector field supported
on an ensemble of straight lines crossing the origin. Only one particular
line is shown.  Right: the inverse functional $\mathcal{W}^{-1}$,
(\ref{eq:functional_Winv}), is represented by a vector field supported
on a plane.
Here, only a cone with a limited slope is shown instead of the whole
half-plane. \label{fig:simulations}}
\end{figure}

Now, let us come back to our original question, which now can be formulated
as follows: what is the geometric object in our 2D space the inverse
functional (\ref{eq:U_functional}) corresponds to? 

Let us consider the
$n$-th term from the (\ref{eq:U_functional}):
\begin{equation}
\mathcal{U}\left[\hat{B}^{\bullet}\right]^{\left(n\right)}=\int ds_{1}d\alpha_{1}\,\hat{B}^{\bullet}\left(\mathbf{x}+s_{1}\mathbf{e}_{\alpha_{1}}\right)\prod_{i=2}^{n}\int ds_{i}d\alpha_{i}\int_{-\infty}^{0}d\tau_{i-1}\partial_{-}\hat{B}^{\bullet}\left(\mathbf{x}+\tau_{i-1}\mathbf{e}_{\alpha_{i-1}}+s_{i}\mathbf{e}_{\alpha_{i}}\right)\,.\label{eq:U_functional-nth_1}
\end{equation}
First, note that the expression does look asymmetric as the first field does not
have a derivative and the $\tau$-dependent part. It can be however
easily put into a symmetric form as follows:
\begin{multline}
\mathcal{U}\left[\hat{B}^{\bullet}\right]^{\left(n\right)}=\frac{i}{2\pi}\int d\alpha\,\,\int ds_{1}d\alpha_{1}\int_{-\infty}^{\infty}d\tau_{1}\,\partial_{-}\hat{B}^{\bullet}\left(\mathbf{x}+\tau_{1}\mathbf{e}_{\alpha}+s_{1}\mathbf{e}_{\alpha_{1}}\right)\\
\prod_{i=2}^{n}\int ds_{i}d\alpha_{i}\int_{-\infty}^{0}d\tau_{i}\partial_{-}\hat{B}^{\bullet}
\left(\mathbf{x}+\tau_{i}\mathbf{e}_{\alpha_{i-1}}+s_{i}\mathbf{e}_{\alpha_{i}}\right)\,.\label{eq:U_functional-nth_1-1}
\end{multline}
Let us represent it in terms of the vectors $\mathfrak{p}$. Similar
as in (\ref{eq:Wilson_pvec}) we will fix the $\alpha$ parameter
(in (\ref{eq:Wilson_pvec}) as well as here it will be integrated
over in the end). Denoting the corresponding chain of $\mathfrak{p}$ vectors by  $\mathfrak{L}_{\alpha}^{\left(n\right)}$ we have 
\begin{multline}
\mathfrak{L}_{\alpha}^{\left(n\right)}=\int d\alpha_{1}\dots d\alpha_{n}\,\int_{-\infty}^{\infty}d\tau_{1}\int_{-\infty}^{0}d\tau_{2}\dots\int_{-\infty}^{0}d\tau_{n}\, \\ \hat{\mathfrak{p}}_{\alpha_{1}}\left(\tau_{1},\alpha\right)\hat{\mathfrak{p}}_{\alpha_{2}}\left(\tau_{2},\alpha_{1}\right)\hat{\mathfrak{p}}_{\alpha_{3}}\left(\tau_{3},\alpha_{2}\right)\dots\hat{\mathfrak{p}}_{\alpha_{n}}\left(\tau_{n},\alpha_{n-1}\right)\,.
\end{multline}
The above integrand is represented in right of Fig.~\ref{fig:l_line}. 

There is one comment in order here. The integrals over positions $\tau$,
are not ordered here (compare Eq.~(\ref{eq:Wilson_pvec})). This
is in fact not required to maintain the order, as each vector has
different direction and there is no risk of misplacing them.

To summarize, the functionals $\mathcal{W}$ and $\mathcal{W}^{-1}$,
(\ref{eq:Functional_W}), (\ref{eq:functional_Winv}), can be characterized
by two entirely different supports and ordering of the fields. The functional $\mathcal{W}$
which contains the Wilson line can be characterized by a vector field
supported on an ensemble of lines crossing the origin in the 2D space,
see Fig.~\ref{fig:simulations} left. The orientations of the vectors have to be integrated over. 
The order of vectors is given by the parameter along the line. On the other hand, the inverse
functional, $\mathcal{W}^{-1}$, is characterized by a vector field
scattered across the plane. The positions and orientations of vectors have to be integrated over. 
The order is defined by the orientations of the vectors: the preceding vector orientation determines 
one coordinate of the position of the following vector, see Fig.~\ref{fig:simulations}
right. 

\section{Summary and conclusions}

\label{sec:Summary}

This work was motivated by several interesting results obtained previously within
the light-front perturbation theory. First, by the solution to the light-front
recursion relations for the amplitudes with MHV helicity configuration \citep{Cruz-Santiago2013}.
It was observed that, in the off-shell current which is the solution to the recursion, 
appeared an object which had exactly the MHV form despite the fact that it was off-shell.
Thus it  resembled the MHV vertex which appeared in the CSW formalism.
 In \citep{Cruz-Santiago2015}
it was demonstrated that in fact this object can be calculated from a 
matrix element of straight infinite Wilson line and thus it is gauge
invariant (see Section~\ref{sec:GaugeInvAmps}). Therefore it was a natural expectation that the transformation \eqref{eq:Transformation_A+} leading to the MHV Lagrangian contained Wilson line as well.
In Section ~\ref{sec:WilsonLineSolution} we have proven that this is indeed the case.
The straight infinite Wilson line solving
the transformation is however  different than what one can
typically find in the literature. Namely, the slope of the line is
defined by a polarization-like vector $\varepsilon_{\alpha}^{+}$,
where the parameter $\alpha$ sets the longitudinal component of the
vector, see (\ref{eq:SlopeDef0}). This parameter is integrated over
in the Wilson line (actually its derivative).
 In the momentum space the integration
over $\alpha$ effectively sets $\varepsilon_{\alpha}^{+}$ to be equal to
the polarization vector for momentum of the Wilson line.

Another direction of study, motivated by the light-front perturbation
theory, in particular results of \citep{Motyka2009}, was to investigate the diagrammatic content of the field transformation (\ref{eq:Transformation_A+}).
As demonstrated in Section~\ref{sec:Diagramma}  it turns out that the Wilson line solution and the inverse solution differ only
by the definition of the light-front energy denominators, containing otherwise
the same gluon cascade. In \citep{Motyka2009}
the same observation distinguishes the gluon wave function (with on-shell initial state) and the
gluon fragmentation function (with on-shell final states). In the latter work it was also observed
that, under certain kinematic assumption, both objects are dual, in
the sense that if one transforms one of them to position space, the
resulting expression will equal the second if positions are simply
replaced by momenta. This motivated a study of the inverse solution,  i.e. $A^{\bullet}[B^{\bullet}]$,
 in the position space. Such inverse
solution was constructed in Section~\ref{sec:Inverse_transf} and it is given by a simple power expansion in fields with shifting arguments
along certain directions, see Eq.~(\ref{eq:U_functional}). Unlike
the Wilson line, which is supported on an ensemble of infinite straight lines in a 2D
space, the inverse solution is scattered over the whole plane (see Section~\ref{sec:Geometry}). 

The inverse solution to the one given by the Wilson line is interesting
also on its own. The coefficients of its momentum space expansion in powers of fields can be identified with the  off-shell
currents for all-like helicity gluons. The latter  are the solutions to the
self-dual sector of the Yang-Mills equations \citep{Bardeen1996}.
In fact, the transformation leading to the MHV action transforms the
self-dual part into the kinetic part of the new action (see also \citep{Gorsky2006}).
On the other hand the self-dual Yang-Mills equations are closely connected
to integrable systems \citep{Ablowitz1993}. Thus, it might be interesting
to investigate whether the position space forms of the solution --
the one that generates the self-dual solutions and its inverse given
by the Wilson line -- might be useful in that matter.

There are several further potential directions of study  which we briefly outline  below.

One is a question related to the high energy limit of QCD. Namely,
in \citep{Lipatov:1995pn} a gauge-invariant effective action for
high-energy QCD was proposed. The crucial element of this action are
new fields, which are in fact Wilson lines when the equations of motion
are utilized. Those Wilson lines have different slope though, then
in the present case (being simply the $\pm$ light-cone direction).
However, the structure of the MHV vertices with Wilson lines (\ref{eq:WilsonLineAnsatz})
resembles the structure of vertices in that theory. Thus, it would be interesting to investigate if the high-energy limit of the MHV action could be taken and how it would be related to the high-energy effective action proposed in \citep{Lipatov:1995pn}.

Another potential direction of study concerns loop corrections. It
is known that the MHV action (\ref{eq:MHV_action}) is not complete
at loop level. As recalled in the Introduction, two solutions have
been proposed. One based on the dimensional regularization and evasion
of the S-matrix equivalence theorem \citep{Ettle2007}, and another
using world-sheet regularization \citep{Brandhuber2007} which leads
to additional term in the light-cone action (\ref{eq:actionLC}) and
thus additional terms in the MHV action. The question is whether the
information that the solution to transformation is given by the Wilson
line might be helpful, in the sense, that the renormalization of Wilson
lines is a known issue from many aspects.

\section*{Acknowledgements}

We are grateful to Lance Dixon and Mirko Serino for turning our attention
to the CSW construction. We also acknowledge discussions with Radu Roiban
and Leszek Motyka.

The work was supported by the Department of Energy Grants No. DE-SC-0002145,
DE-FG02-93ER40771 as well as the National Science Center, Poland, Grant No. 2015/17/B/ST2/01838. The diagrams were drawn using the Jaxodraw \citep{Binosi:2003yf}.

\newpage 

\appendix

\section{Useful identities}

\label{sec:App_ident}

For reader's convenience we collect some of the properties of the
symbols $\tilde{v}_{ij}$ introduced in (\ref{eq:vtild_def}).
\begin{enumerate}
\item $v_{ij}=-v_{ji}$, but $\tilde{v}_{ij}=-\frac{p_{i}^{+}}{p_{j}^{+}}\,\tilde{v}_{ji}$
\label{enu:vtild_prop_1}
\item $\tilde{v}_{ii}=0$ \label{enu:vtild_prop_2}
\item $\tilde{v}_{ij}-\tilde{v}_{il}=\frac{p_{i}^{+}}{p_{l}^{+}}\,\tilde{v}_{lj}$
\label{enu:vtild_prop_3}
\item $\tilde{v}_{\left(ij\right)k}=\tilde{v}_{ik}+\tilde{v}_{jk}$ \label{enu:vtild_prop_4}
\item $\tilde{v}_{\left(1\dots i\right)\left(1\dots n\right)}=-\tilde{v}_{\left(i+1\dots n\right)\left(1\dots n\right)}$
for $i<n$ \label{enu:vtild_prop_5}
\end{enumerate}
For more relations and their proofs see \citep{Cruz-Santiago2015}.
Plenty of useful identities for $v_{ij}$ symbols are contained in
\citep{Motyka2009,Cruz-Santiago2013}, which of course can be translated
to $\tilde{v}_{ij}$ symbols.

In particular, in the present work we shall need the following identity
(adapted from \citep{Motyka2009}):
\begin{equation}
\sum_{i=1}^{n-1}\frac{\tilde{v}_{i\left(i+1\right)}}{p_{i}^{+}}\,\tilde{v}_{\left(1\dots i\right)\left(1\dots n\right)}^{*}=\sum_{i=1}^{n-1}v_{\left(i+1\right)i}\,\tilde{v}_{\left(1\dots i\right)\left(1\dots n\right)}^{*}=-\frac{1}{2}D_{1\dots n}\,,\label{eq:MotykaStasto_rel1}
\end{equation}
where the energy denominator $D_{1\dots n}$ (see Eq.~\eqref{eq:Energy_denom}) has been expressed
in terms of $\tilde{v}_{ij}$ symbols, see Eq.~(\ref{eq:D_vivj}).

\section{Recursion relation for $B^{\bullet}\left[A^{\bullet}\right]$ field}

\label{sec:App_B[A]}

In this appendix we derive and solve the recursion (\ref{eq:Gamma_n_recrel-1}).
We start the calculation in the position space, and later we shall convert to the momentum
space. The field $B^{\bullet}$ has the following power expansion in $A^{\bullet}$
fields:
\begin{multline}
B_{a}^{\bullet}\left(\mathbf{x}\right)=A_{a}^{\bullet}\left(\mathbf{x}\right)+\int d^{3}\mathbf{y}_{1}d^{3}\mathbf{y}_{2}\,\Gamma_{2}^{a\left\{ b_{1}b_{2}\right\} }\left(\mathbf{x};\left\{ \mathbf{y}_{1},\mathbf{y}_{2}\right\} \right)A_{b_{1}}^{\bullet}\left(\mathbf{y}_{1}\right)A_{b_{2}}^{\bullet}\left(\mathbf{y}_{2}\right)\\
+\int d^{3}\mathbf{y}_{1}d^{3}\mathbf{y}_{2}d^{3}\mathbf{y}_{3}\,\Gamma_{3}^{a\left\{ b_{1}b_{2}b_{3}\right\} }\left(\mathbf{x};\left\{ \mathbf{y}_{1},\mathbf{y}_{2},\mathbf{y}_{3}\right\} \right)A_{b_{1}}^{\bullet}\left(\mathbf{y}_{1}\right)A_{b_{2}}^{\bullet}\left(\mathbf{y}_{2}\right)A_{b_{3}}^{\bullet}\left(\mathbf{y}_{3}\right)+\dots\,.\label{eq:Bplus_expansion}
\end{multline}
The symbols $\Gamma^{a\left\{ b_{1}\dots n_{n}\right\} }\left(\mathbf{x};\left\{ \mathbf{y}_{1},\dots,\mathbf{y}_{n}\right\} \right)$
are symmetric in pairs of indices $\left(b_{i},\mathbf{y}_{i}\right)$. 

We shall start with the l.h.s. of (\ref{eq:Transformation_A+}). The
functional derivative reads
\begin{multline}
\frac{\delta B_{a}^{\bullet}\left(\mathbf{x}\right)}{\delta A_{c}^{\bullet}\left(\mathbf{z}\right)}=\delta^{ac}\delta^{3}\left(\mathbf{x}-\mathbf{z}\right)+2\int d^{3}\mathbf{y}\,\Gamma_{2}^{a\left\{ bc\right\} }\left(\mathbf{x};\left\{ \mathbf{y},\mathbf{z}\right\} \right)A_{b}^{\bullet}\left(\mathbf{y}\right)\\
+3\int d^{3}\mathbf{y}_{1}d^{3}\mathbf{y}_{2}\,\Gamma_{3}^{a\left\{ b_{1}b_{2}c\right\} }\left(\mathbf{x};\left\{ \mathbf{y}_{1},\mathbf{y}_{2},\mathbf{z}\right\} \right)A_{b_{1}}^{\bullet}\left(\mathbf{y}_{1}\right)A_{b_{2}}^{\bullet}\left(\mathbf{y}_{2}\right)+\dots\label{eq:funct_deriv}
\end{multline}
Using the expansion (\ref{eq:Bplus_expansion}) and (\ref{eq:funct_deriv})
in the l.h.s. of (\ref{eq:Transformation_A+}) we have
\begin{multline}
\int d^{3}\mathbf{y}\left[-gf^{cbd}A_{b}^{\bullet}\left(\mathbf{y}\right)\gamma_{\mathbf{y}}A_{d}^{\bullet}\left(\mathbf{y}\right)+\omega_{\mathbf{y}}A_{c}^{\bullet}\left(\mathbf{y}\right)\right]\frac{\delta B_{a}^{\bullet}\left(\mathbf{x}\right)}{\delta A_{c}^{\bullet}\left(\mathbf{y}\right)}\\
=\int d^{3}\mathbf{y}\,\left[-gf^{cbd}A_{b}^{\bullet}\left(\mathbf{y}\right)\gamma_{\mathbf{y}}A_{d}^{\bullet}\left(\mathbf{y}\right)+\omega_{\mathbf{y}}A_{c}^{\bullet}\left(\mathbf{y}\right)\right]\Big\{\delta^{ac}\delta^{3}\left(\mathbf{x}-\mathbf{y}\right)+\int d^{3}\mathbf{y}_{1}\,2\Gamma_{2}^{a\left\{ b_{1}c\right\} }\left(\mathbf{x};\left\{ \mathbf{y}_{1},\mathbf{y}\right\} \right)A_{b_{1}}^{\bullet}\left(\mathbf{y}_{1}\right)\\
+\int d^{3}\mathbf{y}_{1}d^{3}\mathbf{y}_{2}\,3\Gamma_{3}^{a\left\{ b_{1}b_{2}c\right\} }\left(\mathbf{x};\left\{ \mathbf{y}_{1},\mathbf{y}_{2},\mathbf{y}\right\} \right)A_{b_{1}}^{\bullet}\left(\mathbf{y}_{1}\right)A_{b_{2}}^{\bullet}\left(\mathbf{y}_{2}\right)+\dots\Big\}\,.
\end{multline}
From this we derive the $n$-th term in the expansion:
\begin{multline}
\int d^{3}\mathbf{y}_{1}\dots d^{3}\mathbf{y}_{n-1}\,\left(n-1\right)\Gamma_{n-1}^{a\left\{ b_{1}\dots b_{n-1}c\right\} }\left(\mathbf{x};\left\{ \mathbf{y}_{1},\dots,\mathbf{y}_{n-1}\right\} \right)A_{b_{1}}^{\bullet}\left(\mathbf{y}_{1}\right)\dots A_{b_{n-2}}^{\bullet}\left(\mathbf{y}_{n-2}\right)\\
\times\left(-g\right)f^{cbd}A_{b}^{\bullet}\left(\mathbf{y}_{n-1}\right)\gamma_{\mathbf{y}_{n-1}}A_{d}^{\bullet}\left(\mathbf{y}_{n-1}\right)\\
+\int d^{3}\mathbf{y}_{1}\dots d^{3}\mathbf{y}_{n}\, n\Gamma_{n}^{a\left\{ b_{1}\dots b_{n}\right\} }\left(\mathbf{x};\left\{ \mathbf{y}_{1},\dots,\mathbf{y}_{n}\right\} \right)A_{b_{1}}^{\bullet}\dots A_{b_{n-1}}^{\bullet}\left(\mathbf{y}_{n-1}\right)\omega_{\mathbf{y}_{n}}A_{b_{n}}^{\bullet}\left(\mathbf{y}_{n}\right)\,.
\end{multline}
Comparing this with the r.h.s of (\ref{eq:Transformation_A+}) with (\ref{eq:Bplus_expansion})
we have for the same power of $A^{\bullet}$ field
\begin{multline}
\int d^{3}\mathbf{y}_{1}\dots d^{3}\mathbf{y}_{n-1}\,\left(n-1\right)\Gamma_{n-1}^{a\left\{ b_{1}\dots b_{n-2}c\right\} }\left(\mathbf{x};\left\{ \mathbf{y}_{1},\dots,\mathbf{y}_{n-1}\right\} \right)A_{b_{1}}^{\bullet}\left(\mathbf{y}_{1}\right)\dots A_{b_{n-2}}^{\bullet}\left(\mathbf{y}_{n-2}\right)\\
\times\left(-g\right)f^{cbd}A_{b}^{\bullet}\left(\mathbf{y}_{n-1}\right)\gamma_{\mathbf{y}_{n-1}}A_{d}^{\bullet}\left(\mathbf{y}_{n-1}\right)\\
+\int d^{3}\mathbf{y}_{1}\dots d^{3}\mathbf{y}_{n}\, n\Gamma_{n}^{a\left\{ b_{1}\dots b_{n}\right\} }\left(\mathbf{x};\left\{ \mathbf{y}_{1},\dots,\mathbf{y}_{n}\right\} \right)A_{b_{1}}^{\bullet}\left(\mathbf{y}_{1}\right)\dots A_{b_{n-1}}^{\bullet}\left(\mathbf{y}_{n-1}\right)\omega_{\mathbf{y}_{n}}A_{b_{n}}^{\bullet}\left(\mathbf{y}_{n}\right)\\
=\int d^{3}\mathbf{y}_{1}\dots d^{3}\mathbf{y}_{n}\,\omega_{\mathbf{x}}\Gamma_{n}^{a\left\{ b_{1}\dots b_{n}\right\} }\left(\mathbf{x};\left\{ \mathbf{y}_{1},\dots,\mathbf{y}_{n}\right\} \right)A_{b_{1}}^{\bullet}\dots A_{b_{n}}^{\bullet}\left(\mathbf{y}_{n}\right)\,.\label{eq:MasterEq_pos}
\end{multline}

We shall now make Fourier transform to the momentum space. The new coefficients $\tilde{\Gamma}_n$ are defined as the Fourier transform of functions $\Gamma_n$ in the following way

\begin{equation}
\Gamma_{n}^{a\left\{ b_{1}\dots b_{n}\right\} }\left(\mathbf{x};\left\{ \mathbf{y}_{1},\dots,\mathbf{y}_{n}\right\} \right)=\int d^{3}\mathbf{p}d^{3}\mathbf{q}_{1}\dots d^{3}\mathbf{q}_{n}\,\tilde{\Gamma}_{n}^{a\left\{ b_{1}\dots b_{n}\right\} }\left(\mathbf{p};\left\{ \mathbf{q}_{1},\dots,\mathbf{q}_{n}\right\} \right)\, e^{i\left(-\mathbf{x}\cdot\mathbf{p}+\mathbf{y}_{1}\cdot\mathbf{q}_{1}+\dots+\mathbf{y}_{n}\cdot\mathbf{q}_{n}\right)}\,.\label{eq:Gamma_mom}
\end{equation}
The transform of gluon fields is given in (\ref{eq:FT_def}). The
expansion of $B^{\bullet}$  in terms of $A^{\bullet}$ fields in the momentum space is given by (\ref{eq:BplusFT-1}).
We arrive at
\begin{multline}
\int d^{3}\mathbf{p}_{1}\dots d^{3}\mathbf{p}_{n}\,\frac{p_{n}^{\star}}{p_{n}^{+}}\left(n-1\right)\tilde{\Gamma}_{n-1}^{a\left\{ b_{1}\dots b_{n-2}c\right\} }\left(\mathbf{P};\left\{ \mathbf{p}_{1},\dots,\mathbf{p}_{n-1}+\mathbf{p}_{n}\right\} \right)\left(-g\right)f^{cb_{n-1}b_{n}}\tilde{A}_{b_{1}}^{\bullet}\left(\mathbf{p}_{1}\right)\dots\tilde{A}_{b_{n}}^{\bullet}\left(\mathbf{p}_{n}\right)\\
+i\int d^{3}\mathbf{p}_{1}\dots d^{3}\mathbf{p}_{n}\, E_{p_{n}}n\tilde{\Gamma}_{n}^{a\left\{ b_{1}\dots b_{n}\right\} }\left(\mathbf{P};\left\{ \mathbf{p}_{1},\dots,\mathbf{p}_{n}\right\} \right)\tilde{A}_{b_{1}}^{\bullet}\left(\mathbf{p}_{1}\right)\dots\tilde{A}_{b_{n}}^{\bullet}\left(\mathbf{p}_{n}\right)\\
=iE_{P}\int d^{3}\mathbf{p}_{1}\dots d^{3}\mathbf{p}_{n}\tilde{\Gamma}_{n}^{a\left\{ b_{1}\dots b_{n}\right\} }\left(\mathbf{P};\left\{ \mathbf{p}_{1},\dots,\mathbf{p}_{n}\right\} \right)\tilde{A}_{b_{1}}^{\bullet}\left(\mathbf{p}_{1}\right)\dots\tilde{A}_{b_{n}}^{\bullet}\left(\mathbf{p}_{n}\right)\,,\label{eq:MasterEq_mom2}
\end{multline}
where $E_{p}$ is defined in (\ref{eq:Ep_def}). Let symmterize the
first term as follows:
\begin{multline}
\int d^{3}\mathbf{p}_{1}\dots d^{3}\mathbf{p}_{n}\,\frac{p_{n}^{\star}}{p_{n}^{+}}\left(n-1\right)\tilde{\Gamma}_{n-1}^{a\left\{ b_{1}\dots b_{n-2}c\right\} }\left(\mathbf{P};\left\{ \mathbf{p}_{1},\dots,\mathbf{p}_{n-1}+\mathbf{p}_{n}\right\} \right)\left(-g\right)f^{cb_{n-1}b_{n}}\tilde{A}_{b_{1}}^{\bullet}\left(\mathbf{p}_{1}\right)\dots\tilde{A}_{b_{n}}^{\bullet}\left(\mathbf{p}_{n}\right)\\
=-\frac{1}{2}i\int d^{3}\mathbf{p}_{1}\dots d^{3}\mathbf{p}_{n}\,\tilde{V}_{++-}^{cb_{n-1}b_{n}}\left(\mathbf{p}_{n-1},\mathbf{p}_{n},-\mathbf{p}_{n-1}-\mathbf{p}_{n}\right)\frac{1}{p_{n-1}^{+}+p_{n}^{+}}\\
\left(n-1\right)\tilde{\Gamma}_{n-1}^{a\left\{ b_{1}\dots b_{n-2}c\right\} }\left(\mathbf{P};\left\{ \mathbf{p}_{1},\dots,\mathbf{p}_{n-1}+\mathbf{p}_{n}\right\} \right)\tilde{A}_{b_{1}}^{\bullet}\left(\mathbf{p}_{1}\right)\dots\tilde{A}_{b_{n}}^{\bullet}\left(\mathbf{p}_{n}\right)\,,
\end{multline}
where we have used the vertex 
\begin{equation}
\tilde{V}_{++-}^{a_{1}a_{2}a_{3}}\left(\mathbf{p}_{1},\mathbf{p}_{2},\mathbf{p}_{3}\right)=-igf^{a_{1}a_{2}a_{3}}\left(\frac{p_{1}^{\star}}{p_{1}^{+}}-\frac{p_{2}^{\star}}{p_{2}^{+}}\right)p_{3}^{+}\,,
\end{equation}
with $\mathbf{p}_{1}+\mathbf{p}_{2}+\mathbf{p}_{3}=0$. In what follows
we shall use shorthand notation
\begin{equation}
\tilde{V}_{++-}^{a_{1}a_{2}a_{3}}\left(\mathbf{p}_{1},\mathbf{p}_{2}\right)\equiv\tilde{V}_{++-}^{a_{1}a_{2}a_{3}}\left(\mathbf{p}_{1},\mathbf{p}_{2},-\mathbf{p}_{1}-\mathbf{p}_{2}\right)\,.
\end{equation}
Next, we symmetrize the second term of the l.h.s of (\ref{eq:MasterEq_mom2}):
\begin{multline}
i\int d^{3}\mathbf{p}_{1}\dots d^{3}\mathbf{p}_{n}\, E_{p_{n}}n\tilde{\Gamma}_{n}^{a\left\{ b_{1}\dots b_{n}\right\} }\left(\mathbf{P};\left\{ \mathbf{p}_{1},\dots,\mathbf{p}_{n}\right\} \right)\tilde{A}_{b_{1}}^{\bullet}\left(\mathbf{p}_{1}\right)\dots\tilde{A}_{b_{n}}^{\bullet}\left(\mathbf{p}_{n}\right)\\
=i\int d^{3}\mathbf{p}_{1}\dots d^{3}\mathbf{p}_{n}\,\left(E_{p_{1}}+\dots+E_{p_{n}}\right)\tilde{\Gamma}_{n}^{a\left\{ b_{1}\dots b_{n}\right\} }\left(\mathbf{P};\left\{ \mathbf{p}_{1},\dots,\mathbf{p}_{n}\right\} \right)\tilde{A}_{b_{1}}^{\bullet}\left(\mathbf{p}_{1}\right)\dots\tilde{A}_{b_{n}}^{\bullet}\left(\mathbf{p}_{n}\right)\,.
\end{multline}
Thus, the equation for the $\tilde{\Gamma}_{n}$ reads
\begin{multline}
\int d^{3}\mathbf{p}_{1}\dots d^{3}\mathbf{p}_{n}\tilde{\Gamma}_{n}^{a\left\{ b_{1}\dots b_{n}\right\} }\left(\mathbf{P};\left\{ \mathbf{p}_{1},\dots,\mathbf{p}_{n}\right\} \right)\tilde{A}_{b_{1}}^{\bullet}\left(\mathbf{p}_{1}\right)\dots\tilde{A}_{b_{n}}^{\bullet}\left(\mathbf{p}_{n}\right)\\
=-\int d^{3}\mathbf{p}_{1}\dots d^{3}\mathbf{p}_{n}\,\frac{1}{2\left(E_{P}-E_{p_{1}}-\dots-E_{p_{n}}\right)}\,\tilde{V}_{++-}^{cb_{n-1}b_{n}}\left(\mathbf{p}_{n-1},\mathbf{p}_{n}\right)\\
\frac{1}{p_{n-1}^{+}+p_{n}^{+}}\left(n-1\right)\tilde{\Gamma}_{n-1}^{a\left\{ b_{1}\dots b_{n-2}c\right\} }\left(\mathbf{P};\left\{ \mathbf{p}_{1},\dots,\mathbf{p}_{n-1}+\mathbf{p}_{n}\right\} \right)\tilde{A}_{b_{1}}^{\bullet}\left(\mathbf{p}_{1}\right)\dots\tilde{A}_{b_{n}}^{\bullet}\left(\mathbf{p}_{n}\right)\,\label{eq:MasterEq_mom3}
\end{multline}
or simply
\begin{multline}
\tilde{\Gamma}_{n}^{a\left\{ b_{1}\dots b_{n}\right\} }\left(\mathbf{P};\left\{ \mathbf{p}_{1},\dots,\mathbf{p}_{n}\right\} \right)=-\frac{1}{2\left(E_{P}-E_{p_{1}}-\dots-E_{p_{n}}\right)}\,\tilde{V}_{++-}^{cb_{n-1}b_{n}}\left(\mathbf{p}_{n-1},\mathbf{p}_{n}\right)\frac{1}{p_{n-1}^{+}+p_{n}^{+}}\\
\left(n-1\right)\tilde{\Gamma}_{n-1}^{a\left\{ b_{1}\dots b_{n-2}c\right\} }\left(\mathbf{P};\left\{ \mathbf{p}_{1},\dots,\mathbf{p}_{n-1}+\mathbf{p}_{n}\right\} \right)\,.\label{eq:Gamma_n_recrel0}
\end{multline}
Above, we understand the equality in a `weak' sense, that is we keep
in mind that we can ultimately integrate it with the $A^{\bullet}$
fields.

Let us note, that we can write the r.h.s. as a sum of terms with the
vertex insertion for all combinations of the momenta. This is simply done
by writing it as a sum with  an appropriate symmetry factor and performing a
suitable exchange of the integration momenta and color indices. This procedure
is necessary if we want to  apply the color decomposition,   and is illustrated graphically in (\arabic{Gamma_diag1}).  We observe that the planar contributions
give $n-1$ terms on the r.h.s of (\ref{eq:Gamma_n_recrel0}). Including in addition 
the non-planar contributions we get finally the following symmetry factors
\begin{equation}
s_{n}=\frac{1}{n-1}\,\binom{n}{2}=\frac{n}{2}\,.
\end{equation}
The final result for the recursion reads
\begin{multline}
\tilde{\Gamma}_{n}^{a\left\{ b_{1}\dots b_{n}\right\} }\left(\mathbf{P};\left\{ \mathbf{p}_{1},\dots,\mathbf{p}_{n}\right\} \right)=-\frac{1}{2\left(E_{P}-E_{p_{1}}-\dots-E_{p_{n}}\right)}\,\frac{1}{s_{n}}\\
\Bigg\{\tilde{V}_{++-}^{cb_{n-1}b_{n}}\left(\mathbf{p}_{n-1},\mathbf{p}_{n}\right)\frac{1}{p_{n-1}^{+}+p_{n}^{+}}\tilde{\Gamma}_{n-1}^{a\left\{ b_{1}\dots b_{n-2}c\right\} }\left(\mathbf{P};\left\{ \mathbf{p}_{1},\dots,\mathbf{p}_{n-1}+\mathbf{p}_{n}\right\} \right)\\
+\tilde{V}_{++-}^{cb_{n-2}b_{n}}\left(\mathbf{p}_{n-2},\mathbf{p}_{n}\right)\frac{1}{p_{n-2}^{+}+p_{n}^{+}}\tilde{\Gamma}_{n-1}^{a\left\{ b_{1}\dots b_{n-3}b_{n-1}c\right\} }\left(\mathbf{P};\left\{ \mathbf{p}_{1},\dots,\mathbf{p}_{n-1},\mathbf{p}_{n-2}+\mathbf{p}_{n}\right\} \right)+\dots\\
+\tilde{V}_{++-}^{cb_{n-2}b_{n-1}}\left(\mathbf{p}_{n-2},\mathbf{p}_{n-1}\right)\frac{1}{p_{n-2}^{+}+p_{n-1}^{+}}\tilde{\Gamma}_{n-1}^{a\left\{ b_{1}\dots b_{n-3}cb_{n}\right\} }\left(\mathbf{P};\left\{ \mathbf{p}_{1},\dots,\mathbf{p}_{n-2}+\mathbf{p}_{n-1},\mathbf{p}_{n}\right\} \right)+\dots\Bigg\}\label{eq:Gamma_n_recrel}
\end{multline}

The first two coefficients of the expansion read
\begin{equation}
\tilde{\Gamma}_{2}^{a\left\{ b_{1}b_{2}\right\} }\left(\mathbf{P};\left\{ \mathbf{p}_{1},\mathbf{p}_{2}\right\} \right)=-\frac{1}{s_{2}}\,\frac{1}{D_{12}}\,\frac{1}{p_{12}^{+}}\,\tilde{V}_{++-}^{ab_{1}b_{2}}\left(\mathbf{p}_{1},\mathbf{p}_{2}\right)\delta^{3}\left(\mathbf{p}_{1}+\mathbf{p}_{2}-\mathbf{P}\right)\,,
\end{equation}
\begin{multline}
\tilde{\Gamma}_{3}^{a\left\{ b_{1}b_{2}b_{3}\right\} }\left(\mathbf{P};\left\{ \mathbf{p}_{1},\mathbf{p}_{2},\mathbf{p}_{3}\right\} \right)\\
=\frac{1}{s_{2}s_{3}}\,\frac{1}{D_{123}}\,\frac{1}{p_{12}^{+}}\,\tilde{V}_{++-}^{cb_{1}b_{2}}\left(\mathbf{p}_{1},\mathbf{p}_{2}\right)\frac{1}{D_{\left(12\right)3}}\,\frac{1}{p_{123}^{+}}\,\tilde{V}_{++-}^{acb_{3}}\left(\mathbf{p}_{1}+\mathbf{p}_{2},\mathbf{p}_{3}\right)\delta^{3}\left(\mathbf{p}_{1}+\mathbf{p}_{2}+\mathbf{p}_{3}-\mathbf{P}\right)\\
+\frac{1}{s_{2}s_{3}}\,\frac{1}{D_{123}}\,\frac{1}{p_{23}^{+}}\,\tilde{V}_{++-}^{cb_{2}b_{3}}\left(\mathbf{p}_{2},\mathbf{p}_{3}\right)\frac{1}{D_{1\left(23\right)}}\,\frac{1}{p_{123}^{+}}\,\tilde{V}_{++-}^{ab_{1}c}\left(\mathbf{p}_{1},\mathbf{p}_{2}+\mathbf{p}_{3}\right)\delta^{3}\left(\mathbf{p}_{1}+\mathbf{p}_{2}+\mathbf{p}_{3}-\mathbf{P}\right)\\
+\frac{1}{s_{2}s_{3}}\,\frac{1}{D_{123}}\,\frac{1}{p_{13}^{+}}\,\tilde{V}_{++-}^{cb_{1}b_{3}}\left(\mathbf{p}_{1},\mathbf{p}_{3}\right)\frac{1}{D_{2\left(13\right)}}\,\frac{1}{p_{123}^{+}}\,\tilde{V}_{++-}^{ab_{2}c}\left(\mathbf{p}_{2},\mathbf{p}_{1}+\mathbf{p}_{3}\right)\delta^{3}\left(\mathbf{p}_{1}+\mathbf{p}_{2}+\mathbf{p}_{3}-\mathbf{P}\right)\,,
\end{multline}
where we used the shorthand notation for the sum of the momenta (\ref{eq:momentum_sum_def})
and (\ref{eq:Energy_denom}). The notation $D_{i\left(kl\right)}$
means that $kl$ has to be treated as an intermediate state with the energy
$E_{p_{kl}}$. The diagrammatic expansion (\arabic{B+_diags}) follows  from the above results.

The recursion relation for the color ordered coefficients $\tilde{\Gamma}_{n}$ can
be derived in the following way. One can use (\ref{eq:Gamma_n_recrel})
and decompose $\tilde{\Gamma}_{n}$, $\tilde{\Gamma}_{n-1}$ into
color ordered amplitudes. Consider the order $\left(1,2\dots n\right)$.
We pick up the term $\mathrm{Tr}\left(t^{b_{1}}\dots t^{b_{i-3}}t^{c}t^{b_{i}}\dots t^{b_{n-1}}\right)$
from $\tilde{\Gamma}_{n-1}\left(\mathbf{P};\left\{ \mathbf{p}_{1},\dots,\mathbf{p}_{i-2}+\mathbf{p}_{i-1},\dots,\mathbf{p}_{n}\right\} \right)$
and observe that the multiplying vertex contains $f^{cb_{i-2}b_{i-1}}$
which fills the trace thanks to the identity $t^{c}f^{cb_{i-2}b_{i-1}}=\left(t^{b_{i-2}}t^{b_{i-1}}-t^{b_{i-1}}t^{b_{i-2}}\right)/i\sqrt{2}$.
Moreover
\begin{equation}
\tilde{V}_{++-}^{a_{1}a_{2}a_{3}}\left(\mathbf{p}_{1},\mathbf{p}_{2},\mathbf{p}_{3}\right)=i\sqrt{2}f^{a_{1}a_{2}a_{3}}\,\tilde{V}_{++-}\left(\mathbf{p}_{1},\mathbf{p}_{2},\mathbf{p}_{3}\right)\,,
\end{equation}
where
\begin{equation}
\tilde{V}_{++-}\left(\mathbf{p}_{1},\mathbf{p}_{2},\mathbf{p}_{3}\right)=-g'\,\left(\frac{p_{1}^{\star}}{p_{1}^{+}}-\frac{p_{2}^{\star}}{p_{2}^{+}}\right)p_{3}^{+}\,,
\end{equation}
is the color-ordered vertex. Thus, for the color-ordered amplitudes the
recursion \eqref{eq:Gamma_n_recrel} has the form
\begin{multline}
\tilde{\Gamma}_{n}\left(\mathbf{P};\mathbf{p}_{1},\dots,\mathbf{p}_{n}\right)=-\frac{1}{D_{1\dots n}}\,\frac{1}{s_{n}}\Bigg\{\tilde{V}_{++-}\left(\mathbf{p}_{n-1},\mathbf{p}_{n}\right)\frac{1}{p_{\left(n-1\right)n}^{+}}\tilde{\Gamma}_{n-1}\left(\mathbf{P};\mathbf{p}_{1},\dots,\mathbf{p}_{n-1}+\mathbf{p}_{n}\right)\\
+\tilde{V}_{++-}\left(\mathbf{p}_{n-2},\mathbf{p}_{n-1}\right)\frac{1}{p_{\left(n-2\right)\left(n-1\right)}^{+}}\tilde{\Gamma}_{n-1}\left(\mathbf{P};\mathbf{p}_{1},\dots,\mathbf{p}_{n-2}+\mathbf{p}_{n-1},\mathbf{p}_{n}\right)+\dots\Bigg\}\,.\label{eq:Gamma_n_recrel_colorord}
\end{multline}

We can easily find a few first terms from this recursion
\begin{equation}
\tilde{\Gamma}_{2}\left(\mathbf{P};\mathbf{p}_{1},\mathbf{p}_{2}\right)=\frac{1}{s_{2}}\,\frac{1}{2}g'\,\frac{p_{12}^{+}}{p_{1}^{+}}\,\frac{1}{\tilde{v}_{21}^{*}}\,\delta^{3}\left(\mathbf{p}_{1}+\mathbf{p}_{2}-\mathbf{P}\right)=-\frac{g'}{2}\,\frac{1}{\tilde{v}_{1\left(12\right)}^{*}}\,\delta^{3}\left(\mathbf{p}_{1}+\mathbf{p}_{2}-\mathbf{P}\right)\,,
\end{equation}
where we used
\begin{equation}
\frac{p_{12}^{+}}{p_{1}^{+}}\,\frac{1}{\tilde{v}_{21}^{*}}=\frac{p_{12}^{+}}{p_{1}^{+}}\,\frac{1}{\tilde{v}_{\left(12\right)1}^{*}}=\frac{p_{12}^{+}}{p_{1}^{+}}\,\frac{1}{-\frac{p_{12}^{+}}{p_{1}^{+}}\tilde{v}_{1\left(12\right)}^{*}}=\frac{-1}{\tilde{v}_{1\left(12\right)}^{*}}\,.
\end{equation}
Next term reads
\begin{multline}
\tilde{\Gamma}_{3}\left(\mathbf{P};\mathbf{p}_{1},\mathbf{p}_{2},\mathbf{p}_{3}\right)=\delta^{3}\left(\mathbf{p}_{1}+\mathbf{p}_{2}+\mathbf{p}_{3}-\mathbf{P}\right)\frac{1}{s_{2}s_{3}}\,\frac{1}{4}\,\frac{1}{\tilde{v}_{12}\tilde{v}_{21}^{*}+\tilde{v}_{13}\tilde{v}_{31}^{*}+\tilde{v}_{23}\tilde{v}_{32}^{*}}\\
\Bigg\{\,\frac{1}{\tilde{v}_{12}\tilde{v}_{21}^{*}}\,\left(-g'\,\frac{p_{12}^{+}}{p_{1}^{+}}\tilde{v}_{12}\right)\left(-g'\,\frac{p_{123}^{+}}{p_{12}^{+}}\tilde{v}_{\left(12\right)3}\right)\\
+\frac{1}{\tilde{v}_{23}\tilde{v}_{32}^{*}}\,\left(-g'\,\frac{p_{23}^{+}}{p_{2}^{+}}\tilde{v}_{23}\right)\left(-g'\,\frac{p_{123}^{+}}{p_{23}^{+}}\tilde{v}_{\left(23\right)1}\right)\Bigg\}\\
=\frac{1}{s_{2}s_{3}}\left(\frac{g'}{2}\right)^{2}\,\frac{1}{\tilde{v}_{1\left(123\right)}^{*}\tilde{v}_{\left(12\right)\left(123\right)}^{*}}\,\delta^{3}\left(\mathbf{p}_{1}+\mathbf{p}_{2}+\mathbf{p}_{3}-\mathbf{P}\right)\,.
\end{multline}
In a similar way we can find that
\begin{equation}
\tilde{\Gamma}_{4}\left(\mathbf{P};\mathbf{p}_{1},\mathbf{p}_{2},\mathbf{p}_{3},\mathbf{p}_{4}\right)=\frac{-1}{s_{2}s_{3}s_{4}}\left(\frac{g'}{2}\right)^{3}\,\frac{1}{\tilde{v}_{1\left(1234\right)}^{*}\tilde{v}_{\left(12\right)\left(1234\right)}^{*}\tilde{v}_{\left(123\right)\left(1234\right)}^{*}}\,\delta^{3}\left(\mathbf{p}_{1}+\dots+\mathbf{p}_{4}-\mathbf{P}\right).
\end{equation}
We postulate that the solution has the form
\begin{equation}
\tilde{\Gamma}_{n}\left(\mathbf{P};\mathbf{p}_{1},\dots,\mathbf{p}_{n}\right)=\frac{1}{s_{2}\dots s_{n}}\,\left(-\frac{g'}{2}\right)^{n-1}\,\frac{1}{\tilde{v}_{1\left(1\dots n\right)}^{*}\tilde{v}_{\left(12\right)\left(1\dots n\right)}^{*}\dots\tilde{v}_{\left(1\dots n-1\right)\left(1\dots n\right)}^{*}}\,\delta^{3}\left(\mathbf{p}_{1}+\dots+\mathbf{p}_{n}-\mathbf{P}\right)\,.\label{eq:Gamma_n_sol}
\end{equation}

One can prove the above result in a way similar to the Wilson line solution, except that here we are dealing with color-ordered objects, which simplifies the entire procedure.
We insert (\ref{eq:Gamma_n_sol}) into the color-ordered version of
(\ref{eq:Gamma_n_recrel}). First, note that according to the ansatz
(\ref{eq:Gamma_n_sol}) we have (we skip momentum conservation delta
functions in what follows)
\begin{multline}
\tilde{\Gamma}_{n-1}\left(\mathbf{P};\mathbf{p}_{1},\dots,\mathbf{p}_{i-2}+\mathbf{p}_{i-1},\dots,\mathbf{p}_{n}\right)=\frac{1}{s_{2}\dots s_{n-1}}\,\left(-\frac{g'}{2}\right)^{n-2}\,\\
\frac{1}{\tilde{v}_{1\left(1\dots n\right)}^{*}\dots\tilde{v}_{\left(1\dots i-3\right)\left(1\dots n\right)}^{*}\tilde{v}_{\left(1\dots i-1\right)\left(1\dots n\right)}^{*}\dots\tilde{v}_{\left(1\dots n-1\right)\left(1\dots n\right)}^{*}}\\
=\frac{1}{s_{2}\dots s_{n-1}}\,\left(-\frac{g'}{2}\right)^{n-2}\,\frac{\tilde{v}_{\left(1\dots i-2\right)\left(1\dots n\right)}^{*}}{\tilde{v}_{1\left(1\dots n\right)}^{*}\dots\tilde{v}_{\left(1\dots n-1\right)\left(1\dots n\right)}^{*}}\,.
\end{multline}
Using this, we have for the r.h.s. of (\ref{eq:Gamma_n_recrel})
\begin{multline}
\frac{1}{s_{2}\dots s_{n-1}s_{n}}\,\left(-\frac{g'}{2}\right)^{n-2}\left(g'\right)\frac{1}{2\left(E_{P}-E_{p_{1}}-\dots-E_{p_{n}}\right)}\,\frac{1}{\tilde{v}_{1\left(1\dots n\right)}^{*}\dots\tilde{v}_{\left(1\dots n-1\right)\left(1\dots n\right)}^{*}}\\
\Bigg\{\frac{\tilde{v}_{\left(n-1\right)n}}{p_{n-1}^{+}}\,\tilde{v}_{\left(1\dots n-1\right)\left(1\dots n\right)}^{*}+\frac{\tilde{v}_{\left(n-2\right)\left(n-1\right)}}{p_{n-2}^{+}}\,\tilde{v}_{\left(1\dots n-2\right)\left(1\dots n\right)}^{*}+\dots+\frac{\tilde{v}_{12}}{p_{1}^{+}}\,\tilde{v}_{1\left(1\dots n\right)}^{*}\Bigg\}\,.
\end{multline}
 Noticing that
\begin{equation}
\frac{1}{s_{2}\dots s_{n}}=\frac{2^{n}}{2\cdot3\cdot\dots\cdot n}=\frac{2^{n}}{n!}\,,
\end{equation}
and utilizing \eqref{eq:MotykaStasto_rel1} the color ordered solution becomes
\begin{equation}
\tilde{\Gamma}_{n}\left(\mathbf{P};\mathbf{p}_{1},\dots,\mathbf{p}_{n}\right)=\frac{1}{n!}\,\left(-g'\right)^{n-1}\,\frac{1}{\tilde{v}_{1\left(1\dots n\right)}^{*}\tilde{v}_{\left(12\right)\left(1\dots n\right)}^{*}\dots\tilde{v}_{\left(1\dots n-1\right)\left(1\dots n\right)}^{*}}\,\delta^{3}\left(\mathbf{p}_{1}+\dots+\mathbf{p}_{n}-\mathbf{P}\right)\,.\label{eq:Gamma_n_sol-1}
\end{equation}
which completes the proof. 

Finally we note that, using (\ref{eq:Gamma_colordecomp-1}), integrating
it with the $A^{\bullet}$ fields and changing the pair of color/momentum
variables we can write the $B^{\bullet}$ field expansion coefficient
simply as
\begin{multline}
\tilde{\Gamma}_{n}^{a\left\{ b_{1}\dots b_{n}\right\} }\left(\mathbf{P};\left\{ \mathbf{p}_{1},\dots,\mathbf{p}_{n}\right\} \right)=\left(-g'\right)^{n-1}\,\frac{1}{\tilde{v}_{1\left(1\dots n\right)}^{*}\tilde{v}_{\left(12\right)\left(1\dots n\right)}^{*}\dots\tilde{v}_{\left(1\dots n-1\right)\left(1\dots n\right)}^{*}}\,\\
\delta^{3}\left(\mathbf{p}_{1}+\dots+\mathbf{p}_{n}-\mathbf{P}\right)\,\mathrm{Tr}\left(t^{a}t^{b_{1}}\dots t^{b_{n}}\right)\,.\label{eq:Gamma_solution}
\end{multline}
which is identical to $\tilde{\Theta}_{n}$.

\section{Rederivation of the MHV action}

\label{sec:App_MHVaction}

In this appendix we collect together the elements  necessary for the derivation of the MHV action \citep{Mansfield2006}.
 We have organized it in several parts. The first
one deals with the light-front quantization of the Yang-Mills action,
while the remaining parts are devoted to the canonical field transformation,
its solution, and the derivation of  the sample MHV vertex.

\subsection{The Yang-Mills action on the light-front}

We start with the Yang-Mills (Y-M) action
\begin{equation}
S_{\mathrm{Y-M}}=\int d^{4}x\,\mathrm{Tr}\left\{ -\frac{1}{4}\hat{F}_{\mu\nu}\hat{F}^{\mu\nu}\right\} \,,
\end{equation}
where the field strength tensor is
\begin{equation}
\hat{F}^{\mu\nu}=\partial^{\mu}\hat{A}^{\nu}-\partial^{\nu}\hat{A}^{\mu}-ig'\left[\hat{A}^{\mu},\hat{A}^{\nu}\right]\,.
\end{equation}
(See Section~\ref{sec:Notation} for our conventions and notation.)
Expanding the action we get explicitly
\begin{multline}
S_{\mathrm{Y-M}}=\int d^{4}x\,\mathrm{Tr}\Bigg\{-\frac{1}{2}\left(\partial^{\mu}\hat{A}_{\nu}\right)^{2}+\frac{1}{2}\partial^{\mu}\hat{A}^{\nu}\partial_{\nu}\hat{A}_{\mu}+ig'\partial^{\mu}\hat{A}^{\nu}\left[\hat{A}_{\mu},\hat{A}_{\nu}\right]\\
-\frac{1}{4}\left(ig'\right)^{2}\left[\hat{A}^{\mu},\hat{A}^{\nu}\right]\left[\hat{A}_{\mu},\hat{A}_{\nu}\right]\Bigg\}\,.\label{eq:YM_action_explicit}
\end{multline}

Using the LC variables, we rewrite each of the terms in (\ref{eq:YM_action_explicit})
and impose the light-cone gauge condition:
\begin{equation}
A\cdot\eta=A^{+}=0\,.
\end{equation}
We divide the Lagrangian into the kinetic part, triple-coupling part,
and four-gluon part
\begin{equation}
\mathcal{L}_{\mathrm{Y-M}}=\mathcal{L}_{2}+\mathcal{L}_{3}+\mathcal{L}_{4}\,.\label{eq:LagrangianFull}
\end{equation}

We start with the kinetic term. After some algebra and integration
by parts (assuming all fields components vanish at infinity in any
direction) we get
\begin{multline}
\mathcal{L}_{2}=\int d^{3}\mathbf{x}\,\frac{1}{2}\mathrm{Tr}\Bigg\{-\hat{A}^{-}\partial_{-}^{2}\hat{A}^{-}-\hat{A}^{\bullet}\partial_{\bullet}^{2}\hat{A}^{\bullet}-\hat{A}^{\star}\partial_{\star}^{2}\hat{A}^{\star}\\
-2\hat{A}^{-}\partial_{-}\partial_{\bullet}\hat{A}^{\bullet}-2\hat{A}^{-}\partial_{-}\partial_{\star}\hat{A}^{\star}-2\hat{A}^{\star}\partial_{\bullet}\partial_{\star}\hat{A}^{\bullet}\\
-4\hat{A}^{\bullet}\partial_{-}\partial_{+}\hat{A}^{\star}+4\hat{A}^{\bullet}\partial_{\bullet}\partial_{\star}\hat{A}^{\star}\Bigg\}\,.
\end{multline}
Noticing that
\begin{equation}
\square=\partial^{\mu}\partial_{\mu}=2\left(\partial_{+}\partial_{-}-\partial_{\bullet}\partial_{\star}\right)\,,
\end{equation}
we can write the kinetic term  as
\begin{multline}
\mathcal{L}_{2}=\int d^{3}\mathbf{x}\,\mathrm{Tr}\Bigg\{-\hat{A}^{\bullet}\square\hat{A}^{\star}-\frac{1}{2}\hat{A}^{-}\partial_{-}^{2}\hat{A}^{-}-\frac{1}{2}\hat{A}^{\bullet}\partial_{\bullet}^{2}\hat{A}^{\bullet}-\frac{1}{2}\hat{A}^{\star}\partial_{\star}^{2}\hat{A}^{\star}\\
-\hat{A}^{-}\partial_{-}\partial_{\bullet}\hat{A}^{\bullet}-\hat{A}^{-}\partial_{-}\partial_{\star}\hat{A}^{\star}-\hat{A}^{\star}\partial_{\bullet}\partial_{\star}\hat{A}^{\bullet}\Bigg\}\,.
\end{multline}
For the triple-gluon coupling we get:
\begin{equation}
\mathcal{L}_{3}=-ig'\,\int d^{3}\mathbf{x}\,\mathrm{Tr}\Bigg\{\partial_{-}\hat{A}^{\bullet}\left[\hat{A}^{-},\hat{A}^{\star}\right]+\partial_{-}\hat{A}^{\star}\left[\hat{A}^{-},\hat{A}^{\bullet}\right]+\partial_{\bullet}\hat{A}^{\bullet}\left[\hat{A}^{\bullet},\hat{A}^{\star}\right]+\partial_{\star}\hat{A}^{\star}\left[\hat{A}^{\star},\hat{A}^{\bullet}\right]\Bigg\}\,.
\end{equation}
Finally, the  four-gluon coupling term reads
\begin{equation}
\mathcal{L}_{4}=\frac{1}{2}g'^{2}\,\int d^{3}\mathbf{x}\,\mathrm{Tr}\Bigg\{\left[\hat{A}^{\bullet},\hat{A}^{\star}\right]\left[\hat{A}^{\star},\hat{A}^{\bullet}\right]\Bigg\}\,.
\end{equation}

The Lagrangian (\ref{eq:LagrangianFull}) is quadratic in $A^{-}$
so following \citep{Scherk1975,Mansfield2006} one can integrate this
field out. Let us note, that actually the Lagrangian is quadratic
in all fields; however, the quadratic terms for the transverse fields
contain field-dependent operators, which in turn would introduce the
field dependent determinant (after integrating over paths).

Let us collect the part of the action dependent on the $A^{-}$ field:
\begin{multline}
S'_{\mathrm{Y-M}}=\int d^{4}x\,\mathrm{Tr}\Bigg\{-\frac{1}{2}\hat{A}^{-}\partial_{-}^{2}\hat{A}^{-}-\hat{A}^{-}\partial_{-}\partial_{\bullet}\hat{A}^{\bullet}-\hat{A}^{-}\partial_{-}\partial_{\star}\hat{A}^{\star}\\
-ig'\left(\partial_{-}A^{\bullet}\left[\hat{A}^{-},\hat{A}^{\star}\right]+\partial_{-}A^{\star}\left[\hat{A}^{-},\hat{A}^{\bullet}\right]\right)\Bigg\}\,.
\end{multline}
This can be written as
\begin{equation}
S'_{\mathrm{Y-M}}=\int d^{4}x\,\mathrm{Tr}\Bigg\{-\frac{1}{2}\hat{A}^{-}\partial_{-}^{2}\hat{A}^{-}-\hat{\Phi}\hat{A}^{-}\Bigg\}\,,
\end{equation}
where
\begin{equation}
\Phi^{a}=\partial_{-}\partial_{\bullet}A^{\bullet a}+\partial_{-}\partial_{\star}A^{\star a}-gf^{abc}\left(\partial_{-}A_{c}^{\bullet}A_{b}^{\star}+\partial_{-}A_{b}^{\bullet}A_{c}^{\star}\right)\,.
\end{equation}
Let us define a new field
\begin{equation}
\hat{W}=\hat{A}^{-}+\partial_{-}^{-2}\hat{\Phi}\,.
\end{equation}
The operator $\partial_{-}^{-1}$ is can be realized by an antiderivative.
The action $S'$ in terms of new field reads (after some algebra and
integrating by parts)
\begin{equation}
S'_{\mathrm{Y-M}}=\int d^{4}x\,\mathrm{Tr}\Bigg\{-\frac{1}{2}\hat{W}\partial_{-}^{2}\hat{W}+\frac{1}{2}\hat{\Phi}\partial_{-}^{-2}\hat{\Phi}\Bigg\}\,,
\end{equation}
The path integral over $W$ fields is Gaussian and thus can be readily
performed:
\begin{equation}
\int\left[dW\right]\,\exp\int d^{4}x\,\mathrm{Tr}\Bigg\{ i\frac{1}{2}\hat{W}\partial_{-}^{2}\hat{W}\Bigg\}=\mathcal{N}\,\left(\det\partial_{-}^{2}\right)^{-1/2}=\mathcal{N}'\,.
\end{equation}
This is a field independent infinite constant which can be discarded.

Thus, we end up with the following Yang-Mills Lagrangian, expressed
in terms of two transverse field components:
\begin{multline}
S_{\mathrm{Y-M}}^{\left(\mathrm{LC}\right)}=\int d^{4}x\,\mathrm{Tr}\Bigg\{-\hat{A}^{\bullet}\square\hat{A}^{\star}-\frac{1}{2}\hat{A}^{\bullet}\partial_{\bullet}^{2}\hat{A}^{\bullet}-\frac{1}{2}\hat{A}^{\star}\partial_{\star}^{2}\hat{A}^{\star}-\hat{A}^{\star}\partial_{\bullet}\partial_{\star}\hat{A}^{\bullet}\\
-ig'\left(\partial_{\bullet}\hat{A}^{\bullet}\left[\hat{A}^{\bullet},\hat{A}^{\star}\right]+\partial_{\star}\hat{A}^{\star}\left[\hat{A}^{\star},\hat{A}^{\bullet}\right]\right)\\
+\frac{1}{2}g'^{2}\left[\hat{A}^{\bullet},\hat{A}^{\star}\right]\left[\hat{A}^{\star},\hat{A}^{\bullet}\right]+\frac{1}{2}\hat{\Phi}\partial_{-}^{-2}\hat{\Phi}\Bigg\}\,.\label{eq:Action_LC}
\end{multline}

The above action mixes various terms between the ones in the  original formulation and
the terms appearing due to the integration over $A^{-}$ field. In
order to find the Feynman rules one needs to rewrite slightly this action. Let us start
with the kinetic term. Collecting all bilinear terms and integrating
by parts we get 
\begin{equation}
\mathcal{L}_{2}^{\left(\mathrm{LC}\right)}=\int d^{3}\mathbf{x}\,\mathrm{Tr}\Bigg\{-\hat{A}^{\bullet}\square\hat{A}^{\star}\Bigg\}\,.
\end{equation}
The trilinear contribution reads
\begin{multline}
\mathcal{L}_{3}^{\left(\mathrm{LC}\right)}=\int d^{3}\mathbf{x}\,\Bigg\{-\frac{1}{2}gf^{abc}\left(\partial_{-}A_{c}^{\bullet}A_{b}^{\star}+\partial_{-}A_{c}^{\star}A_{b}^{\bullet}\right)\partial_{-}^{-2}\left(\partial_{-}\partial_{\bullet}A_{a}^{\bullet}+\partial_{-}\partial_{\star}A_{a}^{\star}\right)\\
-\frac{1}{2}gf^{abc}\left(\partial_{-}\partial_{\bullet}A_{a}^{\bullet}+\partial_{-}\partial_{\star}A_{a}^{\star}\right)\partial_{-}^{-2}\left(\partial_{-}A_{c}^{\bullet}A_{b}^{\star}+\partial_{-}A_{c}^{\star}A_{b}^{\bullet}\right)\\
+gf^{abc}\left(\partial_{\star}A_{a}^{\star}A_{b}^{\star}A_{c}^{\bullet}+\partial_{\bullet}A_{a}^{\bullet}A_{b}^{\bullet}A_{c}^{\star}\right)\Bigg\}\,.\label{eq:L3LC}
\end{multline}
After some tedious algebra, integration by parts and usage of the
following relation (applicable for functions $f,g$ vanishing at infinity)
\begin{equation}
\int dx\,\partial^{-1}f\left(x\right)\partial g\left(x\right)=\int dx\,\partial f\left(x\right)\partial^{-1}g\left(x\right)\,.
\end{equation}
it simplifies to
\begin{equation}
\mathcal{L}_{3}^{\left(\mathrm{LC}\right)}=\int d^{3}\mathbf{x}\, gf^{abc}\Bigg\{\gamma_{\mathbf{x}}A_{a}^{\bullet}\left(\partial_{-}A_{b}^{\star}A_{c}^{\bullet}-\partial_{-}A_{c}^{\star}A_{b}^{\bullet}\right)+\overline{\gamma}_{\mathbf{x}}A_{a}^{\star}\left(\partial_{-}A_{b}^{\bullet}A_{c}^{\star}-\partial_{-}A_{c}^{\bullet}A_{b}^{\star}\right)\Bigg\}\,,
\end{equation}
where we defined the operators
\begin{equation}
\gamma_{\mathbf{x}}=\partial_{-}^{-1}\partial_{\bullet},\,\,\,\,\,\overline{\gamma}_{\mathbf{x}}=\partial_{-}^{-1}\partial_{\star}\,.
\end{equation}
In terms of algebra-valued fields this reads
\begin{equation}
\mathcal{L}_{3}^{\left(\mathrm{LC}\right)}=-i\sqrt{2}g\,\int d^{3}\mathbf{x}\,\mathrm{Tr}\Bigg\{\gamma_{\mathbf{x}}\hat{A}^{\bullet}\left[\partial_{-}\hat{A}^{\star},\hat{A}^{\bullet}\right]+\overline{\gamma}_{\mathbf{x}}\hat{A}^{\star}\left[\partial_{-}\hat{A}^{\bullet},\hat{A}^{\star}\right]\Bigg\}\,.
\end{equation}

In order to find the vertex one needs to perform the Fourier transform to the momentum space. For the $A^{\bullet}A^{\bullet}A^{\star}$
field coupling, which corresponds to the `$++-$' configuration, we get (after a proper change of indices and integration
variables)
\begin{equation}
\mathcal{L}_{++-}^{\left(\mathrm{LC}\right)}=-2igf^{abc}\int d^{3}\mathbf{p}_{1}d^{3}\mathbf{p}_{2}d^{3}\mathbf{p}_{3}\,\delta^{3}\left(\mathbf{p}_{1}+\mathbf{p}_{2}+\mathbf{p}_{3}\right)\frac{p_{1}^{\star}}{p_{1}^{+}}p_{3}^{+}\,\tilde{A}_{a}^{\bullet}\left(\mathbf{p}_{1}\right)\tilde{A}_{b}^{\bullet}\left(\mathbf{p}_{2}\right)\tilde{A}_{c}^{\star}\left(\mathbf{p}_{3}\right)\,.
\end{equation}
This can also be written as
\begin{equation}
\mathcal{L}_{++-}^{\left(\mathrm{LC}\right)}=\int d^{3}\mathbf{p}_{1}d^{3}\mathbf{p}_{2}d^{3}\mathbf{p}_{3}\,\delta^{3}\left(\mathbf{p}_{1}+\mathbf{p}_{2}+\mathbf{p}_{3}\right)\tilde{V}_{++-}^{abc}\left(\mathbf{p}_{1},\mathbf{p}_{2},\mathbf{p}_{3}\right)\,\tilde{A}_{a}^{\bullet}\left(\mathbf{p}_{1}\right)\tilde{A}_{b}^{\bullet}\left(\mathbf{p}_{2}\right)\tilde{A}_{c}^{\star}\left(\mathbf{p}_{3}\right)\,,
\end{equation}
with
\begin{equation}
\tilde{V}_{++-}^{abc}\left(\mathbf{p}_{1},\mathbf{p}_{2},\mathbf{p}_{3}\right)=-igf^{abc}\left(\frac{p_{1}^{\star}}{p_{1}^{+}}-\frac{p_{2}^{\star}}{p_{2}^{+}}\right)p_{3}^{+}\,.
\end{equation}
The mirror symmetric contribution reads
\begin{equation}
\mathcal{L}_{--+}^{\left(\mathrm{LC}\right)}=\int d^{3}\mathbf{p}_{1}d^{3}\mathbf{p}_{2}d^{3}\mathbf{p}_{3}\,\delta^{3}\left(\mathbf{p}_{1}+\mathbf{p}_{2}+\mathbf{p}_{3}\right)\tilde{V}_{--+}^{abc}\left(\mathbf{p}_{1},\mathbf{p}_{2},\mathbf{p}_{3}\right)\,\tilde{A}_{a}^{\star}\left(\mathbf{p}_{1}\right)\tilde{A}_{b}^{\star}\left(\mathbf{p}_{2}\right)\tilde{A}_{c}^{\bullet}\left(\mathbf{p}_{3}\right)\,,
\end{equation}
with
\begin{equation}
\tilde{V}_{--+}^{abc}\left(\mathbf{p}_{1},\mathbf{p}_{2},\mathbf{p}_{3}\right)=-igf^{abc}\left(\frac{p_{1}^{\bullet}}{p_{1}^{+}}-\frac{p_{2}^{\bullet}}{p_{2}^{+}}\right)p_{3}^{+}\,.\label{eq:App_vertex3g}
\end{equation}
Let us proceed to the term with the four-gluon coupling. The original quartic coupling
present in the action (\ref{eq:YM_action_explicit}) reads
\begin{equation}
\frac{1}{2}g'^{2}\mathrm{Tr}\Bigg\{\left[\hat{A}^{\bullet},\hat{A}^{\star}\right]\left[\hat{A}^{\star},\hat{A}^{\bullet}\right]\Bigg\}=-\frac{1}{2}g{}^{2}f^{abc}f^{ade}A_{b}^{\bullet}A_{c}^{\star}A_{d}^{\star}A_{e}^{\bullet}\,.
\end{equation}
The term originating from the integration over $A^{-}$ contributes:
\begin{equation}
\frac{1}{2}g^{2}f^{abc}f^{ade}\int d^{3}\mathbf{x}\,\left(\partial_{-}A_{c}^{\bullet}A_{b}^{\star}+\partial_{-}A_{c}^{\star}A_{b}^{\bullet}\right)\partial_{-}^{-2}\left(\partial_{-}A_{e}^{\bullet}A_{d}^{\star}+\partial_{-}A_{e}^{\star}A_{d}^{\bullet}\right)\,.
\end{equation}
Thus total quartic coupling is
\begin{equation}
\mathcal{L}_{++--}^{\left(\mathrm{LC}\right)}=\frac{1}{2}g{}^{2}f^{abc}f^{ade}\int d^{3}\mathbf{x}\,\Bigg\{\left(\partial_{-}A_{c}^{\bullet}A_{b}^{\star}+\partial_{-}A_{c}^{\star}A_{b}^{\bullet}\right)\partial_{-}^{-2}\left(\partial_{-}A_{e}^{\bullet}A_{d}^{\star}+\partial_{-}A_{e}^{\star}A_{d}^{\bullet}\right)-A_{b}^{\bullet}A_{c}^{\star}A_{d}^{\star}A_{e}^{\bullet}\Bigg\}\,.
\end{equation}
This form accommodates the so-called Coulomb instantaneous interactions.
After a little algebra we get in the momentum space
\begin{multline}
\mathcal{L}_{++--}^{\left(\mathrm{LC}\right)}=\int d^{3}\mathbf{p}_{1}d^{3}\mathbf{p}_{2}d^{3}\mathbf{p}_{3}d^{3}\mathbf{p}_{4}\delta^{4}\left(\mathbf{p}_{1}+\mathbf{p}_{2}+\mathbf{p}_{3}+\mathbf{p}_{4}\right)\\
\tilde{A}_{b_{1}}^{\bullet}\left(\mathbf{p}_{1}\right)\tilde{A}_{b_{2}}^{\star}\left(\mathbf{p}_{2}\right)\tilde{A}_{b_{3}}^{\star}\left(\mathbf{p}_{3}\right)\tilde{A}_{b_{4}}^{\bullet}\left(\mathbf{p}_{4}\right)\,\tilde{V}_{+--+}^{b_{1}b_{2}b_{3}b_{4}}\left(\mathbf{p}_{1},\mathbf{p}_{2},\mathbf{p}_{3},\mathbf{p}_{4}\right)\,,
\end{multline}
where
\begin{equation}
\tilde{V}_{+--+}^{b_{1}b_{2}b_{3}b_{4}}\left(\mathbf{p}_{1},\mathbf{p}_{2},\mathbf{p}_{3},\mathbf{p}_{4}\right)=g{}^{2}f^{ab_{1}b_{2}}f^{ab_{3}b_{4}}\frac{p_{1}^{+}p_{3}^{+}+p_{2}^{+}p_{4}^{+}}{\left(p_{3}^{+}+p_{4}^{+}\right)^{2}}\,.\label{eq:fourgluonvertex}
\end{equation}
One can also show that in position space in terms of matrix-valued
fields the quartic coupling can be written in a compact way as
\begin{equation}
\mathcal{L}_{++--}^{\left(\mathrm{LC}\right)}=-g^{2}\int d^{3}\mathbf{x}\,\mathrm{Tr}\left\{ \left[\partial_{-}\hat{A}^{\bullet},\hat{A}^{\star}\right]\partial_{-}^{-2}\left[\partial_{-}\hat{A}^{\star},\hat{A}^{\bullet}\right]\right\} \,.
\end{equation}

\subsection{The field transformation}

The idea is to introduce new fields $B^{\bullet},B^{\star}$ in place
of the fields $A^{\bullet},A^{\star}$ such that the new kinetic term accomodates the `$++-$' vertex which does not appear in CSW action \citep{Mansfield2006}
\begin{equation}
\mathcal{L}_{2}^{\left(\mathrm{LC}\right)}\left[A^{\bullet},A^{\star}\right]+\mathcal{L}_{++-}^{\left(\mathrm{LC}\right)}\left[A^{\bullet},A^{\star}\right]=\mathcal{L}_{2}^{\left(\mathrm{LC}\right)}\left[B^{\bullet},B^{\star}\right]\,.
\end{equation}
Writing this explicitly we have
\begin{equation}
\int d^{3}\mathbf{x}\,\mathrm{Tr}\Bigg\{-\hat{A}^{\bullet}\square\hat{A}^{\star}-i\sqrt{2}g\gamma_{\mathbf{x}}\hat{A}^{\bullet}\left[\partial_{-}\hat{A}^{\star},\hat{A}^{\bullet}\right]\Bigg\}=\int d^{3}\mathbf{x}\,\mathrm{Tr}\Bigg\{-\hat{B}^{\bullet}\square\hat{B}^{\star}\Bigg\}\,.\label{eq:Transf1}
\end{equation}
The transformation is assumed to be canonical, which leads to a simple
choice
\begin{equation}
B^{\bullet}=B^{\bullet}\left[A^{\bullet}\right]\,,
\end{equation}
and
\begin{equation}
\partial_{-}A_{a}^{\star}\left(\mathbf{x}\right)=\int d^{3}\mathbf{y}\,\frac{\delta B_{c}^{\bullet}\left(\mathbf{y}\right)}{\delta A_{a}^{\bullet}\left(\mathbf{x}\right)}\partial_{-}B_{c}^{\star}\left(\mathbf{y}\right).\label{eq:Bmindef-2-1}
\end{equation}
We will use it in (\ref{eq:Transf1}) to eliminate the $B^{\star}$
field. Then,  after some algebra, the relation (\ref{eq:Transf1})  becomes
\begin{multline}
\int d^{3}\mathbf{x}\,\mathrm{Tr}\Bigg\{2\partial_{+}\hat{A}^{\bullet}\partial_{-}\hat{A}^{\star}+2\partial_{\bullet}\partial_{\star}\partial_{-}^{-1}\partial_{-}\hat{A}^{\bullet}\hat{A}^{\star}-i\sqrt{2}g\gamma_{\mathbf{x}}\hat{A}^{\bullet}\left[\partial_{-}\hat{A}^{\star},\hat{A}^{\bullet}\right]\Bigg\}\\
=\int d^{3}\mathbf{x}\,\mathrm{Tr}\Bigg\{2\partial_{+}\hat{B}^{\bullet}\partial_{-}\hat{B}^{\star}+2\partial_{\bullet}\partial_{\star}\partial_{-}^{-1}\partial_{-}\hat{B}^{\bullet}\hat{B}^{\star}\Bigg\}\,.
\end{multline}
Integrating by parts and using
\begin{equation}
\omega_{\mathbf{x}}=\partial_{\bullet}\partial_{\star}\partial_{-}^{-1}\,,
\end{equation}
we get
\begin{equation}
\int d^{3}\mathbf{x}\,\mathrm{Tr}\Bigg\{2\left(\partial_{+}-\omega_{\mathbf{x}}\right)\hat{A}^{\bullet}\partial_{-}\hat{A}^{\star}-i\sqrt{2}g\left[\hat{A}^{\bullet},\gamma_{\mathbf{x}}\hat{A}^{\bullet}\right]\partial_{-}\hat{A}^{\star}\Bigg\}=\int d^{3}\mathbf{x}\,\mathrm{Tr}\Bigg\{2\left(\partial_{+}-\omega_{\mathbf{x}}\right)\hat{B}^{\bullet}\partial_{-}\hat{B}^{\star}\Bigg\}\,.
\end{equation}
Using (\ref{eq:Bmindef-2-1}) we get
\begin{multline}
\int d^{3}\mathbf{x}\, d^{3}\mathbf{y}\,\mathrm{Tr}\Bigg\{\left(2\left(\partial_{+}-\omega_{\mathbf{x}}\right)\hat{A}^{\bullet}\left(\mathbf{x}\right)-i\sqrt{2}g\left[\hat{A}^{\bullet}\left(\mathbf{x}\right),\gamma_{\mathbf{x}}\hat{A}^{\bullet}\left(\mathbf{x}\right)\right]\right)t^{a}\Bigg\}\frac{\delta B_{c}^{\bullet}\left(\mathbf{y}\right)}{\delta A_{a}^{\bullet}\left(\mathbf{x}\right)}\partial_{-}B_{c}^{\star}\left(\mathbf{y}\right)\\
=\int d^{3}\mathbf{x}\,\mathrm{Tr}\Bigg\{2\left(\partial_{+}-\omega_{\mathbf{x}}\right)\hat{B}^{\bullet}\left(\mathbf{x}\right)\partial_{-}\hat{B}^{\star}\left(\mathbf{x}\right)\Bigg\}\,.
\end{multline}
The field $B^{\star}$ is an arbitrary function here, thus we can
write
\begin{equation}
\int d^{3}\mathbf{y}\,\mathrm{Tr}\Bigg\{\left(2\left(\partial_{+}-\omega_{\mathbf{y}}\right)\hat{A}^{\bullet}\left(\mathbf{y}\right)-i\sqrt{2}g\left[\hat{A}^{\bullet}\left(\mathbf{y}\right),\gamma_{\mathbf{y}}\hat{A}^{\bullet}\left(\mathbf{y}\right)\right]\right)t^{c}\Bigg\}\frac{\delta B_{a}^{\bullet}\left(\mathbf{x}\right)}{\delta A_{c}^{\bullet}\left(\mathbf{y}\right)}=2\left(\partial_{+}-\omega_{\mathbf{x}}\right)B_{a}^{\bullet}\left(\mathbf{x}\right)\,.
\end{equation}
Let us next notice, that if
\begin{equation}
B^{\bullet}\left(y^{+},\mathbf{y}\right)=B^{\bullet}\left(A^{\bullet}\left(y^{+},\mathbf{y}\right)\right)\,,
\end{equation}
i.e. $B^{\bullet}$ depends on $y^{+}$ only through $A^{\bullet}$
it follows that
\begin{equation}
\int d^{3}\mathbf{y}\,\partial_{+}A_{c}^{\bullet}\left(\mathbf{y}\right)\frac{\delta B_{a}^{\bullet}\left(\mathbf{x}\right)}{\delta A_{c}^{\bullet}\left(\mathbf{y}\right)}=\partial_{+}B_{a}^{\bullet}\left(\mathbf{x}\right)\,,
\end{equation}
as can be easily seen. Therefore we get
\begin{equation}
\int d^{3}\mathbf{y}\,\mathrm{Tr}\Bigg\{\left(\omega_{\mathbf{y}}\hat{A}^{\bullet}\left(\mathbf{y}\right)+ig'\left[\hat{A}^{\bullet}\left(\mathbf{y}\right),\gamma_{\mathbf{y}}\hat{A}^{\bullet}\left(\mathbf{y}\right)\right]\right)t^{c}\Bigg\}\frac{\delta B_{a}^{\bullet}\left(\mathbf{x}\right)}{\delta A_{c}^{\bullet}\left(\mathbf{y}\right)}=\omega_{\mathbf{x}}B_{a}^{\bullet}\left(\mathbf{x}\right)\,,\label{eq:Transformation0}
\end{equation}
which can be compactly written as
\begin{equation}
\int d^{3}\mathbf{y}\,\mathrm{Tr}\Bigg\{\left[D_{\star},\gamma_{\mathbf{y}}\hat{A}^{\bullet}\left(\mathbf{y}\right)\right]t^{c}\Bigg\}\frac{\delta B_{a}^{\bullet}\left(\mathbf{x}\right)}{\delta A_{c}^{\bullet}\left(\mathbf{y}\right)}=\omega_{\mathbf{x}}B_{a}^{\bullet}\left(\mathbf{x}\right)\,.\label{eq:Transformation}
\end{equation}

\subsection{Solutions to field transformations}

In order to construct the MHV action one needs to replace $A^{\bullet}$,
$A^{\star}$ fields in the remaining interaction terms to express
them through $B^{\bullet}$, $B^{\star}$. Thus we need solutions
to (\ref{eq:Transformation}) and (\ref{eq:Bmindef-2-1}), $A^{\bullet}=A^{\bullet}\left[B^{\bullet}\right]$,
$A^{\star}=A^{\star}\left[B^{\bullet},B^{\star}\right]$. The solution
$A^{\bullet}\left[B^{\bullet}\right]$ has been  discussed in Section~\ref{sec:MHV_action}
so we do not repeat it here.

We will find $A^{\star}=A^{\star}\left[B^{\bullet},B^{\star}\right]$.
The relation between $A^{\star}$ and $B^{\star}$ fields in given
by (\ref{eq:Bmindef-2-1}). We assume the following expansion 
\begin{multline}
A_{a}^{\star}\left(\mathbf{x}\right)=B_{a}^{\star}\left(\mathbf{x}\right)+\int d^{3}\mathbf{y}_{1}d^{3}\mathbf{y}_{2}\,\Omega_{2}^{ab_{1}b_{2}}\left(\mathbf{x};\mathbf{y}_{1},\mathbf{y}_{2}\right)B_{b_{1}}^{\star}\left(\mathbf{y}_{1}\right)B_{b_{2}}^{\bullet}\left(\mathbf{y}_{2}\right)\\
+\int d^{3}\mathbf{y}_{1}d^{3}\mathbf{y}_{2}d^{3}\mathbf{y}_{2}\,\Omega_{3}^{ab_{1}\left\{ b_{2}b_{3}\right\} }\left(\mathbf{x};\mathbf{y}_{1},\left\{ \mathbf{y}_{2},\mathbf{y}_{3}\right\} \right)B_{b_{1}}^{\star}\left(\mathbf{y}_{1}\right)B_{b_{2}}^{\bullet}\left(\mathbf{y}_{2}\right)B_{b_{3}}^{\bullet}\left(\mathbf{y}_{3}\right)+\dots
\end{multline}
The goal is to find the coefficient functions $\Omega_{n}$. We shall
find a few first terms and generalize the result. The induction proof
for any $n$ can be found in \citep{Ettle2006b}.

First, the r.h.s. of (\ref{eq:Bmindef-2-1}) reads
\begin{multline}
\int d^{3}\mathbf{y}\,\frac{\delta B_{c}^{\bullet}\left(\mathbf{y}\right)}{\delta A_{a}^{\bullet}\left(\mathbf{x}\right)}\partial_{-}B_{c}^{\star}\left(\mathbf{y}\right)=\partial_{-}B_{a}^{\star}\left(\mathbf{x}\right)+2\int d^{3}\mathbf{y}_{1}d^{3}\mathbf{y}_{2}\,\Gamma_{2}^{b_{2}\left\{ ab_{1}\right\} }\left(\mathbf{y}_{2};\left\{ \mathbf{x},\mathbf{y}_{1}\right\} \right)A_{b_{1}}^{\bullet}\left(\mathbf{y}_{1}\right)\partial_{-}B_{b_{2}}^{\star}\left(\mathbf{y}_{2}\right)\\
+3\int d^{3}\mathbf{y}_{1}d^{3}\mathbf{y}_{2}d^{3}\mathbf{y}_{3}\,\Gamma_{3}^{b_{3}\left\{ ab_{1}b_{2}\right\} }\left(\mathbf{y}_{3};\left\{ \mathbf{x},\mathbf{y}_{1},\mathbf{y}_{2}\right\} \right)A_{b_{1}}^{\bullet}\left(\mathbf{y}_{1}\right)A_{b_{2}}^{\bullet}\left(\mathbf{y}_{2}\right)\partial_{-}B_{b_{3}}^{\star}\left(\mathbf{y}_{3}\right)+\dots
\end{multline}
where we have used the symmetry of $\Gamma_{n}$ symbols. Using the
expansion for $A^{\bullet}$ fields we have
\begin{multline}
\int d^{3}\mathbf{y}\,\frac{\delta B_{c}^{\bullet}\left(\mathbf{y}\right)}{\delta A_{a}^{\bullet}\left(\mathbf{x}\right)}\partial_{-}B_{c}^{\star}\left(\mathbf{y}\right)=\partial_{-}B_{a}^{\star}\left(\mathbf{x}\right)\\
+2\int d^{3}\mathbf{y}_{1}\, d^{3}\mathbf{y}_{2}\,\Gamma_{2}^{c\left\{ ab\right\} }\left(\mathbf{y}_{2};\left\{ \mathbf{x},\mathbf{y}_{1}\right\} \right)\partial_{-}B_{c}^{\star}\left(\mathbf{y}_{2}\right)\\
\left[B_{b}^{\bullet}\left(\mathbf{y}_{1}\right)+\int d^{3}\mathbf{z}_{1}d^{3}\mathbf{z}_{2}\,\Psi_{2}^{b\left\{ e_{1}e_{2}\right\} }\left(\mathbf{y}_{1};\mathbf{z}_{1},\mathbf{z}_{2}\right)B_{e_{1}}^{\bullet}\left(\mathbf{z}_{1}\right)B_{e_{2}}^{\bullet}\left(\mathbf{z}_{2}\right)+\dots\right]\\
+3\int d^{3}\mathbf{z}\int d^{3}\mathbf{y}_{1}d^{3}\mathbf{y}_{2}\,\Gamma_{3}^{c\left\{ b_{1}b_{2}a\right\} }\left(\mathbf{z};\left\{ \mathbf{y}_{1},\mathbf{y}_{2},\mathbf{x}\right\} \right)B_{b_{1}}^{\bullet}\left(\mathbf{y}_{1}\right)B_{b_{2}}^{\bullet}\left(\mathbf{y}_{2}\right)\partial_{-}B_{c}^{\star}\left(\mathbf{z}\right)+\dots
\end{multline}
Therefore
\begin{multline}
\int d^{3}\mathbf{y}_{1}d^{3}\mathbf{y}_{2}\,\partial_{-}\Omega_{2}^{ab_{1}b_{2}}\left(\mathbf{x};\mathbf{y}_{1},\mathbf{y}_{2}\right)B_{b_{1}}^{\star}\left(\mathbf{y}_{1}\right)B_{b_{2}}^{\bullet}\left(\mathbf{y}_{2}\right)\\
=2\int d^{3}\mathbf{y}_{1}\, d^{3}\mathbf{y}_{2}\,\Gamma_{2}^{c\left\{ ab\right\} }\left(\mathbf{y}_{2};\left\{ \mathbf{x},\mathbf{y}_{1}\right\} \right)\partial_{-}B_{c}^{\star}\left(\mathbf{y}_{2}\right)B_{b}^{\bullet}\left(\mathbf{y}_{1}\right)\,,
\end{multline}
\begin{multline}
\int d^{3}\mathbf{y}_{1}d^{3}\mathbf{y}_{2}d^{3}\mathbf{y}_{3}\,\partial_{-}\Omega_{3}^{ab_{1}\left\{ b_{2}b_{3}\right\} }\left(\mathbf{x};\mathbf{y}_{1},\left\{ \mathbf{y}_{2},\mathbf{y}_{3}\right\} \right)B_{b_{1}}^{\star}\left(\mathbf{y}_{1}\right)B_{b_{2}}^{\bullet}\left(\mathbf{y}_{2}\right)B_{b_{3}}^{\bullet}\left(\mathbf{y}_{3}\right)\\
=2\int d^{3}\mathbf{y}_{1}\, d^{3}\mathbf{y}_{2}\,\Gamma_{2}^{c\left\{ ab\right\} }\left(\mathbf{y}_{2};\left\{ \mathbf{x},\mathbf{y}_{1}\right\} \right)\partial_{-}B_{c}^{\star}\left(\mathbf{y}_{2}\right)\int d^{3}\mathbf{z}_{1}d^{3}\mathbf{z}_{2}\,\Psi_{2}^{b\left\{ e_{1}e_{2}\right\} }\left(\mathbf{y}_{1};\mathbf{z}_{1},\mathbf{z}_{2}\right)B_{e_{1}}^{\bullet}\left(\mathbf{z}_{1}\right)B_{e_{2}}^{\bullet}\left(\mathbf{z}_{2}\right)\\
+3\int d^{3}\mathbf{y}_{1}d^{3}\mathbf{y}_{2}d^{3}\mathbf{y}_{3}\,\Gamma_{3}^{c\left\{ b_{1}b_{2}a\right\} }\left(\mathbf{y}_{3};\left\{ \mathbf{y}_{1},\mathbf{y}_{2},\mathbf{x}\right\} \right)B_{b_{1}}^{\bullet}\left(\mathbf{y}_{1}\right)B_{b_{2}}^{\bullet}\left(\mathbf{y}_{2}\right)\partial_{-}B_{c}^{\star}\left(\mathbf{y}_{3}\right)\,,
\end{multline}
and so on. Passing to the momentum space we have for $\tilde{\Omega}_{2}$
\begin{multline}
\int d^{3}\mathbf{p}_{1}d^{3}\mathbf{p}_{2}\, P^{+}\tilde{\Omega}_{2}^{ab_{1}b_{2}}\left(\mathbf{P};\mathbf{p}_{1},\mathbf{p}_{2}\right)\tilde{B}_{b_{1}}^{\star}\left(\mathbf{p}_{1}\right)\tilde{B}_{b_{2}}^{\bullet}\left(\mathbf{p}_{2}\right)\\
=2\int d^{3}\mathbf{p}_{1}\, d^{3}\mathbf{p}_{2}\, p_{1}^{+}\tilde{\Gamma}_{2}^{b_{1}\left\{ ab_{2}\right\} }\left(-\mathbf{p}_{1};\left\{ -\mathbf{P},\mathbf{p}_{2}\right\} \right)\tilde{B}_{b_{1}}^{\star}\left(\mathbf{p}_{1}\right)\tilde{B}_{b_{2}}^{\bullet}\left(\mathbf{p}_{2}\right)\,.
\end{multline}
Therefore
\begin{equation}
\tilde{\Omega}_{2}^{ab_{1}b_{2}}\left(\mathbf{P};\mathbf{p}_{1},\mathbf{p}_{2}\right)=-2g'\frac{p_{1}^{+}}{p_{12}^{+}}\,\frac{1}{2!}\Bigg\{\frac{1}{\tilde{v}_{\overline{\left(12\right)}\overline{1}}^{*}}\,\mathrm{Tr}\left(t^{a}t^{b_{2}}t^{b_{1}}\right)+\frac{1}{\tilde{v}_{2\overline{1}}^{*}}\,\mathrm{Tr}\left(t^{a}t^{b_{1}}t^{b_{2}}\right)\Bigg\}\,,
\end{equation}
where the bar sign in $\tilde{v}_{ij}$ symbols denotes the inverted
momentum. From the definition we have the properties:
\begin{equation}
\tilde{v}_{\left(q\right)\left(\overline{p}\right)}=-q_{z}+q^{+}\frac{-p_{z}}{-p^{+}}=\tilde{v}_{\left(q\right)\left(p\right)}\,,\label{eq:vtild_bar1}
\end{equation}
\begin{equation}
\tilde{v}_{\left(\overline{q}\right)\left(p\right)}=-\left(-q_{z}+q^{+}\frac{p_{z}}{p^{+}}\right)=-\tilde{v}_{\left(q\right)\left(p\right)}\,.\label{eq:vtild_bar2}
\end{equation}
Thus
\begin{equation}
\tilde{\Omega}_{2}^{ab_{1}b_{2}}\left(\mathbf{P};\mathbf{p}_{1},\mathbf{p}_{2}\right)=-2g'\frac{p_{1}^{+}}{p_{12}^{+}}\,\frac{1}{2!}\Bigg\{-\frac{1}{\tilde{v}_{21}^{*}}\,\mathrm{Tr}\left(t^{a}t^{b_{2}}t^{b_{1}}\right)+\frac{1}{\tilde{v}_{21}^{*}}\,\mathrm{Tr}\left(t^{a}t^{b_{1}}t^{b_{2}}\right)\Bigg\}\,.
\end{equation}
Also we have
\begin{equation}
\tilde{\Psi}_{2}^{a\left\{ b_{1}b_{2}\right\} }\left(\mathbf{P};\mathbf{p}_{1},\mathbf{p}_{2}\right)=-g'\frac{1}{2!}\Bigg\{\frac{p_{12}^{+}}{p_{1}^{+}}\,\frac{1}{\tilde{v}_{21}^{*}}\mathrm{Tr}\left(t^{a}t^{b_{1}}t^{b_{2}}\right)-\frac{p_{12}^{+}}{p_{1}^{+}}\,\frac{1}{\tilde{v}_{21}^{*}}\mathrm{Tr}\left(t^{a}t^{b_{2}}t^{b_{1}}\right)\Bigg\}\,.
\end{equation}
Therefore we can find the relation between the two coefficients
\begin{equation}
\tilde{\Omega}_{2}^{ab_{1}b_{2}}\left(\mathbf{P};\mathbf{p}_{1},\mathbf{p}_{2}\right)=2\left(\frac{p_{1}^{+}}{p_{12}^{+}}\right)^{2}\tilde{\Psi}_{2}^{a\left\{ b_{1}b_{2}\right\} }\left(\mathbf{P};\mathbf{p}_{1},\mathbf{p}_{2}\right)\,.
\end{equation}
For $\tilde{\Omega}_{3}$ we have
\begin{multline}
\int d^{3}\mathbf{p}_{1}d^{3}\mathbf{p}_{2}d^{3}\mathbf{p}_{3}\, P^{+}\tilde{\Omega}_{3}^{ab_{1}\left\{ b_{2}b_{3}\right\} }\left(\mathbf{P};\mathbf{p}_{1},\left\{ \mathbf{p}_{2},\mathbf{p}_{3}\right\} \right)\tilde{B}_{b_{1}}^{\star}\left(\mathbf{p}_{1}\right)\tilde{B}_{b_{2}}^{\bullet}\left(\mathbf{p}_{2}\right)\tilde{B}_{b_{3}}^{\bullet}\left(\mathbf{p}_{3}\right)\\
=2\int d^{3}\mathbf{p}_{1}d^{3}\mathbf{p}_{2}d^{3}\mathbf{p}_{3}\int d^{3}\mathbf{q}\,\tilde{\Gamma}_{2}^{b_{1}\left\{ ac\right\} }\left(-\mathbf{p}_{1};\left\{ -\mathbf{P},\mathbf{q}\right\} \right)\,\tilde{\Psi}_{2}^{c\left\{ b_{2}b_{3}\right\} }\left(\mathbf{q};\mathbf{p}_{2},\mathbf{p}_{3}\right)p_{1}^{+}\tilde{B}_{b_{1}}^{\star}\left(\mathbf{p}_{1}\right)\tilde{B}_{b_{2}}^{\bullet}\left(\mathbf{p}_{2}\right)\tilde{B}_{b_{3}}^{\bullet}\left(\mathbf{p}_{3}\right)\\
+3\int d^{3}\mathbf{p}_{1}d^{3}\mathbf{p}_{2}d^{3}\mathbf{p}_{3}\,\Gamma_{3}^{b_{1}\left\{ b_{2}b_{3}a\right\} }\left(-\mathbf{p}_{1};\left\{ \mathbf{p}_{2},\mathbf{p}_{3},-\mathbf{P}\right\} \right)p_{1}^{+}B_{b_{1}}^{\star}\left(\mathbf{p}_{1}\right)B_{b_{2}}^{\bullet}\left(\mathbf{p}_{2}\right)B_{b_{3}}^{\bullet}\left(\mathbf{p}_{3}\right)\,,
\end{multline}
which leads to
\begin{multline}
P^{+}\tilde{\Omega}_{3}^{ab_{1}\left\{ b_{2}b_{3}\right\} }\left(\mathbf{P};\mathbf{p}_{1},\left\{ \mathbf{p}_{2},\mathbf{p}_{3}\right\} \right)=2p_{1}^{+}\int d^{3}\mathbf{q}\,\tilde{\Gamma}_{2}^{b_{1}\left\{ ac\right\} }\left(-\mathbf{p}_{1};\left\{ -\mathbf{P},\mathbf{q}\right\} \right)\,\tilde{\Psi}_{2}^{c\left\{ b_{2}b_{3}\right\} }\left(\mathbf{q};\mathbf{p}_{2},\mathbf{p}_{3}\right)\\
+3p_{1}^{+}\Gamma_{3}^{b_{1}\left\{ b_{2}b_{3}a\right\} }\left(-\mathbf{p}_{1};\left\{ \mathbf{p}_{2},\mathbf{p}_{3},-\mathbf{P}\right\} \right)\,.\label{eq:omeg3_1}
\end{multline}
Let us rewrite the $\tilde{\Gamma}_{2}$ appearing above as (taking
into account the convolution in $\mathbf{q}$ and omitting the delta
function)
\begin{multline}
\tilde{\Gamma}_{2}^{b_{1}\left\{ ac\right\} }\left(-\mathbf{p}_{1};\left\{ -\mathbf{P},\mathbf{p}_{23}\right\} \right)=-g'\,\frac{1}{2!}\Bigg\{\frac{1}{\tilde{v}_{\overline{\left(123\right)}\overline{1}}^{*}}\,\mathrm{Tr}\left(t^{b_{1}}t^{a}t^{c}\right)+\frac{1}{\tilde{v}_{\left(23\right)\overline{1}}^{*}}\,\mathrm{Tr}\left(t^{b_{1}}t^{c}t^{a}\right)\Bigg\}\\
=-g'\,\frac{1}{2!}\Bigg\{-\frac{1}{\tilde{v}_{\left(23\right)1}^{*}}\,\mathrm{Tr}\left(t^{b_{1}}t^{a}t^{c}\right)+\frac{1}{\tilde{v}_{\left(23\right)1}^{*}}\,\mathrm{Tr}\left(t^{b_{1}}t^{c}t^{a}\right)\Bigg\}=i\sqrt{2}g'\,\frac{1}{2!}\frac{1}{\tilde{v}_{\left(23\right)1}^{*}}\, f^{b_{1}ac}\,.
\end{multline}
where we used $\mathrm{Tr}\left(t^{a}t^{b}t^{c}-t^{c}t^{b}t^{a}\right)=i\sqrt{2}f^{abc}$.
The $\tilde{\Gamma}_{3}$ appearing in (\ref{eq:omeg3_1}) reads
\begin{multline*}
\Gamma_{3}^{b_{1}\left\{ b_{2}b_{3}a\right\} }\left(-\mathbf{p}_{1};\left\{ \mathbf{p}_{2},\mathbf{p}_{3},-\mathbf{P}\right\} \right)=\left(g'\right)^{2}\frac{1}{3!}\\
\Bigg\{\frac{1}{\tilde{v}_{2\overline{1}}^{*}\tilde{v}_{\left(23\right)\overline{1}}^{*}}\mathrm{Tr}\left(t^{b_{1}}t^{b_{2}}t^{b_{3}}t^{a}\right)+\frac{1}{\tilde{v}_{2\overline{1}}^{*}\tilde{v}_{\left(2\overline{123}\right)\overline{1}}^{*}}\mathrm{Tr}\left(t^{b_{1}}t^{b_{2}}t^{a}t^{b_{3}}\right)\\
+\frac{1}{\tilde{v}_{3\overline{1}}^{*}\tilde{v}_{\left(23\right)\overline{1}}^{*}}\mathrm{Tr}\left(t^{b_{1}}t^{b_{3}}t^{b_{2}}t^{a}\right)+\frac{1}{\tilde{v}_{3\overline{1}}^{*}\tilde{v}_{\left(3\overline{123}\right)\overline{1}}^{*}}\mathrm{Tr}\left(t^{b_{1}}t^{b_{3}}t^{a}t^{b_{2}}\right)\\
+\frac{1}{\tilde{v}_{\overline{123}\overline{1}}^{*}\tilde{v}_{\left(\overline{123}3\right)\overline{1}}^{*}}\mathrm{Tr}\left(t^{b_{1}}t^{a}t^{b_{3}}t^{b_{2}}\right)+\frac{1}{\tilde{v}_{\overline{123}\overline{1}}^{*}\tilde{v}_{\left(\overline{123}2\right)\overline{1}}^{*}}\mathrm{Tr}\left(t^{b_{1}}t^{a}t^{b_{2}}t^{b_{3}}\right)\Bigg\}
\end{multline*}
Using these to calculate $\tilde{\Omega}_{3}$ we get
\begin{multline}
\tilde{\Omega}_{3}^{ab_{1}\left\{ b_{2}b_{3}\right\} }\left(\mathbf{P};\mathbf{p}_{1},\left\{ \mathbf{p}_{2},\mathbf{p}_{3}\right\} \right)=\frac{p_{1}^{+}}{p_{123}^{+}}\delta^{3}\left(\mathbf{p}_{123}-\mathbf{P}\right)\\
\Bigg\{-2i\sqrt{2}g'\,\frac{1}{2!}\frac{1}{\tilde{v}_{\left(23\right)1}^{*}}\, f^{b_{1}ac}\left(g'\right)\frac{p_{23}^{+}}{p_{2}^{+}}\frac{1}{\tilde{v}_{32}^{*}}\,\mathrm{Tr}\left(t^{c}t^{b_{2}}t^{b_{3}}\right)\\
+3\left(g'\right)^{2}\frac{1}{3!}\sum_{\mathrm{perm\,}\left\{ \left(b_{2},\mathbf{p}_{2}\right),\left(b_{3},\mathbf{p}_{3}\right),\left(a,\mathbf{p}_{123}\right)\right\} }\frac{1}{\tilde{v}_{2\overline{1}}^{*}\tilde{v}_{\left(23\right)\overline{1}}^{*}}\mathrm{Tr}\left(t^{b_{1}}t^{b_{2}}t^{b_{3}}t^{a}\right)\Bigg\}\,,
\end{multline}
where the sum in the last line is over permutations of pairs indicated.
The first line in the bracket can be written as
\begin{equation}
-\left(g'\right)^{2}\frac{p_{23}^{+}}{p_{2}^{+}}\,\frac{1}{\tilde{v}_{\left(23\right)1}^{*}\tilde{v}_{32}^{*}}\left[\mathrm{Tr}\left(t^{b_{2}}t^{b_{3}}t^{b_{1}}t^{a}\right)-\mathrm{Tr}\left(t^{b_{2}}t^{b_{3}}t^{a}t^{b_{1}}\right)\right]\,.
\end{equation}
Let us note that this term can be rewritten in $\tilde{\Omega}_{3}$
as follows (in the `weak sense'):
\begin{multline}
\int d^{3}\mathbf{p}_{1}d^{3}\mathbf{p}_{2}d^{3}\mathbf{p}_{3}\frac{p_{1}^{+}}{p_{123}^{+}}\delta^{3}\left(\mathbf{p}_{123}-\mathbf{P}\right)\left(-\left(g'\right)^{2}\frac{p_{23}^{+}}{p_{2}^{+}}\,\frac{1}{\tilde{v}_{\left(23\right)1}^{*}\tilde{v}_{32}^{*}}\right)\\
\left[\mathrm{Tr}\left(t^{b_{2}}t^{b_{3}}t^{b_{1}}t^{a}\right)-\mathrm{Tr}\left(t^{b_{2}}t^{b_{3}}t^{a}t^{b_{1}}\right)\right]\tilde{B}_{b_{1}}^{\star}\left(\mathbf{p}_{1}\right)\tilde{B}_{b_{2}}^{\bullet}\left(\mathbf{p}_{2}\right)\tilde{B}_{b_{3}}^{\bullet}\left(\mathbf{p}_{3}\right)\\
=\frac{1}{2}\int d^{3}\mathbf{p}_{1}d^{3}\mathbf{p}_{2}d^{3}\mathbf{p}_{3}\frac{p_{1}^{+}}{p_{123}^{+}}\delta^{3}\left(\mathbf{p}_{123}-\mathbf{P}\right)\left(-\left(g'\right)^{2}\frac{p_{23}^{+}}{p_{2}^{+}}\,\frac{1}{\tilde{v}_{\left(23\right)1}^{*}\tilde{v}_{23}^{*}}\right)\\
\left[\mathrm{Tr}\left(t^{b_{3}}t^{b_{2}}t^{b_{1}}t^{a}\right)-\mathrm{Tr}\left(t^{b_{3}}t^{b_{2}}t^{a}t^{b_{1}}\right)\right]\tilde{B}_{b_{1}}^{\star}\left(\mathbf{p}_{1}\right)\tilde{B}_{b_{2}}^{\bullet}\left(\mathbf{p}_{2}\right)\tilde{B}_{b_{3}}^{\bullet}\left(\mathbf{p}_{3}\right)\,,
\end{multline}
Now we have
\begin{multline}
\tilde{\Omega}_{3}^{ab_{1}\left\{ b_{2}b_{3}\right\} }\left(\mathbf{P};\mathbf{p}_{1},\left\{ \mathbf{p}_{2},\mathbf{p}_{3}\right\} \right)=\frac{1}{2}\left(g'\right)^{2}\frac{p_{1}^{+}}{p_{123}^{+}}\delta^{3}\left(\mathbf{p}_{123}-\mathbf{P}\right)\\
\Bigg\{\left[-\frac{p_{23}^{+}}{p_{2}^{+}}\,\frac{1}{\tilde{v}_{\left(23\right)1}^{*}\tilde{v}_{32}^{*}}+\frac{1}{\tilde{v}_{\overline{123}\overline{1}}^{*}\tilde{v}_{\left(\overline{123}2\right)\overline{1}}^{*}}\right]\mathrm{Tr}\left(t^{a}t^{b_{2}}t^{b_{3}}t^{b_{1}}\right)+\left[\frac{p_{23}^{+}}{p_{2}^{+}}\,\frac{1}{\tilde{v}_{\left(23\right)1}^{*}\tilde{v}_{32}^{*}}+\frac{1}{\tilde{v}_{2\overline{1}}^{*}\tilde{v}_{\left(23\right)\overline{1}}^{*}}\right]\mathrm{Tr}\left(t^{a}t^{b_{1}}t^{b_{2}}t^{b_{3}}\right)\\
+\left[-\frac{p_{23}^{+}}{p_{2}^{+}}\,\frac{1}{\tilde{v}_{\left(23\right)1}^{*}\tilde{v}_{23}^{*}}+\frac{1}{\tilde{v}_{\overline{123}\overline{1}}^{*}\tilde{v}_{\left(\overline{123}3\right)\overline{1}}^{*}}\right]\mathrm{Tr}\left(t^{a}t^{b_{3}}t^{b_{2}}t^{b_{1}}\right)+\left[\frac{p_{23}^{+}}{p_{2}^{+}}\,\frac{1}{\tilde{v}_{\left(23\right)1}^{*}\tilde{v}_{23}^{*}}+\frac{1}{\tilde{v}_{3\overline{1}}^{*}\tilde{v}_{\left(23\right)\overline{1}}^{*}}\right]\mathrm{Tr}\left(t^{a}t^{b_{1}}t^{b_{3}}t^{b_{2}}\right)\\
+\frac{1}{\tilde{v}_{2\overline{1}}^{*}\tilde{v}_{\left(2\overline{123}\right)\overline{1}}^{*}}\mathrm{Tr}\left(t^{a}t^{b_{3}}t^{b_{1}}t^{b_{2}}\right)+\frac{1}{\tilde{v}_{3\overline{1}}^{*}\tilde{v}_{\left(3\overline{123}\right)\overline{1}}^{*}}\mathrm{Tr}\left(t^{a}t^{b_{2}}t^{b_{1}}t^{b_{3}}\right)\Bigg\}\,.
\end{multline}
Calculating the terms in square brackets, after some algebra using
(\ref{eq:vtild_bar1}),(\ref{eq:vtild_bar2}) and relations from (\ref{sec:App_ident})
we get for the first term:
\begin{equation}
-\frac{p_{23}^{+}}{p_{2}^{+}}\,\frac{1}{\tilde{v}_{\left(23\right)1}^{*}\tilde{v}_{32}^{*}}+\frac{1}{\tilde{v}_{\overline{123}\overline{1}}^{*}\tilde{v}_{\left(\overline{123}2\right)\overline{1}}^{*}}=\frac{1}{\tilde{v}_{23}^{*}\tilde{v}_{31}^{*}}\,.
\end{equation}
Similar calculation follows for the rest of the terms. We obtain
\begin{multline}
\tilde{\Omega}_{3}^{ab_{1}\left\{ b_{2}b_{3}\right\} }\left(\mathbf{P};\mathbf{p}_{1},\left\{ \mathbf{p}_{2},\mathbf{p}_{3}\right\} \right)=\frac{1}{2}\left(g'\right)^{2}\frac{p_{1}^{+}}{p_{123}^{+}}\delta^{3}\left(\mathbf{p}_{123}-\mathbf{P}\right)\\
\Bigg\{\frac{1}{\tilde{v}_{23}^{*}\tilde{v}_{31}^{*}}\mathrm{Tr}\left(t^{a}t^{b_{2}}t^{b_{3}}t^{b_{1}}\right)+\frac{1}{\tilde{v}_{32}^{*}\tilde{v}_{21}^{*}}\mathrm{Tr}\left(t^{a}t^{b_{1}}t^{b_{2}}t^{b_{3}}\right)\\
+\frac{1}{\tilde{v}_{32}^{*}\tilde{v}_{21}^{*}}\mathrm{Tr}\left(t^{a}t^{b_{3}}t^{b_{2}}t^{b_{1}}\right)+\frac{1}{\tilde{v}_{23}^{*}\tilde{v}_{31}^{*}}\mathrm{Tr}\left(t^{a}t^{b_{1}}t^{b_{3}}t^{b_{2}}\right)\\
-\frac{1}{\tilde{v}_{21}^{*}\tilde{v}_{31}^{*}}\mathrm{Tr}\left(t^{a}t^{b_{3}}t^{b_{1}}t^{b_{2}}\right)-\frac{1}{\tilde{v}_{31}^{*}\tilde{v}_{21}^{*}}\mathrm{Tr}\left(t^{a}t^{b_{2}}t^{b_{1}}t^{b_{3}}\right)\Bigg\}\,.
\end{multline}
This can be again expressed using $\tilde{\Psi}_{3}$. After some
algebra we get
\begin{multline}
\tilde{\Omega}_{3}^{ab_{1}\left\{ b_{2}b_{3}\right\} }\left(\mathbf{P};\mathbf{p}_{1},\left\{ \mathbf{p}_{2},\mathbf{p}_{3}\right\} \right)=\frac{3!}{2}\left(\frac{p_{1}^{+}}{p_{123}^{+}}\right)^{2}\\
\Bigg\{\tilde{\Psi}_{3}\left(\mathbf{P};\mathbf{p}_{2},\mathbf{p}_{3},\mathbf{p}_{1}\right)\mathrm{Tr}\left(t^{a}t^{b_{2}}t^{b_{3}}t^{b_{1}}\right)+\tilde{\Psi}_{3}\left(\mathbf{P};\mathbf{p}_{1},\mathbf{p}_{2},\mathbf{p}_{3}\right)\mathrm{Tr}\left(t^{a}t^{b_{1}}t^{b_{2}}t^{b_{3}}\right)\\
+\tilde{\Psi}_{3}\left(\mathbf{P};\mathbf{p}_{3},\mathbf{p}_{2},\mathbf{p}_{1}\right)\mathrm{Tr}\left(t^{a}t^{b_{3}}t^{b_{2}}t^{b_{1}}\right)+\tilde{\Psi}_{3}\left(\mathbf{P};\mathbf{p}_{1},\mathbf{p}_{3},\mathbf{p}_{2}\right)\mathrm{Tr}\left(t^{a}t^{b_{1}}t^{b_{3}}t^{b_{2}}\right)\\
+\tilde{\Psi}_{3}\left(\mathbf{P};\mathbf{p}_{3},\mathbf{p}_{1},\mathbf{p}_{2}\right)\mathrm{Tr}\left(t^{a}t^{b_{3}}t^{b_{1}}t^{b_{2}}\right)+\tilde{\Psi}_{3}\left(\mathbf{P};\mathbf{p}_{2},\mathbf{p}_{1},\mathbf{p}_{3}\right)\mathrm{Tr}\left(t^{a}t^{b_{2}}t^{b_{1}}t^{b_{3}}\right)\Bigg\}\,.
\end{multline}
So, finally
\begin{equation}
\tilde{\Omega}_{3}^{ab_{1}\left\{ b_{2}b_{3}\right\} }\left(\mathbf{P};\mathbf{p}_{1},\left\{ \mathbf{p}_{2},\mathbf{p}_{3}\right\} \right)=3\left(\frac{p_{1}^{+}}{p_{123}^{+}}\right)^{2}\tilde{\Psi}_{3}^{ab_{1}b_{2}b_{3}}\left(\mathbf{P};\mathbf{p}_{1},\dots,\mathbf{p}_{n}\right)\,.
\end{equation}
Note we have skipped the symmetrization brackets for $\tilde{\Psi}_{3}^{ab_{1}b_{2}b_{3}}$.
This means that it is given by the sum of color ordered amplitudes (\ref{eq:Psi_color_decomp}),
not by the expression (\ref{eq:Psi_n}).

This can be generalized to
\begin{equation}
\tilde{\Omega}_{n}^{ab_{1}\left\{ b_{2}\dots b_{n}\right\} }\left(\mathbf{P};\mathbf{p}_{1},\left\{ \mathbf{p}_{2},\dots,\mathbf{p}_{n}\right\} \right)=n\left(\frac{p_{1}^{+}}{p_{1\dots n}^{+}}\right)^{2}\tilde{\Psi}_{n}^{ab_{1}\dots b_{n}}\left(\mathbf{P};\mathbf{p}_{1},\dots,\mathbf{p}_{n}\right)\,.
\end{equation}

\subsection{The MHV vertices}

In this appendix we shall rederive the two lowest vertices in the MHV action.
 In \citep{Ettle2006b}
five-point vertex was also calculated and the general proof can be found in the
original paper \citep{Mansfield2006}.

The goal is to find
\begin{multline}
\mathcal{L}_{--+}^{\left(\mathrm{LC}\right)}\left[B^{\bullet},B^{\star}\right]+\mathcal{L}_{--++}^{\left(\mathrm{LC}\right)}\left[B^{\bullet},B^{\star}\right]+\mathcal{L}_{--+++}^{\left(\mathrm{LC}\right)}\left[B^{\bullet},B^{\star}\right]+\dots\\
=\mathcal{L}_{+--}^{\left(\mathrm{LC}\right)}\left[A^{\bullet},A^{\star}\right]+\mathcal{L}_{++--}^{\left(\mathrm{LC}\right)}\left[A^{\bullet},A^{\star}\right]\,,
\end{multline}
where $\mathcal{L}_{--+\dots+}^{\left(\mathrm{LC}\right)}\left[B^{\bullet},B^{\star}\right]$
are the interaction terms with the MHV vertices.

The first term is simply
\begin{multline*}
\mathcal{L}_{--+}^{\left(\mathrm{LC}\right)}\left[B^{\bullet},B^{\star}\right]=\\
=\int d^{3}\mathbf{p}_{1}d^{3}\mathbf{p}_{2}d^{3}\mathbf{p}_{3}\,\delta^{3}\left(\mathbf{p}_{1}+\mathbf{p}_{2}+\mathbf{p}_{3}\right)\tilde{V}_{--+}^{b_{1}b_{2}b_{3}}\left(\mathbf{p}_{1},\mathbf{p}_{2},\mathbf{p}_{3}\right)\,\tilde{B}_{b_{1}}^{\star}\left(\mathbf{p}_{1}\right)\tilde{B}_{b_{2}}^{\star}\left(\mathbf{p}_{2}\right)\tilde{B}_{b_{3}}^{\bullet}\left(\mathbf{p}_{3}\right)\,,
\end{multline*}
where
\begin{equation}
\tilde{V}_{--+}^{b_{1}b_{2}b_{3}}\left(\mathbf{p}_{1},\mathbf{p}_{2},\mathbf{p}_{3}\right)=-igf^{b_{1}b_{2}b_{3}}\left(\frac{p_{1}^{\bullet}}{p_{1}^{+}}-\frac{p_{2}^{\bullet}}{p_{2}^{+}}\right)p_{3}^{+}=-igf^{b_{1}b_{2}b_{3}}v_{12}^{*}p_{3}^{+}\,.
\end{equation}
We want to write the vertex using color-ordered amplitudes
\begin{equation}
\tilde{V}_{--+}^{b_{1}b_{2}b_{3}}\left(\mathbf{p}_{1},\mathbf{p}_{2},\mathbf{p}_{3}\right)=\mathcal{V}_{--+}\left(\mathbf{p}_{1},\mathbf{p}_{2},\mathbf{p}_{3}\right)\mathrm{Tr}\left(t^{b_{1}}t^{b_{2}}t^{b_{3}}\right)+\mathcal{V}_{--+}\left(\mathbf{p}_{3},\mathbf{p}_{2},\mathbf{p}_{1}\right)\mathrm{Tr}\left(t^{b_{1}}t^{b_{2}}t^{b_{3}}\right)\,.
\end{equation}
We find
\begin{equation}
\mathcal{V}_{--+}\left(\mathbf{p}_{1},\mathbf{p}_{2},\mathbf{p}_{3}\right)=-g'\left(\frac{p_{1}^{+}}{p_{2}^{+}}\right)^{2}\frac{\tilde{v}_{21}^{*4}}{\tilde{v}_{13}^{*}\tilde{v}_{32}^{*}\tilde{v}_{21}^{*}}\,.
\end{equation}

To get the second term we have to expand the $A$ fields. Retaining
only the four-field component we get from the three-point part:
\begin{multline}
\int d^{3}\mathbf{p}_{1}d^{3}\mathbf{p}_{2}d^{3}\mathbf{p}_{3}\,\delta^{3}\left(\mathbf{p}_{1}+\mathbf{p}_{2}+\mathbf{p}_{3}\right)\tilde{V}_{--+}^{b_{1}b_{2}b_{3}}\left(\mathbf{p}_{1},\mathbf{p}_{2},\mathbf{p}_{3}\right)\,\\
\int d^{3}\mathbf{q}_{1}d^{3}\mathbf{q}_{2}\tilde{\Omega}_{2}^{b_{1}c_{1}c_{2}}\left(\mathbf{p}_{1};\mathbf{q}_{1},\mathbf{q}_{2}\right)\tilde{B}_{c_{1}}^{\star}\left(\mathbf{q}_{1}\right)\tilde{B}_{c_{2}}^{\bullet}\left(\mathbf{q}_{2}\right)\tilde{B}_{b_{2}}^{\star}\left(\mathbf{p}_{2}\right)\tilde{B}_{b_{3}}^{\bullet}\left(\mathbf{p}_{3}\right)\\
+\int d^{3}\mathbf{p}_{1}d^{3}\mathbf{p}_{2}d^{3}\mathbf{p}_{3}\,\delta^{3}\left(\mathbf{p}_{1}+\mathbf{p}_{2}+\mathbf{p}_{3}\right)\tilde{V}_{--+}^{b_{1}b_{2}b_{3}}\left(\mathbf{p}_{1},\mathbf{p}_{2},\mathbf{p}_{3}\right)\,\\
\int d^{3}\mathbf{q}_{1}d^{3}\mathbf{q}_{2}\tilde{\Omega}_{2}^{b_{2}c_{1}c_{2}}\left(\mathbf{p}_{2};\mathbf{q}_{1},\mathbf{q}_{2}\right)\tilde{B}_{c_{1}}^{\star}\left(\mathbf{q}_{1}\right)\tilde{B}_{c_{2}}^{\bullet}\left(\mathbf{q}_{2}\right)\tilde{B}_{b_{1}}^{\star}\left(\mathbf{p}_{1}\right)\tilde{B}_{b_{3}}^{\bullet}\left(\mathbf{p}_{3}\right)\\
+\int d^{3}\mathbf{p}_{1}d^{3}\mathbf{p}_{2}d^{3}\mathbf{p}_{3}\,\delta^{3}\left(\mathbf{p}_{1}+\mathbf{p}_{2}+\mathbf{p}_{3}\right)\tilde{V}_{--+}^{b_{1}b_{2}b_{3}}\left(\mathbf{p}_{1},\mathbf{p}_{2},\mathbf{p}_{3}\right)\\
\int d^{3}\mathbf{q}_{1}d^{3}\mathbf{q}_{2}\tilde{\Psi}_{2}^{b_{3}\left\{ c_{1}c_{2}\right\} }\left(\mathbf{p}_{3};\left\{ \mathbf{q}_{1},\mathbf{q}_{2}\right\} \right)\tilde{B}_{b_{1}}^{\star}\left(\mathbf{p}_{1}\right)\tilde{B}_{b_{2}}^{\star}\left(\mathbf{p}_{2}\right)\tilde{B}_{c_{1}}^{\bullet}\left(\mathbf{q}_{1}\right)\tilde{B}_{c_{2}}^{\bullet}\left(\mathbf{q}_{2}\right)
\end{multline}
Changing the integration variables to correspond to the `external' momenta
$p_{1},\dots,p_{4}$ and helicity $--++$ and symmetrizing to involve
all the color orderings 
\begin{multline}
\int d^{3}\mathbf{p}_{1}d^{3}\mathbf{p}_{2}d^{3}\mathbf{p}_{3}d^{3}\mathbf{p}_{4}\int d^{3}\mathbf{q}\,\tilde{B}_{b_{1}}^{\star}\left(\mathbf{p}_{1}\right)\tilde{B}_{b_{2}}^{\star}\left(\mathbf{p}_{2}\right)\tilde{B}_{b_{3}}^{\bullet}\left(\mathbf{p}_{3}\right)\tilde{B}_{b_{4}}^{\bullet}\left(\mathbf{p}_{4}\right)\,\\
\Bigg\{\frac{1}{2}\delta^{3}\left(\mathbf{q}+\mathbf{p}_{2}+\mathbf{p}_{3}\right)\tilde{V}_{--+}^{ab_{2}b_{3}}\left(\mathbf{q},\mathbf{p}_{2},\mathbf{p}_{3}\right)\tilde{\Omega}_{2}^{ab_{1}b_{4}}\left(\mathbf{q};\mathbf{p}_{1},\mathbf{p}_{4}\right)\\
+\frac{1}{2}\delta^{3}\left(\mathbf{q}+\mathbf{p}_{2}+\mathbf{p}_{4}\right)\tilde{V}_{--+}^{ab_{2}b_{4}}\left(\mathbf{q},\mathbf{p}_{2},\mathbf{p}_{4}\right)\tilde{\Omega}_{2}^{ab_{1}b_{3}}\left(\mathbf{q};\mathbf{p}_{1},\mathbf{p}_{3}\right)\\
+\frac{1}{2}\delta^{3}\left(\mathbf{q}+\mathbf{p}_{1}+\mathbf{p}_{3}\right)\tilde{V}_{--+}^{b_{1}ab_{3}}\left(\mathbf{p}_{1},\mathbf{q},\mathbf{p}_{3}\right)\tilde{\Omega}_{2}^{ab_{2}b_{4}}\left(\mathbf{q};\mathbf{p}_{2},\mathbf{p}_{4}\right)\\
+\frac{1}{2}\delta^{3}\left(\mathbf{q}+\mathbf{p}_{1}+\mathbf{p}_{4}\right)\tilde{V}_{--+}^{b_{1}ab_{4}}\left(\mathbf{p}_{1},\mathbf{q},\mathbf{p}_{3}\right)\tilde{\Omega}_{2}^{ab_{2}b_{3}}\left(\mathbf{q};\mathbf{p}_{2},\mathbf{p}_{3}\right)\\
+\delta^{3}\left(\mathbf{q}+\mathbf{p}_{1}+\mathbf{p}_{2}\right)\tilde{V}_{--+}^{b_{1}b_{2}a}\left(\mathbf{p}_{1},\mathbf{p}_{2},\mathbf{q}\right)\tilde{\Psi}_{2}^{ab_{3}b_{4}}\left(\mathbf{q};\mathbf{p}_{3},\mathbf{p}_{4}\right)\Bigg\}\label{eq:MHV_4_1}
\end{multline}
The four-gluon part reads (at lowest order in $B$ fields)
\begin{multline}
\mathcal{L}_{--++}^{\left(\mathrm{LC}\right)}\left[B^{\bullet},B^{\star}\right]=\int d^{3}\mathbf{p}_{1}d^{3}\mathbf{p}_{2}d^{3}\mathbf{p}_{3}d^{3}\mathbf{p}_{4}\delta^{4}\left(\mathbf{p}_{1}+\mathbf{p}_{2}+\mathbf{p}_{3}+\mathbf{p}_{4}\right)\\
\tilde{B}_{b_{1}}^{\star}\left(\mathbf{p}_{1}\right)\tilde{B}_{b_{2}}^{\star}\left(\mathbf{p}_{2}\right)\tilde{B}_{b_{3}}^{\bullet}\left(\mathbf{p}_{3}\right)\tilde{B}_{b_{4}}^{\bullet}\left(\mathbf{p}_{4}\right)\,\tilde{V}_{+--+}^{b_{3}b_{2}b_{1}b_{4}}\left(\mathbf{p}_{3},\mathbf{p}_{2},\mathbf{p}_{1},\mathbf{p}_{4}\right)\,,
\end{multline}
where
\begin{equation}
\tilde{V}_{+--+}^{b_{1}b_{2}b_{3}b_{4}}\left(\mathbf{p}_{1},\mathbf{p}_{2},\mathbf{p}_{3},\mathbf{p}_{4}\right)=g{}^{2}f^{ab_{1}b_{2}}f^{ab_{3}b_{4}}\frac{p_{1}^{+}p_{3}^{+}+p_{2}^{+}p_{4}^{+}}{\left(p_{3}^{+}+p_{4}^{+}\right)^{2}}\,.
\end{equation}
This should be symmetrized as follows:
\begin{multline}
\mathcal{L}_{--++}^{\left(\mathrm{LC}\right)}\left[B^{\bullet},B^{\star}\right]=\int d^{3}\mathbf{p}_{1}d^{3}\mathbf{p}_{2}d^{3}\mathbf{p}_{3}d^{3}\mathbf{p}_{4}\delta^{4}\left(\mathbf{p}_{1}+\mathbf{p}_{2}+\mathbf{p}_{3}+\mathbf{p}_{4}\right)\tilde{B}_{b_{1}}^{\star}\left(\mathbf{p}_{1}\right)\tilde{B}_{b_{2}}^{\star}\left(\mathbf{p}_{2}\right)\tilde{B}_{b_{3}}^{\bullet}\left(\mathbf{p}_{3}\right)\tilde{B}_{b_{4}}^{\bullet}\left(\mathbf{p}_{4}\right)\\
\Bigg\{\frac{1}{2}\tilde{V}_{+--+}^{b_{3}b_{2}b_{1}b_{4}}\left(\mathbf{p}_{3},\mathbf{p}_{2},\mathbf{p}_{1},\mathbf{p}_{4}\right)+\frac{1}{2}\tilde{V}_{+--+}^{b_{3}b_{1}b_{2}b_{4}}\left(\mathbf{p}_{3},\mathbf{p}_{1},\mathbf{p}_{2},\mathbf{p}_{4}\right)\Bigg\}\,,\label{eq:4pointcontribtomhv}
\end{multline}
We are looking for the following form of the new effective MHV coupling:
\begin{equation}
\int d^{3}\mathbf{p}_{1}d^{3}\mathbf{p}_{2}d^{3}\mathbf{p}_{3}d^{3}\mathbf{p}_{4}\,\tilde{\mathcal{V}}_{--++}^{b_{1}b_{2}b_{3}b_{4}}\left(\mathbf{p}_{1},\mathbf{p}_{2},\mathbf{p}_{3},\mathbf{p}_{4}\right)\tilde{B}_{b_{1}}^{\star}\left(\mathbf{p}_{1}\right)\tilde{B}_{b_{2}}^{\star}\left(\mathbf{p}_{2}\right)\tilde{B}_{b_{3}}^{\bullet}\left(\mathbf{p}_{3}\right)\tilde{B}_{b_{4}}^{\bullet}\left(\mathbf{p}_{4}\right)\,,
\end{equation}
where we will use the color decomposition to present the new effective
vertex $\tilde{\mathcal{V}}$ in a simplest form:
\begin{equation}
\tilde{\mathcal{V}}_{--++}^{b_{1}b_{2}b_{3}b_{4}}\left(\mathbf{p}_{1},\mathbf{p}_{2},\mathbf{p}_{3},\mathbf{p}_{4}\right)=\sum_{\sigma\in S_{4}/Z_{4}}\mathrm{Tr}\left(t^{b_{\sigma\left(1\right)}}\dots t^{b_{\sigma\left(4\right)}}\right)\tilde{\mathcal{V}}_{--++}\left(\mathbf{p}_{\sigma\left(1\right)},\mathbf{p}_{\sigma\left(2\right)},\mathbf{p}_{\sigma\left(3\right)},\mathbf{p}_{\sigma\left(4\right)}\right)\,.
\end{equation}
Let us calculate the $\left(1234\right)$ contribution. Converting
(\ref{eq:MHV_4_1}) into the color traces we find that only the first,
fourth and fifth term contributes here. The first term reads
\begin{multline}
\frac{1}{2}\tilde{V}_{--+}^{ab_{2}b_{3}}\left(\mathbf{p}_{\overline{23}},\mathbf{p}_{2},\mathbf{p}_{3}\right)\tilde{\Omega}_{2}^{ab_{1}b_{4}}\left(\mathbf{p}_{\overline{23}};\mathbf{p}_{1},\mathbf{p}_{4}\right)\\
=\frac{1}{2}\left(-ig\right)\, v_{\overline{23}2}^{*}p_{3}^{+}f^{ab_{2}b_{3}}2g'\left(-\frac{p_{1}^{+}}{p_{14}^{+}}\right)\,\frac{1}{2!}\Bigg\{-\frac{1}{\tilde{v}_{41}^{*}}\,\mathrm{Tr}\left(t^{a}t^{b_{4}}t^{b_{1}}\right)+\frac{1}{\tilde{v}_{41}^{*}}\,\mathrm{Tr}\left(t^{a}t^{b_{1}}t^{b_{4}}\right)\Bigg\}\,.
\end{multline}
Since
\begin{equation}
f^{ab_{2}b_{3}}\mathrm{Tr}\left(t^{a}t^{b_{4}}t^{b_{1}}\right)=\frac{1}{i\sqrt{2}}\,\left\{ \mathrm{Tr}\left(t^{b_{1}}t^{b_{2}}t^{b_{3}}t^{b_{4}}\right)-\mathrm{Tr}\left(t^{b_{1}}t^{b_{3}}t^{b_{2}}t^{b_{4}}\right)\right\} 
\end{equation}
we see that the contribution to $\left(1234\right)$ from that term
is
\begin{equation}
d_{1}=\frac{1}{2}\,\frac{1}{i\sqrt{2}}\,\left(-ig\right)\, v_{\overline{23}2}^{*}p_{3}^{+}2\left(\frac{p_{1}^{+}}{p_{14}^{+}}\right)^{2}g^{'}\frac{1}{2!}\,\left(-\frac{p_{14}^{+}}{p_{4}^{+}}\right)\,\frac{1}{\tilde{v}_{14}^{*}}=\frac{1}{2}\left(g'\right)^{2}\,\frac{\left(p_{1}^{+}\right)^{2}p_{3}^{+}}{p_{2}^{+}p_{4}^{+}p_{14}^{+}}\,\frac{\tilde{v}_{2\left(23\right)}^{*}}{\tilde{v}_{14}^{*}}\,.
\end{equation}
Similar, we find from the remaining terms
\begin{gather}
d_{2}=\frac{1}{2}\left(g'\right)^{2}\,\frac{p_{2}^{+}p_{4}^{+}}{p_{14}^{+}p_{23}^{+}}\,\frac{\tilde{v}_{\left(14\right)1}^{*}}{\tilde{v}_{32}^{*}}\,,\\
d_{3}=\frac{1}{i\sqrt{2}}\,\left(-ig\right)\, v_{12}^{*}p_{\overline{12}}^{+}\,\frac{1}{2!}\left(-\frac{p_{34}^{+}}{p_{3}^{+}}\right)\frac{1}{\tilde{v}_{43}^{*}}=-\frac{1}{2}\left(g'\right)^{2}\,\frac{p_{12}^{+}p_{34}^{+}}{p_{2}^{+}p_{3}^{+}}\,\frac{\tilde{v}_{21}^{*}}{\tilde{v}_{43}^{*}}\,.
\end{gather}
The quartic contribution to $\left(1234\right)$ order comes from
the first term in (\ref{eq:4pointcontribtomhv}). Decomposing the
chain $f^{ab_{3}b_{2}}f^{ab_{1}b_{4}}$ into traces we get 
\begin{equation}
d_{4}=-\frac{1}{2}\left(g'\right)^{2}\frac{p_{1}^{+}p_{3}^{+}+p_{2}^{+}p_{4}^{+}}{\left(p_{1}^{+}+p_{4}^{+}\right)^{2}}\,.
\end{equation}
Now, the total MHV vertex reads
\begin{equation}
\tilde{\mathcal{V}}_{--++}\left(\mathbf{p}_{1},\mathbf{p}_{2},\mathbf{p}_{3},\mathbf{p}_{4}\right)=d_{1}+d_{2}+d_{3}+d_{4}\,.
\end{equation}
After some algebra we obtain
\begin{equation}
\tilde{\mathcal{V}}_{--++}\left(\mathbf{p}_{1},\mathbf{p}_{2},\mathbf{p}_{3},\mathbf{p}_{4}\right)=\frac{1}{2}\left(g'\right)^{2}\left(\frac{p_{1}^{+}}{p_{2}^{+}}\right)^{2}\frac{\tilde{v}_{21}^{*4}}{\tilde{v}_{14}^{*}\tilde{v}_{43}^{*}\tilde{v}_{32}^{*}\tilde{v}_{21}^{*}}\,.
\end{equation}
We have also calculated some other color orderings and have similar
structure.

This can be further generalized to any number of $B^{\bullet}$ fields:
\begin{equation}
\tilde{\mathcal{V}}_{--+\dots+}\left(\mathbf{p}_{1},\dots,\mathbf{p}_{n}\right)=\frac{1}{n!}\left(g'\right)^{2}\left(\frac{p_{1}^{+}}{p_{2}^{+}}\right)^{2}\frac{\tilde{v}_{21}^{*4}}{\tilde{v}_{1n}^{*}\tilde{v}_{n\left(n-1\right)}^{*}\tilde{v}_{\left(n-1\right)\left(n-2\right)}^{*}\dots\tilde{v}_{21}^{*}}\,.
\end{equation}

\bibliographystyle{JHEP}
\bibliography{library}

\end{document}